\documentclass[draftcls, onecolumn, journal, letterpaper]{IEEEtran}
\usepackage{color}
\usepackage[latin1]{inputenc}
\usepackage{amsmath}
\usepackage{amsfonts}
\usepackage{amssymb}
\usepackage{epsfig}
\usepackage{units}
\usepackage{nicefrac}
\usepackage[noadjust]{cite}
\usepackage{subfigure}
\usepackage{bm}
\usepackage{verbatim}
\usepackage{pstricks, pst-plot, pst-grad, pstricks-add, pst-node, pstricks-add}
\usepackage[printonlyused]{acronym}	    

\newcommand{\Prob}[0]{\text{Pr}}
\newcommand{\Expected}[1]{\text{E}\left\{#1\right\}}

\newcommand{\D}[1]{\text{d}#1}

\newcommand{\Set}[1]{\mathcal{#1}}
\newcommand{\StronglyTypicalSet}[0]{\Set{A}_\epsilon^{*(n)}}
\newcommand{\Indicator}[0]{1}

\newcommand{\SNRSymb}[0]{\rho}

\newcommand{\RV}[1]{\mathrm{#1}}
\newcommand{\Event}[1]{{#1}}
\newcommand{\Vector}[1]{\underline{#1}}

\newcommand{\Markov}[0]{\leftrightarrow}

\newcommand{\I}[0]{\mathrm{I}}
\newcommand{\h}[0]{\mathrm{h}}
\newcommand{\C}[0]{\mathrm{C}}

\newcommand{\drawniid}[0]{\stackrel{n}{\sim}}

\newcommand{\Vfactor}[0]{\alpha}
\newcommand{\Wfactor}[0]{\beta}

\newcommand{\Vpower}[0]{\Gamma}
\newcommand{\VpowerCov}[0]{{\tilde \Gamma}}

\newcommand{\Matrix}[1]{\mathrm{\mathbf{#1}}}
\newcommand{\MatrixElement}[3]{\left[#3\right]_{#1, #2}}
\newcommand{\PathlossExponent}{\theta}

\newcommand{\Message}[3]{#1_{#2, #3}}
\newcommand{\RVBlock}[3]{#1_{#2}^{#3}}
\newcommand{\MsgBlock}[4]{#1_{#2, #3}^{#4}}
\newcommand{\MsgFactor}[4]{#1_{(#2, #3), #4}}
\newcommand{\MsgPower}[4]{#1_{(#2, #3)}^{#4}}
\newcommand{\MsgCorrelation}[5]{#1_{#2, (#3, #4)}^{#5}}

\newcommand{\DF}[0]{\text{DF}}
\newcommand{\CF}[0]{\text{CF}}
\newcommand{\CN}[0]{\mathcal{CN}}
\newcommand{\bpcu}[0]{bpcu}
\newcommand{\Entropy}[0]{\mathrm{H}}

\newcommand{\ShowFigureMain}[1]{~}
\newcommand{\ShowFigureEnd}[1]{#1}

\newtheorem{lemma}{Lemma}
\newtheorem{theorem}{Theorem}

\newtheorem{corollary}{Corollary}

\newrgbcolor{darkgreen}{0.2 0.6 0.2}
\newrgbcolor{darkred}{0.6 0.2 0.2}
\newrgbcolor{verylightgray}{0.85 0.85 0.85}
\newrgbcolor{phase1}{0.2 0.6 0.2}
\newrgbcolor{phase2}{0.6 0.2 0.2}

\usepackage[english]{babel}

\acrodef{CDMA}[CDMA]{code-division multiple access}
\acrodef{CF}[CF]{compress-and-forward}
\acrodef{DF}[DF]{decode-and-forward}
\acrodef{OFDM}[OFDM]{orthogonal frequency division multiplex}
\acrodef{SINR}[SINR]{signal-to-interference-and-noise ratio}
\acrodef{SNR}[SNR]{signal-to-noise ratio}

\title{Protocols and Performance Limits for Half-Duplex Relay Networks}
\author{
  Peter~Rost,~\IEEEmembership{Member,~IEEE,}~and~Gerhard~Fettweis,~\IEEEmembership{Fellow,~IEEE}%
  \thanks{Manuscript submitted \today. Part of this work has been published in the proceedings of 2008 IEEE Global Communications
  Conference and presented at Asilomar Conference on Signals, Systems, and Computers 2008.}}
\pubid{~}
\begin{document}
\maketitle
  \begin{abstract}
  In this paper, protocols for the half-duplex relay channel are introduced and performance limits are analyzed.
  Relay nodes underly an orthogonality constraint, which prohibits simultaneous receiving and transmitting on the
  same time-frequency resource. Based upon this practical consideration, different protocols are discussed and
  evaluated using a Gaussian system model. 
  For the considered scenarios compress-and-forward based protocols dominate for a wide range of parameters 
  decode-and-forward protocols. In this paper, a protocol with one compress-and-forward and one decode-and-forward based relay is introduced.
  Just as the cut-set bound, which operates in a mode where relays transmit alternately, both relays support each other. 
  Furthermore, it is shown that in practical systems a random channel access provides only marginal performance gains if any.
  \end{abstract}
  \section{Introduction}
      In \cite{Cover.Gamal.TransIT.1979} Cover and El Gamal introduced two basic coding strategies for the three-terminal
      relay channel, which still serve as basis for most relaying protocols today: \ac{DF}, where
      the relay decodes the source message and provides additional, redundant information, and
      \ac{CF}, where the relay node quantizes its channel output and forwards the quantization to the destination.
      Relays using \ac{DF} and \ac{CF} are operating in a \emph{digital} relaying mode, which affects both physical and medium access layer.
      \emph{Analog} relays, by contrast, work as a repeater and simply amplify and forward the received signal. As the 
      former mode offers more flexibility with respect to coding and resource assignment strategies,
      this paper focuses on digital relaying approaches.

      Practical requirements such as power, cost, and space efficiency imply the necessity for small, low-cost terminals, which implement low-complexity protocols.
      These restrictions result in an insufficient separation of transmit and receive path on the same time-frequency resource.
      As the transmission power is usually of much higher order than 
      the received signal power, a severe drop of the \ac{SINR} is caused. Therefore, we assume in the following \emph{half-duplex constrained} relay nodes.
      Generally, the half-duplex constraint requires that each terminal cannot listen and transmit on the same resource simultaneously
      but only on orthogonal resources.

    \subsection{Related Work}
      One of the first works analyzing the capacity of the half-duplex relay channel is \cite{Madsen.VTC.2002},
      which derives upper and lower bounds on the capacity of wireless half-duplex single-relay networks.
      At the same time \cite{Gastpar.Vetterli.INFOCOM.2002} studied
      an upper and lower bound on the capacity of a wireless relay network, where an arbitrary number of relays
      support a single source-destination pair.
      Later, \cite{Khojastepour.Sabharwal.Aazhang.IPSA.2003,Khojastepour.Sabharwal.Aazhang.Globecom.2003}
      derived an upper bound on the achievable rates for general relay networks
      with practical constraints, which are modeled by an arbitrary number of possible states at each node. 
      Kramer \cite{Kramer.2004} introduced the idea of exploiting the randomness of channel states to transmit
      information and hence counteract the half-duplex loss.

      This work particularly discusses alternately transmitting relays nodes in a two-relay network. Alternately transmitting analog
      relay nodes were at first discussed in \cite{Oechtering.Sezgin.2004} and later in \cite{Rankov.Wittneben.2005,Rankov.Wittneben.JSAC.2007},
      where an interference cancellation employed at the destination was introduced. In order to overcome the inter-relay interference,
      \cite{Ribeiro.Cai.Giannakis.ICASSP.2006,Ribeiro.Cai.Giannakis.TransWCom.2006} analyzed alternately
      transmitting relay nodes for systems using \ac{CDMA} under the assumption of
      perfect separation on the inter-relay link. In \cite{Wang.Fan.Krikidis.Thompson.Poor.arxiv.2008} the authors propose a 
      scheme based on superposition coding, which explicitly exploits the inter-relay interference to improve
      diversity and multiplexing gain.

    \subsection{Contribution and Outline of this Work}
      In \cite{Rost.Fettweis.TransIT.2007} a general framework for the full-duplex multiple relay channel has been introduced and combined
      the ideas of partial \ac{DF} and \ac{CF}. 
      This paper extends \cite{Rost.Fettweis.TransIT.2007} by applying known protocols to a half-duplex multiple relay channel
      and introducing new approaches. Among others, we investigate the benefits of a random channel access \cite{Kramer.2004}
      in a network of half-duplex nodes. Additionally, a new regular encoding \ac{CF} strategy is introduced, which overcomes 
      some of the drawbacks of the separation of source and channel coding. In addition, we derive a protocol with alternately transmitting 
      relay nodes, which also considers a direct link compared to the diamond channel \cite{Xue.Sandhu.TransIT.2007}.

      After an introduction of the notation, definitions, and system model in Section \ref{sec:system.model}, we discuss the individual
      approaches in Section \ref{sec:half.duplex:dmc}. Results for these approaches are discussed in Section \ref{sec:half.duplex:results}
      and the paper is concluded in Section \ref{sec:half.duplex:outro}.

  \section{Notations, Definitions, and System Model}\label{sec:system.model}
    In this paper, we use non-italic uppercase letters
    $\RV{X}$ to denote random variables, and italic letters ($N$ or $n$) to denote real or complex-valued scalars.
    Ordered sets are denoted by $\Set{X}$, the cardinality of an ordered set is denoted by $\left\|\Set{X}\right\|$
    and $\left[b : b+k\right]$ is used to denote the ordered set of numbers $b, b+1, \cdots, b+k$.
    Let $\RV{X}_l$ be a random variable parameterized using $l$. Then 
    $\Vector{\RV{X}}_{\Set{C}}$ denotes the vector of all $\RV{X}_l$ with $l\in\Set{C}$ (this applies
    similarly to sets of events). 
    Matrices are denoted by boldface uppercase letters $\Matrix{K}$ and the element in
    the $i$-th row and $j$-th column of matrix $\Matrix{K}$ is denoted by $\MatrixElement{i}{j}{\Matrix{K}}$.
    Furthermore, $\I\left(\RV{X}; \RV{Y} | \RV{Z}\right)$ denotes the mutual information between random variables
    $\RV{X}$ and $\RV{Y}$ given $\RV{Z}$ and $\C\left(x\right)$ abbreviates $\log\left(1 + x\right)$.
    All logarithms are taken to base $2$.

    This paper considers a network of $N+2$ nodes: the source node $s=0$, the set of $N$ relays $\Set{R}:=[1;N]$, and 
    the destination node $d=N+1$.
    With set $\Set{R}$ we express an arbitrary numbering of all relay nodes, which is subject to an optimization 
    (which is not explicitly noted in the following presentation).
    We focus on a Gaussian channel setup where $d_{l', l}$ is the distance between nodes $l'$ and $l\neq l'$ and $\PathlossExponent$ is the path loss exponent,
    such that the gain factor between both nodes is given by $h_{l', l}= d_{l',l}^{-\theta/2}$.
    The channel input at node $l$ and time instances $t\in[1; n]$ is given by the $n$-length sequence of complex Gaussian 
    r.v.s $\left\{\RVBlock{\RV{X}}{l}{t}\right\}_{t=1}^n$ with zero mean and variance $P_l$, denoted by $\RV{X}_{l}\drawniid\CN\left(0, P_l\right)$. 
    Node states are denoted by $\Set{M}_s, \Set{M}_1,\dots\Set{M}_N$ with $\Set{M}_l\in\{L, T\}$, and $L, T$  representing the listening and transmitting state.
    The channel output at node $l\in[1;N+1]$ and time instance $t\in[1;n]$ is given by
    \begin{equation}
      \RVBlock{\RV{Y}}{l}{t} = \Indicator\left(\RVBlock{\RV{M}}{l}{t} = L\right) \cdot \left(\sum\limits_{l'\in[0;N]\setminus l} h_{l', l}\RVBlock{\RV{X}}{l'}{t} + \RVBlock{\RV{Z}}{l}{t}\right),
    \end{equation}
    where $\Indicator(\cdot)$ is the indicator function returning $1$ if its argument is true and $0$ otherwise, and
    $\RV{Z}_l\drawniid\CN\left(0, N_l\right)$ is additive white Gaussian noise.  
    From the orthogonality constraint follows that $\left(\RVBlock{\RV{M}}{l}{t}=L\right)\rightarrow\left(\RVBlock{\RV{X}}{l}{t}=0\right)$ and
    $\left(\RVBlock{\RV{M}}{l}{t}=T\right)\rightarrow\left(\RVBlock{\RV{Y}}{l}{t}=0\right)$.
    This implies that each node $l\in[0;N]$ must fulfill the power constraint 
    $\left(\RVBlock{\RV{M}}{l}{t}=T\right)\rightarrow\Expected{\left|\RVBlock{\RV{X}}{l}{t}\right|^2} = P_l$, which requires that each node
    can switch from transmit to receive state in arbitrarily short time. In contrast to bursty relay approaches where power is concentrated 
    on a small portion of the overall block, we define a peak power and do not normalize the overall spent energy.

  \section{Protocols for the Half-Duplex Channel}\label{sec:half.duplex:dmc}
    For the sake of readability and comprehensibility, this section only treats a network with $N=2$ relays while the Appendix
    extends the results to networks with an arbitrary number of relay nodes.

    \subsection{Decode-and-Forward Protocols}\label{sec:half.duplex:df}
      Assume that the states $\RV{M}_l$ of source and relay nodes are chosen randomly. If they are interpreted as a bit pattern,
      we can be exploit them as an additional information carrier \cite{Kramer.2004}. However, in order to obtain a significant gain,
      the system must provide a high granularity of resources. The first protocol class, which is considered in this paper, is an
      application of \ac{DF} and the idea of randomized channel access to the half-duplex multiple-relay channel.
      The source intends to communicate a message $\RV{W}_s$, which is mapped to the message tuple $\left(\RV{M}_s, \Message{\RV{U}}{s}{1}, \Message{\RV{U}}{s}{2}, 
      \Message{\RV{U}}{s}{3}\right)$ consisting of the source' state $\RV{M}_s$ and three different, superimposed messages with individual
      rates $\Message{R}{s}{k}$. The first relay only decodes the source state and the first message level $\Message{\RV{U}}{s}{1}$,
      while the second relay additionally decodes the second message level $\Message{\RV{U}}{s}{2}$, and finally the destination needs to decode
      the complete tuple in order to correctly reconstruct the source message. Relay $1$ supports the first message level by transmitting additional,
      redundant information represented by the tuple $\left(\RV{M}_1, \Message{\RV{V}}{1}{1}\right)$. 
      If the source has channel knowledge for the complete network, it can coherently support
      the transmission of relay $1$. Relay $2$ exploits this additional information in order to decode $\left(\RV{M}_s, \Message{\RV{U}}{s}{1}\right)$
      and then also provides additional redundant information for the first two source message levels to the destination node. 
      
      Using the previously introduced notation of the considered Gaussian system model, the channel input at the source and both relays is the following
      superposition of signals:
      \begin{eqnarray}
	\RVBlock{\RV{X}}{s}{t} & = & \Indicator\left(\RVBlock{\RV{M}}{s}{t} = T\right) \sqrt{P_s}\Biggl(\sum\limits_{k=1}^{3}\sqrt{\MsgFactor{\Vfactor}{s}{s}{k}}\MsgBlock{\RV{U}}{s}{k}{t} +
	\Indicator\left(\RVBlock{\RV{M}}{1}{t}=T\right)\sqrt{\MsgFactor{\Vfactor}{s}{1}{1}}\MsgBlock{\RV{V}}{1}{1}{t}
	{+}\sum\limits_{k=1}^2 \Indicator\left(\RVBlock{\RV{M}}{2}{t}=T\right)\sqrt{\MsgFactor{\Vfactor}{s}{2}{k}}\MsgBlock{\RV{V}}{2}{k}{t}
	\Biggr)\\
	\RVBlock{\RV{X}}{1}{t} & = & \Indicator\left(\RVBlock{\RV{M}}{1}{t}=T\right)\sqrt{P_1}
	\left(
	\sqrt{\MsgFactor{\Vfactor}{1}{1}{1}}\MsgBlock{\RV{V}}{1}{1}{t} +
	\Indicator\left(\RVBlock{\RV{M}}{2}{t}=T\right)\sqrt{\MsgFactor{\Vfactor}{1}{2}{1}}\MsgBlock{\RV{V}}{2}{1}{t}
	\right) \\
	\RVBlock{\RV{X}}{2}{t} & = & \Indicator\left(\RVBlock{\RV{M}}{2}{t}=T\right)\sqrt{P_2}
	\left(
	\sqrt{\MsgFactor{\Vfactor}{2}{2}{1}}\MsgBlock{\RV{V}}{2}{1}{t} +
	\sqrt{\MsgFactor{\Vfactor}{2}{2}{2}}\MsgBlock{\RV{V}}{2}{2}{t}
	\right),
      \end{eqnarray}
      where $\MsgFactor{\Vfactor}{l'}{l}{k}$ denotes the fraction of power spent by node $l'$ for the support of message level $k$ sent by relay $l$ and
      is assumed to be constant for all transmission phases, which might result in an average transmit power of node $l$ less than $P_l$. 
      An adaptive power fraction results in an enormous parameter space and in case of non-coherent transmission
      (which appears to be more practically relevant) no power savings are obtained anyway.

      The differential entropy for the channel output $\RV{Y}_{l'}$ if the channel states of nodes $\Set{L}$ are known and 
      $\overline{\Set{L}}=[0;N]\setminus\left\{\Set{L}, l\right\}$ are unknown is denoted by 
      $\MsgPower{\h}{l}{l'}{k}\left(\Event{\Vector{m}}_\Set{L}\right)$ and defined in detailed in Appendix \ref{appendix:proof:theorem:halfduplex:gauss:df}.
      Furthermore, $\Set{P_\text{DF}}$ denotes the set of channel input pdfs, which assign the different power levels $\MsgFactor{\Vfactor}{l'}{l}{k}$ such that
      the power constraints in Section \ref{sec:system.model} are satisfied and assign the probabilities to the node states $\RV{M}_l$ (a more detailed definition
      of $\Set{P}_\text{DF}$ is given in Appendix \ref{appendix:proof:theorem:halfduplex:gauss:df}).
      \begin{theorem}\label{theorem:halfduplex:gauss:df}
	The achievable rates for the previously described partial \ac{DF} protocol are given by
	\begin{equation}
	  R = \sup\limits_{p\in\Set{P_\text{DF}}}\left(\Message{R}{s}{1} + \Message{R}{s}{2} + \Message{R}{s}{3}\right),\label{eq:halfduplex:gauss:df:30}
	\end{equation}
	with the individual rate constraints
	\begin{eqnarray}
	  \Message{R}{s}{1} & \leq & \min\Bigl\{\MsgPower{Q}{s}{1}{1}\left(\Set{L}_1\right),
	      \MsgPower{Q}{s}{2}{1}\left(\Set{L}_1\right) + \MsgPower{Q}{1}{2}{1}\left(\Set{L}_{2}\right), \nonumber \\
	      & & \quad\quad\quad \MsgPower{Q}{s}{d}{1}\left(\Set{L}_1\right) + \MsgPower{Q}{1}{d}{1}\left(\Set{L}_{2}\right) + \MsgPower{Q}{2}{d}{1}\left(\Set{L}_{d}\right)
	    \Bigr\},\label{eq:halfduplex:gauss:df:31} \\
	  \Message{R}{s}{2} & \leq & 
	    \min\left\{
	      \MsgPower{Q}{s}{2}{2}\left(\Set{L}_0\right),
	      \MsgPower{Q}{s}{d}{2}\left(\Set{L}_0\right) + \MsgPower{Q}{2}{d}{2}\left(\Set{L}_2\right)
	    \right\} \\
	  \Message{R}{s}{3} & \leq & \MsgPower{Q}{s}{d}{3}\left(\Set{L}_0\right).
	\end{eqnarray}
	where $\Set{L}_l = [l :  2]$ is the set of nodes for which the state is known.
	The mutual information function $\MsgPower{Q}{l}{l'}{k}\left(\Set{L}\right)$ is given by
	\begin{align}
	  \MsgPower{Q}{l}{l'}{1}\left(\Set{L}\right) & = 
	  \sum\limits_{\substack{\Event{\Vector{m}}_{\Set{L}}\in\Set{M}_{\Set{L}}:\\
	  \Event{m}_{l'}=L}} p\left(\Event{\Vector{m}}_{\Set{L}}\right) \Bigl(\MsgPower{\h}{l}{l'}{0}\left(\Event{\Vector{m}}_\Set{L}\right) 
	  {-}\nonumber \\
	  &\quad\sum\limits_{\Event{m}_l\in\Set{M}_l} p\left(\Event{m}_l | \Event{\Vector{m}}_{\Set{L}}\right) \MsgPower{\h}{l}{l'}{1}\left(\Event{\Vector{m}}_{\left\{\Set{L}, l\right\}}\right)\Bigr), \label{eq:halfduplex:gauss:df:32}\\
	  \MsgPower{\mathrm{Q}}{l}{l'}{k}\left(\Set{L}\right) & = 
	  \sum\limits_{\substack{\Event{\Vector{m}}_{\Set{L}}\in\Set{M}_{\Set{L}}:\\ \Event{m}_{l'} = L}}p\left(\Event{\Vector{m}}_{\Set{L}}\right)
	  \left(\MsgPower{\h}{l}{l'}{k-1}\left(\Event{\Vector{m}}_\Set{L}\right) - \MsgPower{\h}{l}{l'}{k}\left(\Event{\Vector{m}}_\Set{L}\right)\right).\label{eq:halfduplex:gauss:df:33}
	\end{align}
      \end{theorem}
      \begin{proof}
	The theorem is an application of the more general Theorem \ref{appendix:theorem:half.duplex:20} given in Appendix \ref{appendix:proof:theorem:halfduplex:gauss:df}
	and describing the achievable rates for an arbitrary number of relay nodes.
      \end{proof}
      Eq. (\ref{eq:halfduplex:gauss:df:31}) is the minimum of the three cuts for the first source message level: 
      from source to relay $1$, from source and relay $1$ to relay $2$, and from source, relay $1$ and $2$ to the destination. 
      However, we can see that the transmit-diversity gain is increasing with the number of nodes, which already decoded the message.
      The function $\MsgPower{Q}{l}{l'}{k}\left(\Set{L}\right)$ gives the mutual information in the half duplex channel between nodes $l$ and $l'$ and
      message level $k$. In case of a fixed channel access the channel state of all nodes is known, hence
      only $\MsgPower{Q}{l}{l'}{k}\left([0;2]\right)$ is used. Nonetheless, in case of a random channel access we face the difficulty
      to evaluate an integral of the form
      \begin{equation}
	\int\limits_0^\infty \left(\sum\limits_k \frac{a_k\lambda_k}{\pi}e^{-\lambda_k y}\right)\log\left(\sum\limits_k \frac{a_k\lambda_k}{\pi}e^{-\lambda_k y}\right)\text{d}y,
	\label{eq:halfduplex:gauss:df:300}
      \end{equation}
      which can only be loosely upper and lower bounded (using log-sum inequality and Jensen's inequality). Therefore, the results presented
      in Section \ref{sec:half.duplex:results} follow from a numerical evaluation of this integral.

      If we use only a subset of $\Set{P}_\text{DF}$, which includes only those input pdfs with deterministic state probabilities, the previous theorem
      gives the achievable rates for a fixed transmission schedule. Such a schedule is preferable as it needs no additional complexity and
      hardware to detect the node states (only wireline based networks can support this detection at reasonable complexity). 
      Furthermore, consider an \ac{OFDM} system with groups of $F_c$ subcarriers, which are assigned to users. 
      Then the actual advantage through a random channel access is reduced by a factor $1/F_c$, which makes a fixed transmission schedule an even more preferable choice.
      Finally, consider a multihopping approach with reuse factor $\nicefrac{1}{k}$ \cite{Herhold.Zimmermann.Fettweis.JCN.2005}. This implies
      that one resource is only occupied by $\nicefrac{1}{k}$-th of all nodes, or that one node only uses $\nicefrac{1}{k}$-th of
      the available resources. Applied to a half-duplex relay network this implies that all $p\in\Set{P}_\text{DF}$ 
     must satisfy
      \begin{equation*}
	\forall l\in[0;N]: \Prob\left(\Event{m}_{l} = T \Biggl| \sum\limits_{j\in[0;N]} \Indicator\left(\Event{m}_{j} = T\right) > \left\lfloor\frac{1}{k}(N+1)\right\rfloor\right) = 0.
      \end{equation*}

    \subsection{Compress-and-Forward Protocols}\label{sec:half.duplex:cf}
      In this section, we discuss a \ac{CF} based approach, where, by contrast to \ac{DF}, relay nodes need not to decode
      the source messages but forward their quantized channel output. Due to the fact that the channel input of each
      relay cannot be predicted, we assume a fixed transmission schedule known at each node. 
      In comparison to previous work, we introduce a \ac{CF} approach using joint source-channel coding to
      overcome the drawbacks of separating both \cite{Gastpar.PhD.2002}.

      More specifically, both relays $l\in[1;2]$ create for \emph{each} possible quantization $\hat{\RV{Y}}_l$ a \emph{corresponding} broadcast message $\RV{X}_l$. 
      Depending on the channel output $\RV{Y}_1$ in block $b$, relay $1$ searches for a jointly typical quantization and then transmits in block $b+2$
      the corresponding broadcast message. Similarly, relay $2$ transmits in block $b+1$ the broadcast message corresponding to the quantization in block $b$.
      This shift of blocks allows the destination to use quantizations of relay $2$ for the decoding of the broadcast message transmitted by relay $1$.
      To decode the quantization index for instance of relay $2$ for block $b$, the destination has to create two sets.
      The first set contains all those indices of broadcast messages which are jointly typical with $\RV{Y}_d$ in block $b+1$ and the second set of those indices
      such that the quantization is jointly typical with $\RV{Y}_d$ in block $b$. Using the intersection of both sets is then the index of the correct quantization
      of relay $2$ in block $b$. Similarly, the destination proceeds to decode the quantization of the first relay node where $\hat{{\RV{Y}}}_{2}$ is exploited 
      to improve the quality of $\hat{{\RV{Y}}}_{1}$ and $\RV{X}_1$.  

      \ShowFigureMain{%
      \begin{figure}
	\centering
	\subfigure[Irregular encoding]{\begingroup
\unitlength=1mm
\begin{picture}(87, 40)(0, 0)
  \psset{xunit=1mm, yunit=1mm, linewidth=0.1mm}
  \psset{arrowsize=3pt 4, arrowlength=1.4, arrowinset=.4}\psset{axesstyle=frame}%
  \psset{shadowcolor=gray, shadowsize=1mm}%

  \rput(-3, -10){%
    \rput(5, 30){\rnode{Source}{$\RV{Y}$}}%
    \rput(15, 30){\rnode{Encoder}{\psframebox[framearc=.2, shadow=true]{\rotateleft{Encoder}}}}%
    \ncline{->}{Source}{Encoder}%
    \rput(35, 30){\rnode{Channel}{\psframebox[framearc=.2, shadow=true]{\rotateleft{$p\left(\Event{y}_{1}, \Event{y}_2 | \Event{w}\right)$}}}}%
    \ncline{->}{Encoder}{Channel}\ncput*{$\RV{W}$}%
    \pnode(38.3, 36){Channel1}%
    \pnode(38.3, 24){Channel2}%
%
    \rput(55, 36){\rnode{G11}{\psframebox[framearc=.2, shadow=true]{$g_{2}$}}}%
    \ncline{->}{Channel1}{G11}\ncput*{$\RV{Y}_{2}$}%
    \rput(75, 36){\rnode{G12}{\psframebox[framearc=.2, shadow=true]{$g_{2}'$}}}%
    \ncline{->}{G11}{G12}\ncput*{$\hat{\RV{W}}$}%
    \rput(75, 48){\rnode{Side1}{$\RV{Y}_2$}}%
    \ncline{->}{Side1}{G12}%
    \rput(88, 36){\rnode{Result}{$\hat{\hat{\RV{Y}}}$}}%
    \ncline{->}{G12}{Result}%
    \rput(55, 24){\rnode{G21}{\psframebox[framearc=.2, shadow=true]{$g_{1}$}}}%
    \ncline{->}{Channel2}{G21}\ncput*{$\RV{Y}_{1}$}%
    \rput(75, 24){\rnode{G22}{\psframebox[framearc=.2, shadow=true]{$g_{1}'$}}}%
    \ncline{->}{G21}{G22}\ncput*{$\hat{\RV{W}}$}%
    \rput(75, 12){\rnode{Side2}{$\RV{Y}_1$}}%
    \ncline{->}{Side2}{G22}%
    \rput(88, 24){\rnode{Result}{$\hat{\hat{\RV{Y}}}$}}%
    \ncline{->}{G22}{Result}%
    }
  
\end{picture}
\endgroup}
\subfigure[Regular encoding]{\begingroup
\unitlength=1mm
\begin{picture}(67, 40)(0, 0)
  \psset{xunit=1mm, yunit=1mm, linewidth=0.1mm}
  \psset{arrowsize=3pt 4, arrowlength=1.4, arrowinset=.4}\psset{axesstyle=frame}%
  \psset{shadowcolor=gray, shadowsize=1mm}%

  \rput(-3, -10){%
    \rput(5, 30){\rnode{Source}{$\RV{Y}$}}%
    \rput(15, 30){\rnode{Encoder}{\psframebox[framearc=.2, shadow=true]{\rotateleft{Encoder}}}}%
    \ncline{->}{Source}{Encoder}%
    \rput(35, 30){\rnode{Channel}{\psframebox[framearc=.2, shadow=true]{\rotateleft{$p\left(\Event{y}_{1}, \Event{y}_2 | \Event{w}\right)$}}}}%
    \ncline{->}{Encoder}{Channel}\ncput*{$\RV{W}$}%
    \pnode(38.3, 36){Channel1}%
    \pnode(38.3, 24){Channel2}%
%
    \rput(55, 36){\rnode{G12}{\psframebox[framearc=.2, shadow=true]{$g_{2}$}}}%
    \ncline{->}{Channel1}{G12}\ncput*{$\RV{Y}_{2}$}%
    \rput(55, 48){\rnode{Side1}{$\RV{Y}_2$}}%
    \ncline{->}{Side1}{G12}%
    \rput(68, 36){\rnode{Result}{$\hat{\hat{\RV{Y}}}$}}%
    \ncline{->}{G12}{Result}%
    \rput(55, 24){\rnode{G22}{\psframebox[framearc=.2, shadow=true]{$g_{1}$}}}%
    \ncline{->}{Channel2}{G22}\ncput*{$\RV{Y}_{1}$}%
    \rput(55, 12){\rnode{Side2}{$\RV{Y}_1$}}%
    \ncline{->}{Side2}{G22}%
    \rput(68, 24){\rnode{Result}{$\hat{\hat{\RV{Y}}}$}}%
    \ncline{->}{G22}{Result}%
    }
  
\end{picture}
\endgroup}
	\caption{Two different strategies for \ac{CF} with multiple receivers.}
	\label{fig:halfduplex:regular.CF}
      \end{figure}
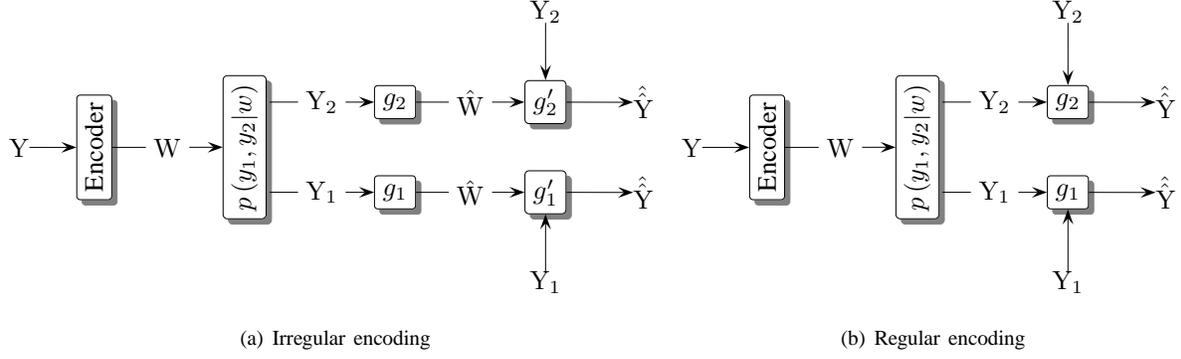}
      Due to the \emph{regular encoding}, we are able to alleviate
      the drawbacks of source-channel coding separation. The difference of irregular and regular encoding is illustrated in Fig. \ref{fig:halfduplex:regular.CF}.
      Consider for instance the multiple-description problem \cite{Gamal.Cover.TransIT.1982} and assume two receivers and an \emph{irregular encoding}.
      In this case, both decoders are forced to decode at first the broadcast and then the quantization messages, where
      the weaker source-to-destination link is the bottleneck for the achievable broadcast message rates. Now consider a regular encoding scheme. This time,
      the worse source-to-destination link can be balanced out with stronger side information while the better
      source-to-destination link allows for weaker side information. 

      Using the previously introduced Gaussian system model, the channel input at both relays using \ac{CF} is given by
      \begin{eqnarray}
	\RVBlock{\RV{X}}{1}{t} & = & \Indicator\left(\RVBlock{\RV{M}}{1}{t}=T\right)\sqrt{P_1\Message{\Wfactor}{1}{1}}\MsgBlock{\RV{W}}{1}{1}{t} \\
	\RVBlock{\RV{X}}{2}{t} & = & \Indicator\left(\RVBlock{\RV{M}}{2}{t}=T\right)\sqrt{P_2\Message{\Wfactor}{2}{1}}\MsgBlock{\RV{W}}{2}{1}{t}
      \end{eqnarray}
      with the broadcast messages $\Message{\RV{W}}{1}{1}, \Message{\RV{W}}{2}{1}\drawniid\mathcal{CN}\left(0, 1\right)$ and
      their fractional power factors $\Message{\Wfactor}{1}{1}, \Message{\Wfactor}{2}{1}$.
      Since we only use one source message level, the channel input at the source node is simply given by
      \begin{equation}
	\RVBlock{\RV{X}}{s}{t} = \Indicator\left(\RVBlock{\RV{M}}{s}{t} = T\right) \sqrt{P_s\MsgFactor{\Vfactor}{s}{s}{1}}\MsgBlock{\RV{U}}{s}{1}{t}.
      \end{equation}
      In addition, we need the following auxiliary variables describing receive power variances:
      \begin{itemize}
	\item The received power at node $l$, which originates from the transmission of nodes $\Set{L}\subset[0;N]$ is given by
	  $\MsgPower{\Vpower}{\Set{L}}{l}{~}\left(\Event{\Vector{m}}_{[0;N]}\right)$. 
	\item The covariance of the channel outputs at nodes $l$ and $l'$ for the transmission sent by nodes $\Set{L}$ is given by
	  $\MsgCorrelation{\VpowerCov}{\Set{L}}{l}{l'}{~}\left(\Event{\Vector{m}}_{[0;N]}\right)$.
	\item
	  Finally, let $\Matrix{K}_{\Set{L}, \Set{L'}}\left(\Event{\Vector{m}}_{[0;N]}\right)$ be the covariance matrix
	  of all quantizations at nodes $l\in\Set{L}'$ and the destination's channel output when all messages from nodes $\Set{L}^c = [0;2]\setminus\Set{L}$
	  are known.
      \end{itemize}
      For the benefit of readability, the arguments of $\MsgPower{\Vpower}{\Set{L}}{l}{~}\left(\Event{\Vector{m}}_{[0;N]}\right)$ and
      $\MsgCorrelation{\VpowerCov}{\Set{L}'}{l}{l'}{~}\left(\Event{\Vector{m}}_{[0;N]}\right)$ are dropped.
      \begin{theorem}\label{theorem:half.duplex.random:19}
	The regular \ac{CF} approach achieves any rates
	\begin{equation}
	  R \leq \sup\limits_{p}
	  \sum\limits_{\Event{\Vector{m}}_{[0 : 2]}\in\Set{M}_{[0 : 2]}}p\left(\Event{\Vector{m}}_{[0 : 2]}\right)
	  \log\left(\frac{\left\| \Matrix{K}_{s, [1 : 2]}\left(\Event{\Vector{m}}_{[0 : 2]}\right)\right\|}{\left\| \Matrix{K}_{\emptyset, [1 : 2]}\left(\Event{\Vector{m}}_{[0 : 2]}\right)\right\|}\right),\label{eq:halfduplex:gauss:cf:10}
	\end{equation}
	subject to 
	\begin{multline}
	  \sum\limits_{{\{\Event{\Vector{m}}_{[0 : 2]}\in\Set{M}_{[0 : 2]}:\Event{m}_{2-0}=L\}}}p\left(\Event{\Vector{m}}_{[0 : 2]}\right)
	  \log\left(\frac{\left\|\Matrix{K}_{[0 :  1], 2}\right\|}{\Message{N}{2}{1} \left\| \Matrix{K}_{[0 :  1], \emptyset} \right\|}\right)
	  \\
	  \leq \sum\limits_{{\{\Event{\Vector{m}}_{[0 : 2]}\in\Set{M}_{[0 : 2]}: \Event{m}_{2}=T\}}}p_{\RV{M}_{[0 : 2]}}\left(\Event{\Vector{m}}_{[0 : 2]}\right)
	  \log\left(\frac{\left\|\Matrix{K}_{[0 :  2], \emptyset}\right\|}
	  {\left\|\Matrix{K}_{[0 :  1], \emptyset}\right\|}\right)\label{eq:halfduplex:gauss:cf:15}
	\end{multline}
	for the quantization at relay $2$ and
	\begin{multline}
	  \sum\limits_{\left\{\Event{\Vector{m}}_{[0 : 2]}\in\Set{M}_{[0 : 2]}: \Event{m}_{2-1}=L\right\}}p\left(\Event{\Vector{m}}_{[0 : 2]}\right)
	  \Biggl[
	  \C\left(\frac{\MsgPower{\Vpower}{2}{1}{~}}{\MsgPower{\Vpower}{0}{1}{~}+\Message{N}{1}{1}+N_{1}}\right) +
	  \log\left(\frac{\left\|\Matrix{K}_{s, [1 : 2]}\right\|}{\Message{N}{1}{1} \left\| \Matrix{K}_{s, 2} \right\|}\right)
	  \Biggr]
	  \\
	  \leq \sum\limits_{\left\{\Event{\Vector{m}}_{[0 : 2]}\in\Set{M}_{[0 : 2]}: \Event{m}_{1}=T\right\}}p_{\RV{M}_{[0 : 2]}}\left(\Event{\Vector{m}}_{[0 : 2]}\right)
	  \log\left(\frac{\left\|\Matrix{K}_{[0 :  1], 2}\right\|}
	  {\left\|\Matrix{K}_{s, 2}\right\|}\right).\label{eq:halfduplex:gauss:cf:20}
	\end{multline}
	for the quantization at relay $1$.
      \end{theorem}
      \begin{proof}
	The theorem is an application of the more general Theorem \ref{appendix:theorem:half.duplex:21} given in Appendix \ref{appendix:proof:theorem:halfduplex:gauss:cf}
	and describing the achievable rates for an arbitrary number of relay nodes.
      \end{proof}
      Eq. (\ref{eq:halfduplex:gauss:cf:15}) and (\ref{eq:halfduplex:gauss:cf:20}) reflect the side condition on the
      quantization quality. The right hand side of both inequalities gives the channel coding constraint, and
      the left hand side gives the source coding constraint. Both quantization noise variances must be determined iteratively
      in descending order, starting with (\ref{eq:halfduplex:gauss:cf:15}).

    \subsection{Alternately Transmitting Relays}\label{sec:half.duplex:yarp}
      This section introduces a protocol for two alternately transmitting relay nodes of which
      one relay node is transmitting while the other relay is listening. By contrast to the previous two protocols, we apply
      a mixed approach where one relay supports the source using \ac{CF} and one node employs \ac{DF}. The major
      bottleneck in such a network is the inter-relay interference, which can, however,  
      be exploited if the destination uses the \ac{CF} transmission to decode not only the source but also the \ac{DF}-relay transmission. 
      Nonetheless, we still face the problem that the \ac{DF} relay is interfered by the \ac{CF} relay, which we mitigate using
      the previously introduced regular encoding approach, i.\,e., both \ac{DF} relay and the destination decode the transmission of the
      \ac{CF} relay but use different side information.

      \ShowFigureMain{%
      \begin{figure}
	\centering
	\begingroup
\unitlength=1mm
\begin{picture}(101, 38)(0, 0)
  \psset{xunit=1mm, yunit=1mm, linewidth=0.1mm}
  \psset{arrowsize=3pt 4, arrowlength=1.4, arrowinset=.4}

  \rput(17, 28)
  {
  \rput[r](15, 0){\textcolor{phase1}{$\begin{array}{c}\text{Phase 1, $[1; n_1]$:}\\\text{($\Event{x}_{s,1}$ with rate $R_{\DF}$)}\end{array}$}}
    \cnodeput(20, 0){S}{$s$}
    \cnodeput(40, 0){R1}{$1$}
    \cnodeput(60, 0){R2}{$2$}
    \cnodeput(80, 0){D}{$d$}
    \ncline[linecolor=phase1]{->}{S}{R1}\nbput{$\Event{x}_{s,1}$}
    \ncline[linestyle=dashed, linecolor=phase2]{<-}{R1}{R2}\nbput{$\Event{x}_2$}
    \ncline[linecolor=phase2]{->}{R2}{D}\nbput{$\Event{x}_2$}
    \ncarc[arcangle=20, linestyle=solid, linecolor=phase1]{->}{S}{D}\naput{$\Event{x}_{s,1}$}
  }

  \rput(17, 8)
  {
  \rput[r](15, 0){\textcolor{phase2}{$\begin{array}{c}\text{Phase 2, $[n_1+1; n]$:}\\\text{($\Event{x}_{s,2}$ with rate $R_{\CF}$)}\end{array}$}}
    \cnodeput(20, 0){S}{$s$}
    \cnodeput(40, 0){R1}{$1$}
    \cnodeput(60, 0){R2}{$2$}
    \cnodeput(80, 0){D}{$d$}
    \ncline[linecolor=phase1]{->}{R1}{R2}\nbput{$\Event{x}_1$}
    \ncarc[arcangle=35, linecolor=phase2, linestyle=solid]{->}{S}{D}\nbput{$\Event{x}_{s,2}$}
    \ncarc[arcangle=20, linecolor=phase2, linestyle=solid]{->}{S}{R2}
    \ncarc[arcangle=-40, linecolor=phase1, linestyle=solid]{->}{R1}{D}
  }
\end{picture}
\endgroup
	\caption[Example for a half-duplex channel with two alternately transmitting relay nodes.]
	{Example for a half-duplex channel with two alternately transmitting relay nodes.
	The solid lines indicate actual information exchange while the dashed line indicates the interfering
	transmission from node $2$ to $1$.}
	\label{figure:halfduplex.random:alternately_transmitting}
      \end{figure}
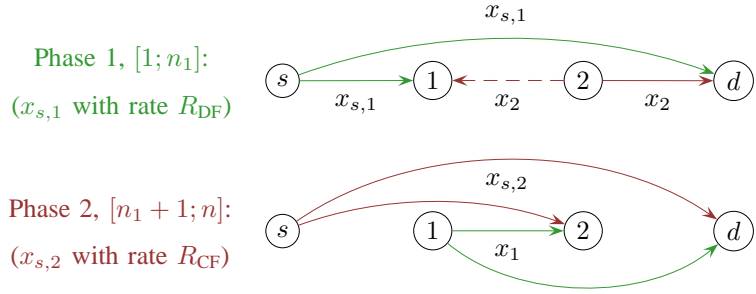}
      Fig. \ref{figure:halfduplex.random:alternately_transmitting} illustrates the setup:
      the overall transmission period is divided into two phases with probabilities $p_1$ (phase $1$ in Fig. \ref{figure:halfduplex.random:alternately_transmitting})
      and $p_2$ (phase $2$ in Fig. \ref{figure:halfduplex.random:alternately_transmitting}) such that
      \begin{align*}
	& \Prob\left(\RV{M}_s = T\right) = 1, &
	& \Prob\left(\RV{M}_1 = T | \RV{M}_2 = L\right) = 1, &
	& \Prob\left(\RV{M}_1 = L | \RV{M}_2 = T\right) = 1, \\
	& p_1 = \Prob\left(\RV{M}_1 = T\right), &
	& p_2 = \Prob\left(\RV{M}_2 = T\right) = 1 - p_1,
      \end{align*}
      with each phase of length $n_1=n\cdot p_1$ and $n_2=n\cdot p_2$, respectively. The source message is divided in two parts
      $\RV{X}_{s,1}$ and $\RV{X}_{s,2}$ with rates $R_\DF$ and $R_\CF$, respectively. 
      \ShowFigureMain{%
      \begin{figure}
	\centering
	\begingroup
\unitlength=1mm
\begin{picture}(112, 49)(0, 0)
  \psset{xunit=1mm, yunit=1mm, linewidth=0.1mm}
  \psset{arrowsize=3pt 4, arrowlength=1.4, arrowinset=.4}

  \rput(0, 7){%
    \rput[c](7, 13){Relay $2$:}
    \rput(15, 8)
    {%
      \psframe(0, 0)(40, 10)
      \rput[c](20, 5){\textcolor{phase2}{$\RV{x}_2(\RVBlock{q}{2}{b})$}}
      \psbrace[rot=90, linewidth=0.1mm, ref=t, nodesepB=-2mm](0, 0)(40, 0){\textcolor{phase1}{$\begin{array}{c}n_1 = n\cdot p_1\\\text{(Phase 1)}\end{array}$}}
      \psframe(42, 0)(97, 10)
      \rput[c](69.5, 5){$\RV{y}_2(b)\mapsto\hat{\RV{y}}_2(\RVBlock{q}{2}{b+1})\mapsto\textcolor{phase2}{\RV{x}_2(\RVBlock{q}{2}{b+1})}$}
      \psbrace[rot=90, linewidth=0.1mm, ref=t, nodesepB=-2mm](42, 0)(97, 0){\textcolor{phase2}{$\begin{array}{c}n_2 = n\cdot p_2\\\text{(Phase 2)}\end{array}$}}
    }

    \rput[c](7, 25){Relay $1$:}
    \rput(15, 20)
    {%
      \psframe(0, 0)(40, 10)
      \rput[c](20, 5){$\RV{y}_1(b)\mapsto\textcolor{phase1}{\RV{x}_1(\RVBlock{q}{1}{b})}$}
      \psframe(42, 0)(97, 10)
      \rput[c](69.5, 5){\textcolor{phase1}{$\RV{x}_1(\RVBlock{q}{1}{b})$}}
    }

    \rput[c](7, 37){Source:}
    \rput(15, 32)
    {%
      \psframe(0, 0)(40, 10)
      \rput[c](20, 5){\textcolor{phase1}{$x_{s,1}(\MsgBlock{q}{s}{1}{b})$}}
      \psframe(42, 0)(97, 10)
      \rput[c](69.5, 5){\textcolor{phase2}{$\RV{x}_{s,2}(\MsgBlock{q}{s}{2}{b})$}}
    }
  }

\end{picture}
\endgroup
	\caption[Coding structure for alternatively transmitting relay nodes.]
	{Coding structure for the combined strategy with $N=2$ alternately transmitting relays.}
	\label{figure:halfduplex.random:encoding_mixed_yarp}
      \end{figure}
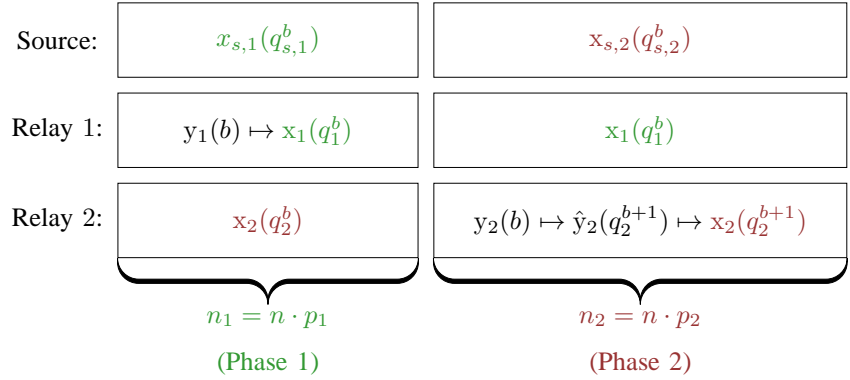}
      Fig. \ref{figure:halfduplex.random:encoding_mixed_yarp} illustrates the coding procedure: relay $2$ supports the source message
      in the second phase of each block using its quantized channel output $\hat{\RV{Y}}_2$ for which in the first phase of the next block the
      corresponding broadcast message is transmitted (again using regular encoding). Relay $1$ decodes this quantization by taking account for
      the fact that it depends on its own transmission signal in the previous block. Alternatively, if the inter-relay channel is weak, relay $1$
      simply treats the transmission as noise. After removing the interference from relay $2$, relay $1$ decodes the source message $\RV{X}_{s,1}$
      and transmits in the next phase redundant information to support it.

      The decoding process starts with decoding the quantization index of relay $2$ for block $b+1$, which contains information for the relay transmission
      supporting the source message of block $b$. It can then use this quantization and its own channel output to decode the source message transmitted in block $b$. 
      After this message is known, the destination decodes $\RV{X}_{s,2}$ for which it uses again the quantization of relay $2$ (after subtracting
      the previously decoded signals) as well as its own channel output.

      Again, we apply this protocol to the Gaussian setup described in Section \ref{sec:system.model}.
      Let the source message use the two individual messages $\RV{X}_{s,1}, \RV{X}_{s,2}\drawniid\CN\left(0, 1\right)$
      of lengths $n_1=p_1\cdot n$ and $n_2=p_2\cdot n$. Then the source channel input is given by
      \begin{equation}
	\RVBlock{\RV{X}}{s}{t} = \sqrt{P_s}\left[1\left(t\leq n_1\right)\sqrt{\MsgFactor{\Vfactor}{s}{s}{1}}\MsgBlock{\RV{X}}{s}{1}{t} +
	    1\left(t>n_1\right)\left(\sqrt{\MsgFactor{\Vfactor}{s}{s}{2}}\MsgBlock{\RV{X}}{s}{2}{t}+\sqrt{\MsgFactor{\Vfactor}{s}{1}{1}}\MsgBlock{\RV{V}}{1}{1}{t}\right)\right].
      \end{equation}
      Relay $1$ supports the source message $\RV{X}_{s,1}$ with $\Message{\RV{V}}{1}{1}\drawniid\CN\left(0, 1\right)$ with rate $R_\DF$,
      such that the channel input is given by
      \begin{equation}
	\RVBlock{\RV{X}}{1}{t} = 1\left(t>n_1\right) \sqrt{P_1\MsgFactor{\Vfactor}{1}{1}{1}}\MsgBlock{\RV{V}}{1}{1}{t}.
      \end{equation}
      Assume that quantization signals at relay $2$ are generated according to $\Message{\hat{\RV{Y}}}{2}{1}\drawniid\CN\left(0, \sigma_{\RV{Y}_2}^2+\Message{N}{2}{1}\right)$.
      Relay $2$ uses the broadcast messages $\Message{\RV{W}}{2}{1}\drawniid\CN\left(0, 1\right)$ (both with codebook size $2^{n\Message{\Delta}{2}{1}}$) such that
      its channel input is given by
      \begin{equation}
	\RVBlock{\RV{X}}{2}{t} = 1\left(t\leq n_1\right) \sqrt{P_2\Message{\Wfactor}{2}{1}}\MsgBlock{\RV{W}}{2}{1}{t}.
      \end{equation}
      In the following theorem, we reuse the definitions of Section \ref{sec:half.duplex:df} and \ref{sec:half.duplex:cf}, e.\,g.,
      the covariance matrices $\Matrix{K}_1 = \Matrix{K}_{\left\{s,1\right\}, 2}\left(\RV{M}_{[1,2]}=\left\{T, L\right\}\right)$
      of the quantized channel output at relay $2$ and the channel output at the destination before the transmission
      of relay $1$ is decoded, and $\Matrix{K}_2 = \Matrix{K}_{s, 2}\left(\RV{M}_{[1,2]}=\left\{T, L\right\}\right)$
      after this transmission is decoded. Now we can define the following corollary of Theorems \ref{theorem:halfduplex:gauss:df} and
      \ref{theorem:half.duplex.random:19}:
      \begin{corollary}\label{theorem:half.duplex.random:21}
	The previously presented combined protocol achieves any rate 
	\begin{equation}
	  R=\sup\limits_{p\in\Set{P}}\left(R_\DF + R_\CF\right)
	\end{equation}
	where the rate achieved by the \ac{DF} phase is constrained by
	\begin{equation}
	  \begin{split}
	    R_{\text{DF}} \leq \min\biggl\{ & p_1\C\left(\frac{\MsgPower{\Vpower}{s}{d}{~}}{N_d}\right) + 
	    p_2\log\left(\frac{\left\|\Matrix{K}_1\right\|}{\left\|\Matrix{K}_2\right\|}\right),
	    p_1\C\left(\frac{\MsgPower{\Vpower}{s}{1}{~}}{N_1}\right)
	    \biggr\},
	  \end{split}\label{eq:theorem:half.duplex.random:2100}
	\end{equation}
	if node $1$ decodes the quantization of node $2$, and
	\begin{equation}
	  \begin{split}
	    R_{\text{DF}} \leq \min\biggl\{ & p_1\C\left(\frac{\MsgPower{\Vpower}{s}{d}{~}}{N_d}\right) + 
	    p_2\log\left(\frac{\left\|\Matrix{K}_1\right\|}{\left\|\Matrix{K}_2\right\|}\right),
	    p_1\C\left(\frac{\MsgPower{\Vpower}{s}{1}{~}}{N_1 + \MsgPower{\Vpower}{2}{1}{~}}\right)
	    \biggr\},
	  \end{split}\label{eq:theorem:half.duplex.random:2110}
	\end{equation}
	otherwise. The rate achieved by the \ac{CF} phase is limited by
	\begin{equation}
	  R_{\text{CF}} \leq p_2 \C\left(\frac{\MsgPower{\Vpower}{s}{2}{~}}{N_2 + \Message{N}{2}{1}} + \frac{\MsgPower{\Vpower}{s}{d}{~}}{N_d}\right),
	  \label{eq:theorem:half.duplex.random:2120}
	\end{equation}
	and subject to a lower bound on the quantization noise:
	\begin{equation}
	  \Message{N}{2}{1} \geq \left(\MsgPower{\Vpower}{\left\{s,1\right\}}{2}{~} + N_2 - \frac{\left(\MsgCorrelation{\VpowerCov}{\left\{s,1\right\}}{2}{d}{~}\right)^2}
	    {\MsgPower{\Vpower}{\left\{s,1\right\}}{d}{~} + N_d}\right)
	  \left(2^{\Message{\hat R}{2}{1}/p_2}-1\right)^{-1}
	  \label{eq:theorem:half.duplex.random:2130}
	\end{equation}
	with the broadcast message rate
	\begin{equation}
	  \Message{\hat R}{2}{1} = p_1 \C\left(\frac{\MsgPower{\Vpower}{2}{d}{~}}{N_d + \MsgPower{\Vpower}{s}{d}{~}}\right), 
	  \label{eq:theorem:half.duplex.random:2140}
	\end{equation}
	and if relay $1$ decodes $\hat{\RV{Y}}_2$ also subject to
	\begin{equation}
	  \Message{N}{2}{1} \geq \left(\MsgPower{\Vpower}{s}{2}{~} + N_2\right)\cdot\left(2^{\Message{\hat R}{2}{2} / p_2} - 1\right)^{-1}
	  \text{ with }
	  \Message{\hat R}{2}{2} = p_1 \C\left(\frac{\MsgPower{\Vpower}{2}{1}{~}}{N_1 + \MsgPower{\Vpower}{s}{1}{~}}\right).
	  \label{eq:theorem:half.duplex.random:2150}
	\end{equation}
      \end{corollary}
      \begin{proof}
	Eq. (\ref{eq:theorem:half.duplex.random:2100}) and (\ref{eq:theorem:half.duplex.random:2110}) are an application of the \ac{DF} rates
	where relay $1$ decodes $\RV{X}_{s, 1}$. The latter of both equations needs to consider the additional interference
	of relay $2$ as its quantization is not decoded. Eq. (\ref{eq:theorem:half.duplex.random:2120})-(\ref{eq:theorem:half.duplex.random:2150})
	follow from an application of the \ac{CF} rates where relay $2$ quantizes the channel output and, in addition to Theorem \ref{theorem:half.duplex.random:19},
	the previously decoded messages of relay $1$ is used as additional side information.
      \end{proof}

  \section{Results and Discussion}\label{sec:half.duplex:results}
    \ShowFigureMain{%
    \begin{figure}
      \centering
      \begingroup
\unitlength=1mm
\begin{picture}(65, 20)(0, 0)

  \psset{xunit=1mm, yunit=1mm, linewidth=0.2mm}

  \rput(2, 5){\cnodeput(0, 0){Source}{$s$}}
  \rput(60, 5){\cnodeput(0, 0){Destination}{$d$}}
  \rput(20, 5){\cnodeput(0, 0){Relay1}{$1$}}
  \rput(42, 5){\cnodeput(0, 0){Relay2}{$2$}}

  \pnode(2, 18){Source1}
  \pnode(2, 11){Source2}
  \pnode(2, 16){Source3}
  \pnode(60, 18){Destination1}
  \pnode(60, 11){Destination2}
  \pnode(60, 16){Destination3}
  \pnode(20, 13){Relay11}
  \pnode(20, 11){Relay12}
  \pnode(42, 13){Relay21}
  \pnode(42, 11){Relay22}

  \ncline{-}{Source}{Source1}
  \ncline{-}{Destination}{Destination1}
  \ncline{-}{Relay1}{Relay11}
  \ncline{-}{Relay2}{Relay21}

  \ncline{<->}{Source2}{Relay12}\naput{$r$}
  \ncline{<->}{Relay22}{Destination2}\naput{$r$}
  \ncline{<->}{Source3}{Destination3}\naput{$1$}
\end{picture}
\endgroup
      \caption{Setup for our analysis.}
      \label{fig:halfduplex:setup}
    \end{figure}
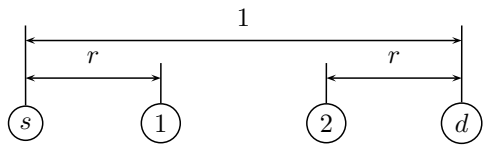}
    In order to evaluate the performance of the previously introduced protocols, we present in this part results for the linear network
    illustrated in Fig. \ref{fig:halfduplex:setup}, i.\,e.,
    we consider a system of $N=2$ relay nodes, equal transmission power $P_s = P_1 = P_2$ and noise power $N_1 = N_2 = N_d$
    such that $\SNRSymb_{s,d} = \nicefrac{P_s}{N_d} = \unit[10]{dB}$.
    Both relays are symmetrically placed such that
    $d_{s,1} = d_{2,d} = \left|r\right|$ and $d_{1,2} = 1 - 2r$, i.\,e., if $r>0$ both relays are placed between source and destination while
    $r<0$ implies that neither relay is placed between source and destination. Unless otherwise noted, we consider a path loss exponent $\PathlossExponent=4$.
    Our analysis presents results for the cut-set bound \cite{Cover.Gamal.TransIT.1979} applied to the half-duplex relay network, \acl{CF},
    Partial Decode-and-Forward (PDF) using all degrees of freedom provided by the \ac{DF} introduced in Section \ref{sec:half.duplex:df},
    a simpler Decode-and-Forward (DF) with one source message level and either \emph{full resource reuse} where all nodes can transmit on all available
    resources or \emph{no resource reuse} where each node transmits on an orthogonal resource, and finally the introduced approach for two alternately
    transmitting relays.

    \subsection{Achievable Rates for Two Relays}
      \ShowFigureMain{%
      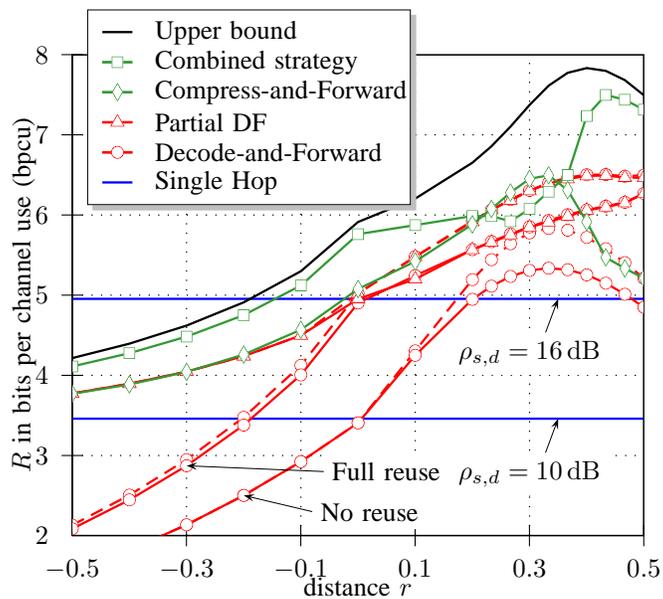
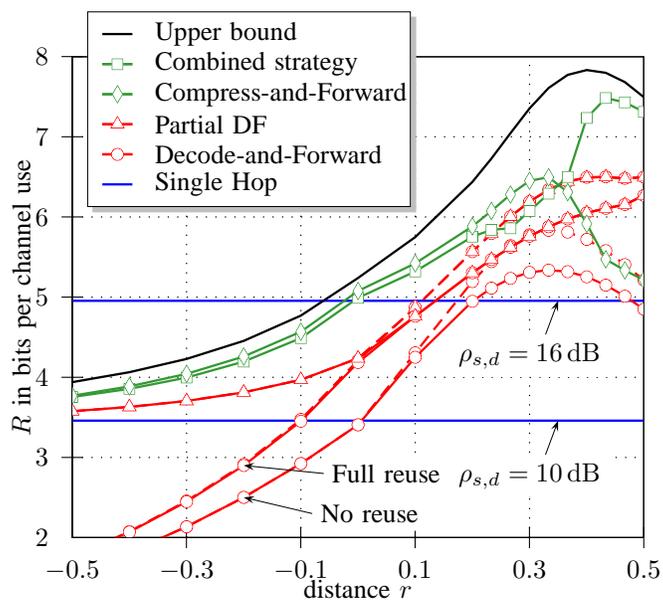
\begin{figure}
	\centering
	\subfigure[Coherent transmission]{\begingroup
\unitlength=1mm
\psset{xunit=76.00000mm, yunit=10.66667mm, linewidth=0.1mm}
\psset{arrowsize=2pt 3, arrowlength=1.4, arrowinset=.4}\psset{axesstyle=frame}
\begin{pspicture}(-0.63158, 0.50000)(0.50000, 8.00000)
\rput(-0.02632, -0.75000){%
\psaxes[subticks=0, labels=all, xsubticks=1, ysubticks=1, Ox=-0.5, Oy=2, Dx=0.2, Dy=1]{-}(-0.50000, 2.00000)(-0.50000, 2.00000)(0.50000, 8.00000)%
\multips(-0.30000, 2.00000)(0.20000, 0.0){4}{\psline[linecolor=black, linestyle=dotted, linewidth=0.2mm](0, 0)(0, 6.00000)}
\multips(-0.50000, 3.00000)(0, 1.00000){5}{\psline[linecolor=black, linestyle=dotted, linewidth=0.2mm](0, 0)(1.00000, 0)}
\rput[b](0.00000, 1.25000){distance $r$}
\rput[t]{90}(-0.60526, 5.00000){$R$ in bits per channel use (bpcu)}
\psclip{\psframe(-0.50000, 2.00000)(0.50000, 8.00000)}
\psline[linecolor=blue, plotstyle=curve, linewidth=0.3mm, showpoints=false, linestyle=solid, dotstyle=square, dotscale=1.3 1.3](-0.50000, 3.45943)(-0.40000, 3.45943)(-0.30000, 3.45943)(-0.20000, 3.45943)(-0.10000, 3.45943)(0.00000, 3.45943)(0.10000, 3.45943)(0.20000, 3.45943)(0.23333, 3.45943)(0.26667, 3.45943)(0.30000, 3.45943)(0.33333, 3.45943)(0.36667, 3.45943)(0.40000, 3.45943)(0.43333, 3.45943)(0.46667, 3.45943)(0.50000, 3.45943)
\psline[linecolor=blue, plotstyle=curve, linewidth=0.3mm, showpoints=false, linestyle=solid, dotstyle=square, dotscale=1.3 1.3](-0.50000, 4.95420)(-0.40000, 4.95420)(-0.30000, 4.95420)(-0.20000, 4.95420)(-0.10000, 4.95420)(0.00000, 4.95420)(0.10000, 4.95420)(0.20000, 4.95420)(0.23333, 4.95420)(0.26667, 4.95420)(0.30000, 4.95420)(0.33333, 4.95420)(0.36667, 4.95420)(0.40000, 4.95420)(0.43333, 4.95420)(0.46667, 4.95420)(0.50000, 4.95420)
\psline[linecolor=red, plotstyle=curve, linewidth=0.3mm, showpoints=true, linestyle=dashed, dotstyle=o, dotscale=1.3 1.3](-0.50000, 2.13197)(-0.40000, 2.50654)(-0.30000, 2.94868)(-0.20000, 3.47666)(-0.10000, 4.12514)(0.00000, 5.02260)(0.10000, 5.48108)(0.20000, 5.92645)(0.23333, 6.06518)(0.26667, 6.19272)(0.30000, 6.30683)(0.33333, 6.40046)(0.36667, 6.46772)(0.40000, 6.50174)(0.43333, 6.50446)(0.46667, 6.48355)(0.50000, 6.49679)
\psline[linecolor=red, plotstyle=curve, linewidth=0.3mm, showpoints=true, linestyle=dashed, dotstyle=triangle, dotscale=1.3 1.3](-0.50000, 3.77410)(-0.40000, 3.89566)(-0.30000, 4.04644)(-0.20000, 4.23944)(-0.10000, 4.49818)(0.00000, 5.02217)(0.10000, 5.47937)(0.20000, 5.91468)(0.23333, 6.06443)(0.26667, 6.18219)(0.30000, 6.30141)(0.33333, 6.39214)(0.36667, 6.44929)(0.40000, 6.48725)(0.43333, 6.49267)(0.46667, 6.46667)(0.50000, 6.46472)
\psline[linecolor=red, plotstyle=curve, linewidth=0.3mm, showpoints=true, linestyle=dashed, dotstyle=o, dotscale=1.3 1.3](-0.50000, 1.54770)(-0.40000, 1.81968)(-0.30000, 2.13606)(-0.20000, 2.50209)(-0.10000, 2.92359)(0.00000, 3.40806)(0.10000, 4.30553)(0.20000, 5.19412)(0.23333, 5.44248)(0.26667, 5.64206)(0.30000, 5.77657)(0.33333, 5.83340)(0.36667, 5.81122)(0.40000, 5.72046)(0.43333, 5.57936)(0.46667, 5.40569)(0.50000, 5.21120)
\psline[linecolor=black, plotstyle=curve, linewidth=0.3mm, showpoints=false, linestyle=solid, dotstyle=square, dotscale=1.3 1.3](-0.50000, 4.21501)(-0.40000, 4.39516)(-0.30000, 4.62073)(-0.20000, 4.91064)(-0.10000, 5.29880)(0.00000, 5.91292)(0.10000, 6.20447)(0.20000, 6.65242)(0.23333, 6.85682)(0.26667, 7.10253)(0.30000, 7.37232)(0.33333, 7.61315)(0.36667, 7.77354)(0.40000, 7.83256)(0.43333, 7.79778)(0.46667, 7.68243)(0.50000, 7.49355)
\psline[linecolor=red, plotstyle=curve, linewidth=0.3mm, showpoints=true, linestyle=solid, dotstyle=o, dotscale=1.3 1.3](-0.50000, 2.08571)(-0.40000, 2.44674)(-0.30000, 2.87219)(-0.20000, 3.37997)(-0.10000, 4.00558)(0.00000, 4.90240)(0.10000, 5.24394)(0.20000, 5.56835)(0.23333, 5.66864)(0.26667, 5.76338)(0.30000, 5.85150)(0.33333, 5.93164)(0.36667, 6.00383)(0.40000, 6.06330)(0.43333, 6.11007)(0.46667, 6.15862)(0.50000, 6.26863)
\psline[linecolor=red, plotstyle=curve, linewidth=0.3mm, showpoints=true, linestyle=solid, dotstyle=triangle, dotscale=1.3 1.3](-0.50000, 3.77412)(-0.40000, 3.89573)(-0.30000, 4.04696)(-0.20000, 4.23928)(-0.10000, 4.49863)(0.00000, 4.95324)(0.10000, 5.20055)(0.20000, 5.56722)(0.23333, 5.65899)(0.26667, 5.75155)(0.30000, 5.84895)(0.33333, 5.91151)(0.36667, 5.98412)(0.40000, 6.05857)(0.43333, 6.09815)(0.46667, 6.15085)(0.50000, 6.26529)
\psline[linecolor=red, plotstyle=curve, linewidth=0.3mm, showpoints=true, linestyle=solid, dotstyle=o, dotscale=1.3 1.3](-0.50000, 1.54770)(-0.40000, 1.81968)(-0.30000, 2.13606)(-0.20000, 2.50209)(-0.10000, 2.92359)(0.00000, 3.40733)(0.10000, 4.24970)(0.20000, 4.94804)(0.23333, 5.11232)(0.26667, 5.23370)(0.30000, 5.30899)(0.33333, 5.33668)(0.36667, 5.31727)(0.40000, 5.25337)(0.43333, 5.14931)(0.46667, 5.01156)(0.50000, 4.84903)
\psline[linecolor=darkgreen, plotstyle=curve, linewidth=0.3mm, showpoints=true, linestyle=solid, dotstyle=square, dotscale=1.3 1.3](-0.50000, 4.11060)(-0.40000, 4.27634)(-0.30000, 4.48330)(-0.20000, 4.75148)(-0.10000, 5.12083)(0.00000, 5.76104)(0.10000, 5.87378)(0.20000, 5.98733)(0.23333, 5.98569)(0.26667, 5.92175)(0.30000, 6.07497)(0.33333, 6.29086)(0.36667, 6.49992)(0.40000, 7.23542)(0.43333, 7.49745)(0.46667, 7.43745)(0.50000, 7.31348)
\psline[linecolor=darkgreen, plotstyle=curve, linewidth=0.3mm, showpoints=true, linestyle=solid, dotstyle=diamond, dotscale=1.3 1.3](-0.50000, 3.76980)(-0.40000, 3.88645)(-0.30000, 4.04323)(-0.20000, 4.25923)(-0.10000, 4.56891)(0.00000, 5.07648)(0.10000, 5.42114)(0.20000, 5.88293)(0.23333, 6.07590)(0.26667, 6.27990)(0.30000, 6.45161)(0.33333, 6.49559)(0.36667, 6.30980)(0.40000, 5.91435)(0.43333, 5.47124)(0.46667, 5.33359)(0.50000, 5.21169)
\endpsclip
\psline[linecolor=black, linestyle=solid, linewidth=0.1mm]{->}(0.30000, 2.50000)(0.35000, 3.45943)
\psline[linecolor=black, linestyle=solid, linewidth=0.1mm]{->}(0.30000, 4.00000)(0.35000, 4.95420)
\psline[linecolor=black, linestyle=solid, linewidth=0.1mm]{->}(-0.06000, 2.80000)(-0.30000, 2.87219)
\psline[linecolor=black, linestyle=solid, linewidth=0.1mm]{->}(-0.08000, 2.30000)(-0.20000, 2.50209)
\rput[b](0.30000, 2.50000){\psframebox[linestyle=none, fillcolor=white, fillstyle=solid]{$\SNRSymb_{s,d}=\unit[10]{dB}$}}
\rput[b](0.30000, 4.00000){\psframebox[linestyle=none, fillcolor=white, fillstyle=solid]{$\SNRSymb_{s,d}=\unit[16]{dB}$}}
\rput[l](-0.06000, 2.80000){\psframebox[linestyle=none, fillcolor=white, fillstyle=solid]{Full reuse}}
\rput[l](-0.08000, 2.30000){\psframebox[linestyle=none, fillcolor=white, fillstyle=solid]{No reuse}}
\psframe[linecolor=black, fillstyle=solid, fillcolor=white, shadowcolor=lightgray, shadowsize=1mm, shadow=true](-0.47368, 6.12500)(0.09211, 8.56250)
\rput[l](-0.35526, 8.28125){Upper bound}
\psline[linecolor=black, linestyle=solid, linewidth=0.3mm](-0.44737, 8.28125)(-0.39474, 8.28125)
\rput[l](-0.35526, 7.90625){Combined strategy}
\psline[linecolor=darkgreen, linestyle=solid, linewidth=0.3mm](-0.44737, 7.90625)(-0.39474, 7.90625)
\psline[linecolor=darkgreen, linestyle=solid, linewidth=0.3mm](-0.44737, 7.90625)(-0.39474, 7.90625)
\psdots[linecolor=darkgreen, linestyle=solid, linewidth=0.3mm, dotstyle=square, dotscale=1.3 1.3, linecolor=darkgreen](-0.42105, 7.90625)
\rput[l](-0.35526, 7.53125){Compress-and-Forward}
\psline[linecolor=darkgreen, linestyle=solid, linewidth=0.3mm](-0.44737, 7.53125)(-0.39474, 7.53125)
\psline[linecolor=darkgreen, linestyle=solid, linewidth=0.3mm](-0.44737, 7.53125)(-0.39474, 7.53125)
\psdots[linecolor=darkgreen, linestyle=solid, linewidth=0.3mm, dotstyle=diamond, dotscale=1.3 1.3, linecolor=darkgreen](-0.42105, 7.53125)
\rput[l](-0.35526, 7.15625){Partial DF}
\psline[linecolor=red, linestyle=solid, linewidth=0.3mm](-0.44737, 7.15625)(-0.39474, 7.15625)
\psline[linecolor=red, linestyle=solid, linewidth=0.3mm](-0.44737, 7.15625)(-0.39474, 7.15625)
\psdots[linecolor=red, linestyle=solid, linewidth=0.3mm, dotstyle=triangle, dotscale=1.3 1.3, linecolor=red](-0.42105, 7.15625)
\rput[l](-0.35526, 6.78125){Decode-and-Forward}
\psline[linecolor=red, linestyle=solid, linewidth=0.3mm](-0.44737, 6.78125)(-0.39474, 6.78125)
\psline[linecolor=red, linestyle=solid, linewidth=0.3mm](-0.44737, 6.78125)(-0.39474, 6.78125)
\psdots[linecolor=red, linestyle=solid, linewidth=0.3mm, dotstyle=o, dotscale=1.3 1.3, linecolor=red](-0.42105, 6.78125)
\rput[l](-0.35526, 6.40625){Single Hop}
\psline[linecolor=blue, linestyle=solid, linewidth=0.3mm](-0.44737, 6.40625)(-0.39474, 6.40625)
}\end{pspicture}
\endgroup
 }
	\subfigure[Non-coherent transmission]{\begingroup
\unitlength=1mm
\psset{xunit=76.00000mm, yunit=10.66667mm, linewidth=0.1mm}
\psset{arrowsize=2pt 3, arrowlength=1.4, arrowinset=.4}\psset{axesstyle=frame}
\begin{pspicture}(-0.63158, 0.50000)(0.50000, 8.00000)
\rput(-0.02632, -0.75000){%
\psaxes[subticks=0, labels=all, xsubticks=1, ysubticks=1, Ox=-0.5, Oy=2, Dx=0.2, Dy=1]{-}(-0.50000, 2.00000)(-0.50000, 2.00000)(0.50000, 8.00000)%
\multips(-0.30000, 2.00000)(0.20000, 0.0){4}{\psline[linecolor=black, linestyle=dotted, linewidth=0.2mm](0, 0)(0, 6.00000)}
\multips(-0.50000, 3.00000)(0, 1.00000){5}{\psline[linecolor=black, linestyle=dotted, linewidth=0.2mm](0, 0)(1.00000, 0)}
\rput[b](0.00000, 1.25000){distance $r$}
\rput[t]{90}(-0.60526, 5.00000){$R$ in bits per channel use}
\psclip{\psframe(-0.50000, 2.00000)(0.50000, 8.00000)}
\psline[linecolor=blue, plotstyle=curve, linewidth=0.3mm, showpoints=false, linestyle=solid, dotstyle=square, dotscale=1.3 1.3](-0.50000, 3.45943)(-0.40000, 3.45943)(-0.30000, 3.45943)(-0.20000, 3.45943)(-0.10000, 3.45943)(0.00000, 3.45943)(0.10000, 3.45943)(0.20000, 3.45943)(0.23333, 3.45943)(0.26667, 3.45943)(0.30000, 3.45943)(0.33333, 3.45943)(0.36667, 3.45943)(0.40000, 3.45943)(0.43333, 3.45943)(0.46667, 3.45943)(0.50000, 3.45943)
\psline[linecolor=blue, plotstyle=curve, linewidth=0.3mm, showpoints=false, linestyle=solid, dotstyle=square, dotscale=1.3 1.3](-0.50000, 4.95420)(-0.40000, 4.95420)(-0.30000, 4.95420)(-0.20000, 4.95420)(-0.10000, 4.95420)(0.00000, 4.95420)(0.10000, 4.95420)(0.20000, 4.95420)(0.23333, 4.95420)(0.26667, 4.95420)(0.30000, 4.95420)(0.33333, 4.95420)(0.36667, 4.95420)(0.40000, 4.95420)(0.43333, 4.95420)(0.46667, 4.95420)(0.50000, 4.95420)
\psline[linecolor=red, plotstyle=curve, linewidth=0.3mm, showpoints=true, linestyle=dashed, dotstyle=o, dotscale=1.3 1.3](-0.50000, 1.75614)(-0.40000, 2.07652)(-0.30000, 2.45763)(-0.20000, 2.91489)(-0.10000, 3.47512)(0.00000, 4.21236)(0.10000, 4.87283)(0.20000, 5.56694)(0.23333, 5.79344)(0.26667, 6.00528)(0.30000, 6.19142)(0.33333, 6.34031)(0.36667, 6.44229)(0.40000, 6.49424)(0.43333, 6.50099)(0.46667, 6.48146)(0.50000, 6.49802)
\psline[linecolor=red, plotstyle=curve, linewidth=0.3mm, showpoints=true, linestyle=dashed, dotstyle=triangle, dotscale=1.3 1.3](-0.50000, 3.57681)(-0.40000, 3.63024)(-0.30000, 3.70509)(-0.20000, 3.81198)(-0.10000, 3.96940)(0.00000, 4.23717)(0.10000, 4.87003)(0.20000, 5.56697)(0.23333, 5.79049)(0.26667, 6.00516)(0.30000, 6.19124)(0.33333, 6.33945)(0.36667, 6.44233)(0.40000, 6.48952)(0.43333, 6.49908)(0.46667, 6.47794)(0.50000, 6.49516)
\psline[linecolor=red, plotstyle=curve, linewidth=0.3mm, showpoints=true, linestyle=dashed, dotstyle=o, dotscale=1.3 1.3](-0.50000, 1.54770)(-0.40000, 1.81968)(-0.30000, 2.13606)(-0.20000, 2.50209)(-0.10000, 2.92359)(0.00000, 3.40806)(0.10000, 4.30571)(0.20000, 5.19412)(0.23333, 5.44250)(0.26667, 5.64230)(0.30000, 5.77661)(0.33333, 5.83343)(0.36667, 5.81122)(0.40000, 5.72051)(0.43333, 5.57941)(0.46667, 5.40570)(0.50000, 5.21118)
\psline[linecolor=black, plotstyle=curve, linewidth=0.3mm, showpoints=false, linestyle=solid, dotstyle=square, dotscale=1.3 1.3](-0.50000, 3.93974)(-0.40000, 4.06370)(-0.30000, 4.22906)(-0.20000, 4.45450)(-0.10000, 4.76977)(0.00000, 5.23917)(0.10000, 5.74879)(0.20000, 6.43731)(0.23333, 6.72874)(0.26667, 7.04503)(0.30000, 7.35421)(0.33333, 7.60892)(0.36667, 7.77212)(0.40000, 7.83244)(0.43333, 7.79778)(0.46667, 7.68219)(0.50000, 7.49355)
\psline[linecolor=red, plotstyle=curve, linewidth=0.3mm, showpoints=true, linestyle=solid, dotstyle=o, dotscale=1.3 1.3](-0.50000, 1.75183)(-0.40000, 2.07005)(-0.30000, 2.44784)(-0.20000, 2.89993)(-0.10000, 3.45231)(0.00000, 4.18447)(0.10000, 4.76239)(0.20000, 5.30362)(0.23333, 5.46905)(0.26667, 5.62196)(0.30000, 5.75939)(0.33333, 5.87813)(0.36667, 5.97571)(0.40000, 6.05164)(0.43333, 6.10786)(0.46667, 6.15600)(0.50000, 6.26856)
\psline[linecolor=red, plotstyle=curve, linewidth=0.3mm, showpoints=true, linestyle=solid, dotstyle=triangle, dotscale=1.3 1.3](-0.50000, 3.57671)(-0.40000, 3.63011)(-0.30000, 3.70493)(-0.20000, 3.81178)(-0.10000, 3.96913)(0.00000, 4.23689)(0.10000, 4.76170)(0.20000, 5.30310)(0.23333, 5.46901)(0.26667, 5.62164)(0.30000, 5.75934)(0.33333, 5.87770)(0.36667, 5.97473)(0.40000, 6.05139)(0.43333, 6.10656)(0.46667, 6.15151)(0.50000, 6.26846)
\psline[linecolor=red, plotstyle=curve, linewidth=0.3mm, showpoints=true, linestyle=solid, dotstyle=o, dotscale=1.3 1.3](-0.50000, 1.54770)(-0.40000, 1.81968)(-0.30000, 2.13606)(-0.20000, 2.50209)(-0.10000, 2.92359)(0.00000, 3.40733)(0.10000, 4.24970)(0.20000, 4.94807)(0.23333, 5.11235)(0.26667, 5.23373)(0.30000, 5.30900)(0.33333, 5.33671)(0.36667, 5.31734)(0.40000, 5.25338)(0.43333, 5.14931)(0.46667, 5.01158)(0.50000, 4.84904)
\psline[linecolor=darkgreen, plotstyle=curve, linewidth=0.3mm, showpoints=true, linestyle=solid, dotstyle=square, dotscale=1.3 1.3](-0.50000, 3.75405)(-0.40000, 3.85584)(-0.30000, 3.99663)(-0.20000, 4.19597)(-0.10000, 4.48913)(0.00000, 4.99060)(0.10000, 5.32020)(0.20000, 5.75417)(0.23333, 5.83978)(0.26667, 5.85842)(0.30000, 6.07445)(0.33333, 6.29051)(0.36667, 6.49900)(0.40000, 7.23438)(0.43333, 7.48330)(0.46667, 7.42949)(0.50000, 7.31029)
\psline[linecolor=darkgreen, plotstyle=curve, linewidth=0.3mm, showpoints=true, linestyle=solid, dotstyle=diamond, dotscale=1.3 1.3](-0.50000, 3.76980)(-0.40000, 3.88645)(-0.30000, 4.04323)(-0.20000, 4.25923)(-0.10000, 4.56891)(0.00000, 5.07648)(0.10000, 5.42114)(0.20000, 5.88293)(0.23333, 6.07590)(0.26667, 6.27990)(0.30000, 6.45161)(0.33333, 6.49559)(0.36667, 6.30980)(0.40000, 5.91435)(0.43333, 5.47124)(0.46667, 5.33359)(0.50000, 5.21169)
\endpsclip
\psline[linecolor=black, linestyle=solid, linewidth=0.1mm]{->}(0.30000, 2.50000)(0.35000, 3.45943)
\psline[linecolor=black, linestyle=solid, linewidth=0.1mm]{->}(0.30000, 4.00000)(0.35000, 4.95420)
\psline[linecolor=black, linestyle=solid, linewidth=0.1mm]{->}(-0.06000, 2.80000)(-0.20000, 2.89993)
\psline[linecolor=black, linestyle=solid, linewidth=0.1mm]{->}(-0.08000, 2.30000)(-0.20000, 2.50209)
\rput[b](0.30000, 2.50000){\psframebox[linestyle=none, fillcolor=white, fillstyle=solid]{$\SNRSymb_{s,d}=\unit[10]{dB}$}}
\rput[b](0.30000, 4.00000){\psframebox[linestyle=none, fillcolor=white, fillstyle=solid]{$\SNRSymb_{s,d}=\unit[16]{dB}$}}
\rput[l](-0.06000, 2.80000){\psframebox[linestyle=none, fillcolor=white, fillstyle=solid]{Full reuse}}
\rput[l](-0.08000, 2.30000){\psframebox[linestyle=none, fillcolor=white, fillstyle=solid]{No reuse}}
\psframe[linecolor=black, fillstyle=solid, fillcolor=white, shadowcolor=lightgray, shadowsize=1mm, shadow=true](-0.47368, 6.12500)(0.09211, 8.56250)
\rput[l](-0.35526, 8.28125){Upper bound}
\psline[linecolor=black, linestyle=solid, linewidth=0.3mm](-0.44737, 8.28125)(-0.39474, 8.28125)
\rput[l](-0.35526, 7.90625){Combined strategy}
\psline[linecolor=darkgreen, linestyle=solid, linewidth=0.3mm](-0.44737, 7.90625)(-0.39474, 7.90625)
\psline[linecolor=darkgreen, linestyle=solid, linewidth=0.3mm](-0.44737, 7.90625)(-0.39474, 7.90625)
\psdots[linecolor=darkgreen, linestyle=solid, linewidth=0.3mm, dotstyle=square, dotscale=1.3 1.3, linecolor=darkgreen](-0.42105, 7.90625)
\rput[l](-0.35526, 7.53125){Compress-and-Forward}
\psline[linecolor=darkgreen, linestyle=solid, linewidth=0.3mm](-0.44737, 7.53125)(-0.39474, 7.53125)
\psline[linecolor=darkgreen, linestyle=solid, linewidth=0.3mm](-0.44737, 7.53125)(-0.39474, 7.53125)
\psdots[linecolor=darkgreen, linestyle=solid, linewidth=0.3mm, dotstyle=diamond, dotscale=1.3 1.3, linecolor=darkgreen](-0.42105, 7.53125)
\rput[l](-0.35526, 7.15625){Partial DF}
\psline[linecolor=red, linestyle=solid, linewidth=0.3mm](-0.44737, 7.15625)(-0.39474, 7.15625)
\psline[linecolor=red, linestyle=solid, linewidth=0.3mm](-0.44737, 7.15625)(-0.39474, 7.15625)
\psdots[linecolor=red, linestyle=solid, linewidth=0.3mm, dotstyle=triangle, dotscale=1.3 1.3, linecolor=red](-0.42105, 7.15625)
\rput[l](-0.35526, 6.78125){Decode-and-Forward}
\psline[linecolor=red, linestyle=solid, linewidth=0.3mm](-0.44737, 6.78125)(-0.39474, 6.78125)
\psline[linecolor=red, linestyle=solid, linewidth=0.3mm](-0.44737, 6.78125)(-0.39474, 6.78125)
\psdots[linecolor=red, linestyle=solid, linewidth=0.3mm, dotstyle=o, dotscale=1.3 1.3, linecolor=red](-0.42105, 6.78125)
\rput[l](-0.35526, 6.40625){Single Hop}
\psline[linecolor=blue, linestyle=solid, linewidth=0.3mm](-0.44737, 6.40625)(-0.39474, 6.40625)
}\end{pspicture}
\endgroup
 }
	\caption[Achievable rates for the Gaussian half-duplex two-relay channel.]
	{Achievable rates for the Gaussian half-duplex two-relay channel. Solid curves indicate a fixed transmission strategy
	and dashed lines indicate a random transmission schedule. $\SNRSymb_{s,d}=\unit[16]{dB}$
	again indicates the power-normalized case.}
	\label{fig:halfduplex:results:two_relays}
      \end{figure}}
      Fig. \ref{fig:halfduplex:results:two_relays} shows the achievable rates for coherent and non-coherent
      transmission as well as for fixed and random transmission schedules. \ac{DF} with a \emph{random} transmission schedule achieves 
      a performance improvement of up to $\unit[0.5]{\bpcu}$ over \ac{DF} with a \emph{fixed} transmission schedule, which is much less than the theoretical maximum of $\unit[2]{\bpcu}$.
      The superposition coding of PDF provides only for $r<0$ gains over the less complex DF, which result from a mode where relay $2$ is turned off.

      Coherent transmission for \ac{DF} does not provide any gains for $r\gtrsim0.33$, which implies that the additional complexity is not beneficial.
      For a large range of $r$, the combined strategy provides the maximum performance close to the cut-set bound for $r\approx0.5$.
      Interestingly, at $r\approx0.5$ the cut-set bound uses two alternately transmitting relay nodes.
      There is a significant performance drop of the combined strategy for $0<r<0.4$ due to the increased interference between
      both relays, which is not strong enough to be decoded and not weak enough to be ignored.

    \subsection{Full vs. Half-Duplex Relaying}
      Fig. \ref{fig:halfduplex:results:singe_relay} shows the achievable rates of a \emph{single-relay} Gaussian half-duplex and full-duplex relay network,
      where we only consider relay $1$ and permanently turn off relay $2$. 
      \ShowFigureMain{%
      \begin{figure}[htb]
	\centering
	\subfigure[Half-duplex Network]{\begingroup
\unitlength=1mm
\psset{xunit=38.00000mm, yunit=15.00000mm, linewidth=0.1mm}
\psset{arrowsize=2pt 3, arrowlength=1.4, arrowinset=.4}\psset{axesstyle=frame}
\begin{pspicture}(-1.26316, 1.33333)(1.00000, 6.00000)
\rput(-0.05263, -0.13333){%
\psaxes[subticks=0, labels=all, xsubticks=1, ysubticks=1, Ox=-1, Oy=2, Dx=0.4, Dy=1]{-}(-1.00000, 2.00000)(-1.00000, 2.00000)(1.00000, 6.00000)%
\multips(-0.60000, 2.00000)(0.40000, 0.0){4}{\psline[linecolor=black, linestyle=dotted, linewidth=0.2mm](0, 0)(0, 4.00000)}
\multips(-1.00000, 3.00000)(0, 1.00000){3}{\psline[linecolor=black, linestyle=dotted, linewidth=0.2mm](0, 0)(2.00000, 0)}
\rput[b](0.00000, 1.46667){distance $r$}
\rput[t]{90}(-1.21053, 4.00000){$R$ in bits per channel use (bpcu)}
\psclip{\psframe(-1.00000, 2.00000)(1.00000, 6.00000)}
\psline[linecolor=blue, plotstyle=curve, linewidth=0.3mm, showpoints=false, linestyle=solid, dotstyle=square, dotscale=1.3 1.3](-1.00000, 3.45943)(-0.80000, 3.45943)(-0.60000, 3.45943)(-0.40000, 3.45943)(-0.20000, 3.45943)(0.00000, 3.45943)(0.20000, 3.45943)(0.40000, 3.45943)(0.50000, 3.45943)(0.60000, 3.45943)(0.80000, 3.45943)(1.00000, 3.45943)
\psline[linecolor=blue, plotstyle=curve, linewidth=0.3mm, showpoints=false, linestyle=solid, dotstyle=square, dotscale=1.3 1.3](-1.00000, 4.39232)(-0.80000, 4.39232)(-0.60000, 4.39232)(-0.40000, 4.39232)(-0.20000, 4.39232)(0.00000, 4.39232)(0.20000, 4.39232)(0.40000, 4.39232)(0.50000, 4.39232)(0.60000, 4.39232)(0.80000, 4.39232)(1.00000, 4.39232)
\psline[linecolor=red, plotstyle=curve, linewidth=0.3mm, showpoints=true, linestyle=dashed, dotstyle=o, dotscale=1.3 1.3](-1.00000, 3.46122)(-0.80000, 3.49350)(-0.60000, 3.54870)(-0.40000, 3.65168)(-0.20000, 3.85723)(0.00000, 4.31435)(0.20000, 4.71473)(0.40000, 5.13139)(0.50000, 5.16883)(0.60000, 5.02622)(0.80000, 4.32902)(1.00000, 3.40806)
\psline[linecolor=red, plotstyle=curve, linewidth=0.3mm, showpoints=true, linestyle=dashed, dotstyle=o, dotscale=1.3 1.3](-1.00000, 3.46035)(-0.80000, 3.46035)(-0.60000, 3.46035)(-0.40000, 3.46035)(-0.20000, 3.46035)(0.00000, 3.50717)(0.20000, 4.38783)(0.40000, 5.06273)(0.50000, 5.14888)(0.60000, 5.02228)(0.80000, 4.32899)(1.00000, 3.40806)
\psline[linecolor=black, plotstyle=curve, linewidth=0.3mm, showpoints=false, linestyle=solid, dotstyle=square, dotscale=1.3 1.3](-1.00000, 3.53439)(-0.80000, 3.57378)(-0.60000, 3.64007)(-0.40000, 3.75636)(-0.20000, 3.96967)(0.00000, 4.38653)(0.20000, 4.91220)(0.40000, 5.37590)(0.50000, 5.43785)(0.60000, 5.34604)(0.80000, 4.85526)(1.00000, 4.33525)
\psline[linecolor=red, plotstyle=curve, linewidth=0.3mm, showpoints=true, linestyle=solid, dotstyle=o, dotscale=1.3 1.3](-1.00000, 3.46063)(-0.80000, 3.49162)(-0.60000, 3.54476)(-0.40000, 3.64309)(-0.20000, 3.83838)(0.00000, 4.28260)(0.20000, 4.55484)(0.40000, 4.79827)(0.50000, 4.81400)(0.60000, 4.72462)(0.80000, 4.22576)(1.00000, 3.40733)
\psline[linecolor=red, plotstyle=curve, linewidth=0.3mm, showpoints=true, linestyle=solid, dotstyle=o, dotscale=1.3 1.3](-1.00000, 3.45943)(-0.80000, 3.45943)(-0.60000, 3.45943)(-0.40000, 3.45943)(-0.20000, 3.45943)(0.00000, 3.50606)(0.20000, 4.29242)(0.40000, 4.74569)(0.50000, 4.79675)(0.60000, 4.72041)(0.80000, 4.22571)(1.00000, 3.40733)
\psline[linecolor=darkgreen, plotstyle=curve, linewidth=0.3mm, showpoints=true, linestyle=solid, dotstyle=diamond, dotscale=1.3 1.3](-1.00000, 3.49057)(-0.80000, 3.52730)(-0.60000, 3.59768)(-0.40000, 3.72213)(-0.20000, 3.94425)(0.00000, 4.37730)(0.20000, 4.78289)(0.40000, 5.00492)(0.50000, 4.96886)(0.60000, 4.84611)(0.80000, 4.48474)(1.00000, 4.18854)
\endpsclip
\psline[linecolor=black, linestyle=solid, linewidth=0.1mm]{->}(0.50000, 2.80000)(0.45000, 3.45943)
\psline[linecolor=black, linestyle=solid, linewidth=0.1mm]{->}(0.50000, 3.60000)(0.45000, 4.39232)
\psline[linecolor=black, linestyle=solid, linewidth=0.1mm]{->}(-0.14000, 3.00000)(-0.28000, 3.76027)
\psline[linecolor=black, linestyle=solid, linewidth=0.1mm]{->}(0.18000, 2.70000)(0.04000, 3.66333)
\rput[b](0.50000, 2.80000){\psframebox[linestyle=none, fillcolor=white, fillstyle=solid]{$\SNRSymb_{s,d}=\unit[10]{dB}$}}
\rput[b](0.50000, 3.60000){\psframebox[linestyle=none, fillcolor=white, fillstyle=solid]{$\SNRSymb_{s,d}=\unit[13]{dB}$}}
\rput[c](-0.14000, 3.00000){\psframebox[linestyle=none, fillcolor=white, fillstyle=solid]{Full reuse}}
\rput[c](0.18000, 2.70000){\psframebox[linestyle=none, fillcolor=white, fillstyle=solid]{No reuse}}
\psframe[linecolor=black, fillstyle=solid, fillcolor=white, shadowcolor=lightgray, shadowsize=1mm, shadow=true](-0.94737, 4.93333)(0.18421, 6.13333)
\rput[l](-0.71053, 5.93333){Upper bound}
\psline[linecolor=black, linestyle=solid, linewidth=0.3mm](-0.89474, 5.93333)(-0.78947, 5.93333)
\rput[l](-0.71053, 5.66667){Compress-and-Forward}
\psline[linecolor=darkgreen, linestyle=solid, linewidth=0.3mm](-0.89474, 5.66667)(-0.78947, 5.66667)
\psline[linecolor=darkgreen, linestyle=solid, linewidth=0.3mm](-0.89474, 5.66667)(-0.78947, 5.66667)
\psdots[linecolor=darkgreen, linestyle=solid, linewidth=0.3mm, dotstyle=diamond, dotscale=1.3 1.3, linecolor=darkgreen](-0.84211, 5.66667)
\rput[l](-0.71053, 5.40000){Decode-and-Forward}
\psline[linecolor=red, linestyle=solid, linewidth=0.3mm](-0.89474, 5.40000)(-0.78947, 5.40000)
\psline[linecolor=red, linestyle=solid, linewidth=0.3mm](-0.89474, 5.40000)(-0.78947, 5.40000)
\psdots[linecolor=red, linestyle=solid, linewidth=0.3mm, dotstyle=o, dotscale=1.3 1.3, linecolor=red](-0.84211, 5.40000)
\rput[l](-0.71053, 5.13333){Single Hop}
\psline[linecolor=blue, linestyle=solid, linewidth=0.3mm](-0.89474, 5.13333)(-0.78947, 5.13333)
}\end{pspicture}
\endgroup
 }
	\subfigure[Full-duplex Network]{\begingroup
\unitlength=1mm
\psset{xunit=38.00000mm, yunit=10.00000mm, linewidth=0.1mm}
\psset{arrowsize=2pt 3, arrowlength=1.4, arrowinset=.4}\psset{axesstyle=frame}
\begin{pspicture}(-1.26316, 1.00000)(1.00000, 8.00000)
\rput(-0.05263, -0.20000){%
\psaxes[subticks=0, labels=all, xsubticks=1, ysubticks=1, Ox=-1, Oy=2, Dx=0.4, Dy=1]{-}(-1.00000, 2.00000)(-1.00000, 2.00000)(1.00000, 8.00000)%
\multips(-0.60000, 2.00000)(0.40000, 0.0){4}{\psline[linecolor=black, linestyle=dotted, linewidth=0.2mm](0, 0)(0, 6.00000)}
\multips(-1.00000, 3.00000)(0, 1.00000){5}{\psline[linecolor=black, linestyle=dotted, linewidth=0.2mm](0, 0)(2.00000, 0)}
\rput[b](0.00000, 1.20000){distance $r$}
\rput[t]{90}(-1.21053, 5.00000){$R$ in bits per channel use (bpcu)}
\psclip{\psframe(-1.00000, 2.00000)(1.00000, 8.00000)}
\psline[linecolor=blue, plotstyle=curve, linewidth=0.3mm, showpoints=false, linestyle=solid, dotstyle=square, dotscale=1.3 1.3](-1.49000, 3.45943)(-1.39000, 3.45943)(-1.29000, 3.45943)(-1.19000, 3.45943)(-1.09000, 3.45943)(-0.99000, 3.45943)(-0.89000, 3.45943)(-0.79000, 3.45943)(-0.69000, 3.45943)(-0.59000, 3.45943)(-0.49000, 3.45943)(-0.39000, 3.45943)(-0.29000, 3.45943)(-0.19000, 3.45943)(-0.09000, 3.45943)(0.01000, 3.45943)(0.11000, 3.45943)(0.21000, 3.45943)(0.31000, 3.45943)(0.41000, 3.45943)(0.51000, 3.45943)(0.61000, 3.45943)(0.71000, 3.45943)(0.81000, 3.45943)(0.91000, 3.45943)(1.01000, 3.45943)(1.11000, 3.45943)(1.21000, 3.45943)(1.31000, 3.45943)(1.41000, 3.45943)
\psline[linecolor=blue, plotstyle=curve, linewidth=0.3mm, showpoints=false, linestyle=solid, dotstyle=square, dotscale=1.3 1.3](-1.49000, 4.39232)(-1.39000, 4.39232)(-1.29000, 4.39232)(-1.19000, 4.39232)(-1.09000, 4.39232)(-0.99000, 4.39232)(-0.89000, 4.39232)(-0.79000, 4.39232)(-0.69000, 4.39232)(-0.59000, 4.39232)(-0.49000, 4.39232)(-0.39000, 4.39232)(-0.29000, 4.39232)(-0.19000, 4.39232)(-0.09000, 4.39232)(0.01000, 4.39232)(0.11000, 4.39232)(0.21000, 4.39232)(0.31000, 4.39232)(0.41000, 4.39232)(0.51000, 4.39232)(0.61000, 4.39232)(0.71000, 4.39232)(0.81000, 4.39232)(0.91000, 4.39232)(1.01000, 4.39232)(1.11000, 4.39232)(1.21000, 4.39232)(1.31000, 4.39232)(1.41000, 4.39232)
\psline[linecolor=red, plotstyle=curve, linewidth=0.3mm, showpoints=true, linestyle=solid, dotstyle=o, dotscale=1.3 1.3](-1.49000, 1.59878)(-1.39000, 1.87924)(-1.29000, 2.20512)(-1.19000, 2.58176)(-1.09000, 3.01511)(-0.99000, 3.51225)(-0.89000, 3.55872)(-0.79000, 3.58184)(-0.69000, 3.61187)(-0.59000, 3.65130)(-0.49000, 3.70364)(-0.39000, 3.77387)(-0.29000, 3.86899)(-0.19000, 3.99880)(-0.09000, 4.17666)(0.01000, 4.42023)(0.11000, 4.75158)(0.21000, 5.19668)(0.31000, 5.78442)(0.41000, 6.54730)(0.51000, 7.21738)(0.61000, 6.19424)(0.71000, 5.33457)(0.81000, 4.59876)(0.91000, 3.96186)(1.01000, 3.40733)(1.11000, 2.92359)(1.21000, 2.50209)(1.31000, 2.13606)(1.41000, 1.81968)
\psline[linecolor=darkgreen, plotstyle=curve, linewidth=0.3mm, showpoints=true, linestyle=solid, dotstyle=diamond, dotscale=1.3 1.3](-1.49000, 3.46463)(-1.39000, 3.46711)(-1.29000, 3.47089)(-1.19000, 3.47664)(-1.09000, 3.48537)(-0.99000, 3.49839)(-0.89000, 3.51735)(-0.79000, 3.54402)(-0.69000, 3.58013)(-0.59000, 3.62736)(-0.49000, 3.68786)(-0.39000, 3.76518)(-0.29000, 3.86534)(-0.19000, 3.99786)(-0.09000, 4.17659)(0.01000, 4.42023)(0.11000, 4.75125)(0.21000, 5.18951)(0.31000, 5.72737)(0.41000, 6.25314)(0.51000, 6.46216)(0.61000, 6.16115)(0.71000, 5.61467)(0.81000, 5.09286)(0.91000, 4.67695)(1.01000, 4.36526)(1.11000, 4.13632)(1.21000, 3.96817)(1.31000, 3.84308)(1.41000, 3.74812)
\psline[linecolor=darkgreen, plotstyle=curve, linewidth=0.3mm, showpoints=true, linestyle=solid, dotstyle=square, dotscale=1.3 1.3](-1.49000, 3.46463)(-1.39000, 3.46711)(-1.29000, 3.47089)(-1.19000, 3.47664)(-1.09000, 3.48537)(-0.99000, 3.51225)(-0.89000, 3.55893)(-0.79000, 3.58208)(-0.69000, 3.61212)(-0.59000, 3.65158)(-0.49000, 3.70396)(-0.39000, 3.77421)(-0.29000, 3.86936)(-0.19000, 3.99920)(-0.09000, 4.17708)(0.01000, 4.42065)(0.11000, 4.75197)(0.21000, 5.19708)(0.31000, 5.78477)(0.41000, 6.54758)(0.51000, 6.50377)(0.61000, 6.16293)(0.71000, 5.61467)(0.81000, 5.09286)(0.91000, 4.67695)(1.01000, 4.36526)(1.11000, 4.13632)(1.21000, 3.96817)(1.31000, 3.84308)(1.41000, 3.74812)
\psline[linecolor=black, plotstyle=curve, linewidth=0.3mm, showpoints=false, linestyle=solid, dotstyle=square, dotscale=1.3 1.3](-1.49000, 3.49315)(-1.39000, 3.49908)(-1.29000, 3.50635)(-1.19000, 3.51535)(-1.09000, 3.52658)(-0.99000, 3.54073)(-0.89000, 3.55872)(-0.79000, 3.58184)(-0.69000, 3.61187)(-0.59000, 3.65130)(-0.49000, 3.70364)(-0.39000, 3.77387)(-0.29000, 3.86899)(-0.19000, 3.99880)(-0.09000, 4.17666)(0.01000, 4.42023)(0.11000, 4.75158)(0.21000, 5.19668)(0.31000, 5.78442)(0.41000, 6.54730)(0.51000, 7.31121)(0.61000, 6.37893)(0.71000, 5.65398)(0.81000, 5.09721)(0.91000, 4.67709)(1.01000, 4.36526)(1.11000, 4.13646)(1.21000, 3.96948)(1.31000, 3.84755)(1.41000, 3.75809)
\endpsclip
\psline[linecolor=black, linestyle=solid, linewidth=0.1mm]{->}(0.50000, 2.30000)(0.45000, 3.45943)
\psline[linecolor=black, linestyle=solid, linewidth=0.1mm]{->}(0.50000, 3.50000)(0.45000, 4.39232)
\rput[b](0.50000, 2.30000){\psframebox[linestyle=none, fillcolor=white, fillstyle=solid]{$\SNRSymb_{s,d}=\unit[10]{dB}$}}
\rput[b](0.50000, 3.50000){\psframebox[linestyle=none, fillcolor=white, fillstyle=solid]{$\SNRSymb_{s,d}=\unit[13]{dB}$}}
\psframe[linecolor=black, fillstyle=solid, fillcolor=white, shadowcolor=lightgray, shadowsize=1mm, shadow=true](-0.94737, 6.00000)(0.18421, 7.80000)
\rput[l](-0.71053, 7.50000){Upper bound}
\psline[linecolor=black, linestyle=solid, linewidth=0.3mm](-0.89474, 7.50000)(-0.78947, 7.50000)
\rput[l](-0.71053, 7.10000){Compress-and-Forward}
\psline[linecolor=darkgreen, linestyle=solid, linewidth=0.3mm](-0.89474, 7.10000)(-0.78947, 7.10000)
\psline[linecolor=darkgreen, linestyle=solid, linewidth=0.3mm](-0.89474, 7.10000)(-0.78947, 7.10000)
\psdots[linecolor=darkgreen, linestyle=solid, linewidth=0.3mm, dotstyle=diamond, dotscale=1.3 1.3, linecolor=darkgreen](-0.84211, 7.10000)
\rput[l](-0.71053, 6.70000){Decode-and-Forward}
\psline[linecolor=red, linestyle=solid, linewidth=0.3mm](-0.89474, 6.70000)(-0.78947, 6.70000)
\psline[linecolor=red, linestyle=solid, linewidth=0.3mm](-0.89474, 6.70000)(-0.78947, 6.70000)
\psdots[linecolor=red, linestyle=solid, linewidth=0.3mm, dotstyle=o, dotscale=1.3 1.3, linecolor=red](-0.84211, 6.70000)
\rput[l](-0.71053, 6.30000){Single Hop}
\psline[linecolor=blue, linestyle=solid, linewidth=0.3mm](-0.89474, 6.30000)(-0.78947, 6.30000)
}\end{pspicture}
\endgroup
 }
	\caption[Achievable rates for the Gaussian single-relay channel.]
	{Achievable rates for the Gaussian single-relay channel with non-coherent transmission. Solid curves indicate fixed transmission strategy
	and dashed lines indicate a random transmission schedule.  $\SNRSymb_{s,d}=\unit[13]{dB}$ again indicates the power-normalized case.}
	\label{fig:halfduplex:results:singe_relay}
      \end{figure}
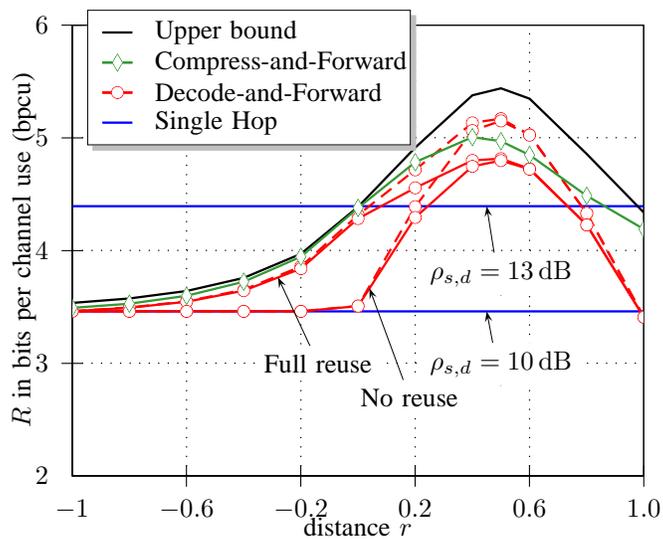
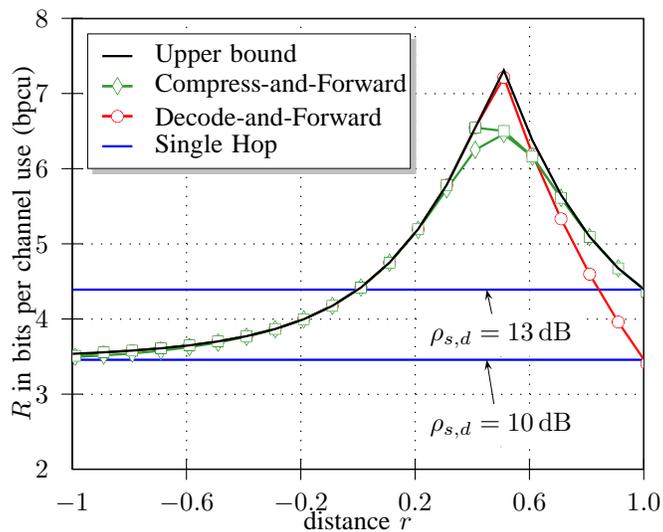}
      Compared to a full-duplex relay network,
      \ac{DF} with a fixed transmission schedule achieves rates which are up to about \unit[2.5]{\bpcu} lower.
      By contrast to full-duplex relaying, in half-duplex relaying \ac{CF} is able to dominate \ac{DF} for all $r$ although the difference is not significant. 
      In addition, none of the protocols is able to achieve the upper bound for $r>0$. For $r>0.4$ a simple multihop protocol without any resource
      reuse achieves the maximum \ac{DF} performance. On the other hand, for $r<0.4$ \ac{DF} with full resource reuse provides a significant gain, which implies that for those scenarios
      where the source-to-relay link is of high quality, it is preferable to form virtual transmit-antenna arrays. Fig. \ref{fig:halfduplex:results:singe_relay}
      does not show the performance of partial \ac{DF} as it does not provide any performance gain over single-level \ac{DF}.
      A comparison of Fig. \ref{fig:halfduplex:results:two_relays}(b) and Fig. \ref{fig:halfduplex:results:singe_relay}(b) reveals that at $r=0.5$
      (mid-way placed relays) two half-duplex \ac{DF} relays are not able to achieve the same performance as one full-duplex \ac{DF} relay. 
      Only the combined strategy with two alternately transmitting relays is able to achieve the same performance as one full-duplex relay.

      \ShowFigureMain{%
      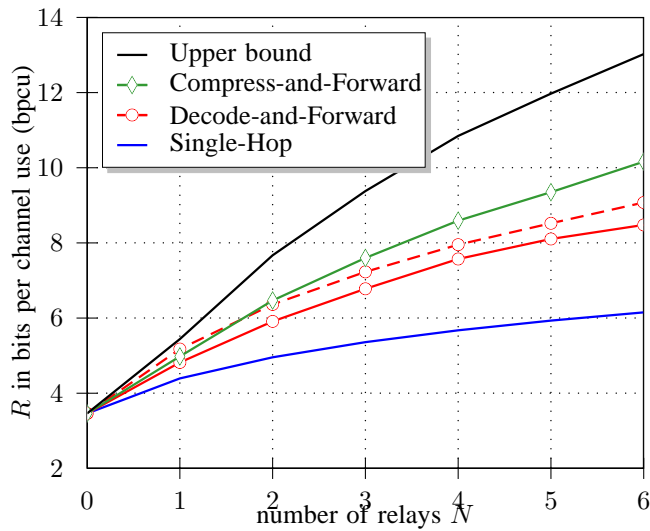
\begin{figure}
	\centering
	\begingroup
\unitlength=1mm
\psset{xunit=12.33333mm, yunit=5.00000mm, linewidth=0.1mm}
\psset{arrowsize=2pt 3, arrowlength=1.4, arrowinset=.4}\psset{axesstyle=frame}
\begin{pspicture}(-0.97297, 0.00000)(6.00000, 14.00000)
\rput(-0.16216, -0.40000){%
\psaxes[subticks=0, labels=all, xsubticks=1, ysubticks=1, Ox=0, Oy=2, Dx=1, Dy=2]{-}(0.00000, 2.00000)(0.00000, 2.00000)(6.00000, 14.00000)%
\multips(1.00000, 2.00000)(1.00000, 0.0){5}{\psline[linecolor=black, linestyle=dotted, linewidth=0.2mm](0, 0)(0, 12.00000)}
\multips(0.00000, 4.00000)(0, 2.00000){5}{\psline[linecolor=black, linestyle=dotted, linewidth=0.2mm](0, 0)(6.00000, 0)}
\rput[b](3.00000, 0.40000){number of relays $N$}
\rput[t]{90}(-0.81081, 8.00000){$R$ in bits per channel use (bpcu)}
\psclip{\psframe(0.00000, 2.00000)(6.00000, 14.00000)}
\psline[linecolor=blue, plotstyle=curve, linewidth=0.3mm, showpoints=false, linestyle=solid, dotstyle=square, dotscale=1.3 1.3](0.00000, 3.45943)(1.00000, 4.39232)(2.00000, 4.95420)(3.00000, 5.35755)(4.00000, 5.67243)(5.00000, 5.93074)(6.00000, 6.14975)
\psline[linecolor=red, plotstyle=curve, linewidth=0.3mm, showpoints=true, linestyle=dashed, dotstyle=o, dotscale=1.3 1.3](0.00000, 3.45943)(1.00000, 5.17304)(2.00000, 6.36074)(3.00000, 7.22769)(4.00000, 7.95141)(5.00000, 8.52024)(6.00000, 9.07311)
\psline[linecolor=red, plotstyle=curve, linewidth=0.3mm, showpoints=true, linestyle=solid, dotstyle=o, dotscale=1.3 1.3](0.00000, 3.45943)(1.00000, 4.81690)(2.00000, 5.90966)(3.00000, 6.77724)(4.00000, 7.57016)(5.00000, 8.10305)(6.00000, 8.47683)
\psline[linecolor=darkgreen, plotstyle=curve, linewidth=0.3mm, showpoints=true, linestyle=solid, dotstyle=diamond, dotscale=1.3 1.3](0.00000, 3.45943)(1.00000, 4.97685)(2.00000, 6.46622)(3.00000, 7.59841)(4.00000, 8.59006)(5.00000, 9.34856)(6.00000, 10.15919)
\psline[linecolor=black, plotstyle=curve, linewidth=0.3mm, showpoints=false, linestyle=solid, dotstyle=square, dotscale=1.3 1.3](0.00000, 3.45943)(1.00000, 5.43864)(2.00000, 7.66850)(3.00000, 9.37718)(4.00000, 10.84493)(5.00000, 11.96995)(6.00000, 13.02263)
\endpsclip
\psframe[linecolor=black, fillstyle=solid, fillcolor=white, shadowcolor=lightgray, shadowsize=1mm, shadow=true](0.16216, 10.00000)(3.64865, 13.60000)
\rput[l](0.89189, 13.00000){Upper bound}
\psline[linecolor=black, linestyle=solid, linewidth=0.3mm](0.32432, 13.00000)(0.64865, 13.00000)
\rput[l](0.89189, 12.20000){Compress-and-Forward}
\psline[linecolor=darkgreen, linestyle=solid, linewidth=0.3mm](0.32432, 12.20000)(0.64865, 12.20000)
\psline[linecolor=darkgreen, linestyle=solid, linewidth=0.3mm](0.32432, 12.20000)(0.64865, 12.20000)
\psdots[linecolor=darkgreen, linestyle=solid, linewidth=0.3mm, dotstyle=diamond, dotscale=1.3 1.3, linecolor=darkgreen](0.48649, 12.20000)
\rput[l](0.89189, 11.40000){Decode-and-Forward}
\psline[linecolor=red, linestyle=solid, linewidth=0.3mm](0.32432, 11.40000)(0.64865, 11.40000)
\psline[linecolor=red, linestyle=solid, linewidth=0.3mm](0.32432, 11.40000)(0.64865, 11.40000)
\psdots[linecolor=red, linestyle=solid, linewidth=0.3mm, dotstyle=o, dotscale=1.3 1.3, linecolor=red](0.48649, 11.40000)
\rput[l](0.89189, 10.60000){Single-Hop}
\psline[linecolor=blue, linestyle=solid, linewidth=0.3mm](0.32432, 10.60000)(0.64865, 10.60000)
}\end{pspicture}
\endgroup
 
	\caption{Achievable rates depending on the network size.  Solid curves indicate fixed transmission strategy a-priori
	known to all nodes, and dashed lines indicate a random transmission schedule which is chosen randomly at each node. Results for single-hop
	are power-normalized such that the power introduced by additional relay nodes is also used in case of single-hop.}
	\label{fig:halfduplex:results:rates_over_noRelays}
      \end{figure}}
      Fig. \ref{fig:halfduplex:results:rates_over_noRelays} shows the achievable rate of the individual protocols for an increasing number
      of relays placed in equal distances. By contrast to the full-duplex channel \cite{Rost.PhD.2009}, \ac{DF} is unable to achieve the cut-set bound for
      an increasing number of relays. The advantage of a random transmission does not increase with the network size, 
      which makes static schedules even more attractive. An open
      challenge is the design of a protocol, which is able to achieve the same performance as the cut-set bound or at least the same within
      a non-increasing interval. \ac{DF} faces the problem that it needs to decode the source message, while
      \ac{CF} increases the effective noise. Hence, the optimal protocol would be a \ac{DF} protocol, which needs not to decode the complete 
      source message but can still provide noise-free redundant information.

      \ShowFigureMain{%
      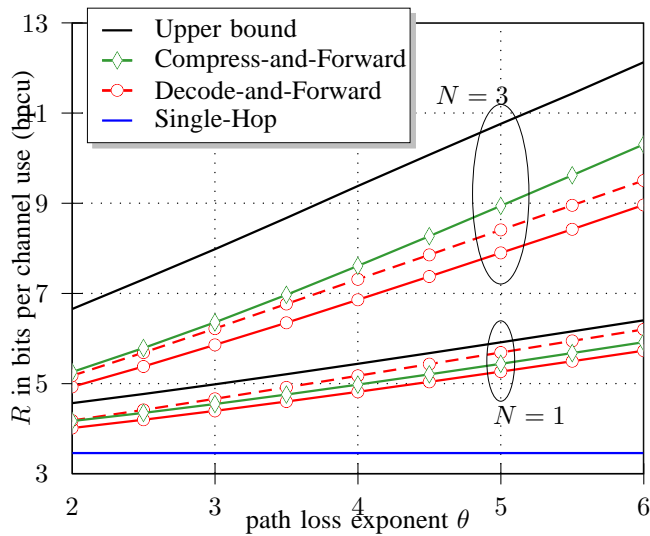
\begin{figure}
	\centering
	\begingroup
\unitlength=1mm
\psset{xunit=19.00000mm, yunit=6.00000mm, linewidth=0.1mm}
\psset{arrowsize=2pt 3, arrowlength=1.4, arrowinset=.4}\psset{axesstyle=frame}
\begin{pspicture}(1.47368, 1.33333)(6.00000, 13.00000)
\rput(-0.10526, -0.33333){%
\psaxes[subticks=0, labels=all, xsubticks=1, ysubticks=1, Ox=2, Oy=3, Dx=1, Dy=2]{-}(2.00000, 3.00000)(2.00000, 3.00000)(6.00000, 13.00000)%
\multips(3.00000, 3.00000)(1.00000, 0.0){3}{\psline[linecolor=black, linestyle=dotted, linewidth=0.2mm](0, 0)(0, 10.00000)}
\multips(2.00000, 5.00000)(0, 2.00000){4}{\psline[linecolor=black, linestyle=dotted, linewidth=0.2mm](0, 0)(4.00000, 0)}
\rput[b](4.00000, 1.66667){path loss exponent $\PathlossExponent$}
\rput[t]{90}(1.57895, 8.00000){$R$ in bits per channel use (bpcu)}
\psclip{\psframe(2.00000, 3.00000)(6.00000, 13.00000)}
\psline[linecolor=blue, plotstyle=curve, linewidth=0.3mm, showpoints=false, linestyle=solid, dotstyle=square, dotscale=1.3 1.3](2.00000, 3.45943)(2.50000, 3.45943)(3.00000, 3.45943)(3.50000, 3.45943)(4.00000, 3.45943)(4.50000, 3.45943)(5.00000, 3.45943)(5.50000, 3.45943)(6.00000, 3.45943)
\psline[linecolor=red, plotstyle=curve, linewidth=0.3mm, showpoints=true, linestyle=solid, dotstyle=o, dotscale=1.3 1.3](2.00000, 4.01428)(2.50000, 4.19805)(3.00000, 4.39471)(3.50000, 4.60171)(4.00000, 4.81690)(4.50000, 5.03851)(5.00000, 5.26512)(5.50000, 5.49563)(6.00000, 5.72921)
\psline[linecolor=red, plotstyle=curve, linewidth=0.3mm, showpoints=true, linestyle=dashed, dotstyle=o, dotscale=1.3 1.3](2.00000, 4.18097)(2.50000, 4.41583)(3.00000, 4.66231)(3.50000, 4.91588)(4.00000, 5.17306)(4.50000, 5.43141)(5.00000, 5.68941)(5.50000, 5.94617)(6.00000, 6.20132)
\psline[linecolor=darkgreen, plotstyle=curve, linewidth=0.3mm, showpoints=true, linestyle=solid, dotstyle=diamond, dotscale=1.3 1.3](2.00000, 4.16950)(2.50000, 4.35000)(3.00000, 4.54681)(3.50000, 4.75671)(4.00000, 4.97685)(4.50000, 5.20478)(5.00000, 5.43855)(5.50000, 5.67666)(6.00000, 5.91794)
\psline[linecolor=black, plotstyle=curve, linewidth=0.3mm, showpoints=false, linestyle=solid, dotstyle=square, dotscale=1.3 1.3](2.00000, 4.56593)(2.50000, 4.76886)(3.00000, 4.98361)(3.50000, 5.20761)(4.00000, 5.43864)(4.50000, 5.67492)(5.00000, 5.91506)(5.50000, 6.15800)(6.00000, 6.40297)
\psline[linecolor=red, plotstyle=curve, linewidth=0.3mm, showpoints=true, linestyle=solid, dotstyle=o, dotscale=1.3 1.3](2.00000, 4.92666)(2.50000, 5.37903)(3.00000, 5.85835)(3.50000, 6.34813)(4.00000, 6.85808)(4.50000, 7.37407)(5.00000, 7.89832)(5.50000, 8.42126)(6.00000, 8.96092)
\psline[linecolor=red, plotstyle=curve, linewidth=0.3mm, showpoints=true, linestyle=dashed, dotstyle=o, dotscale=1.3 1.3](2.00000, 5.17344)(2.50000, 5.68729)(3.00000, 6.21887)(3.50000, 6.76516)(4.00000, 7.31244)(4.50000, 7.85416)(5.00000, 8.40831)(5.50000, 8.95340)(6.00000, 9.50349)
\psline[linecolor=darkgreen, plotstyle=curve, linewidth=0.3mm, showpoints=true, linestyle=solid, dotstyle=diamond, dotscale=1.3 1.3](2.00000, 5.25542)(2.50000, 5.78830)(3.00000, 6.35745)(3.50000, 6.96787)(4.00000, 7.60939)(4.50000, 8.26889)(5.00000, 8.94044)(5.50000, 9.61900)(6.00000, 10.30290)
\psline[linecolor=black, plotstyle=curve, linewidth=0.3mm, showpoints=false, linestyle=solid, dotstyle=square, dotscale=1.3 1.3](2.00000, 6.65233)(2.50000, 7.30983)(3.00000, 7.97993)(3.50000, 8.67260)(4.00000, 9.38101)(4.50000, 10.07435)(5.00000, 10.75966)(5.50000, 11.42964)(6.00000, 12.12546)
\endpsclip
\rput(5, 9.2){\psellipse(0, 0)(0.2, 2)}
\rput(4.8, 11.4){$N=3$}
\rput(5, 5.5){\psellipse(0, 0)(0.1, 0.9)}
\rput(5.2, 4.3){$N=1$}
\psframe[linecolor=black, fillstyle=solid, fillcolor=white, shadowcolor=lightgray, shadowsize=1mm, shadow=true](2.10526, 10.33333)(4.36842, 13.33333)
\rput[l](2.57895, 12.83333){Upper bound}
\psline[linecolor=black, linestyle=solid, linewidth=0.3mm](2.21053, 12.83333)(2.42105, 12.83333)
\rput[l](2.57895, 12.16667){Compress-and-Forward}
\psline[linecolor=darkgreen, linestyle=solid, linewidth=0.3mm](2.21053, 12.16667)(2.42105, 12.16667)
\psline[linecolor=darkgreen, linestyle=solid, linewidth=0.3mm](2.21053, 12.16667)(2.42105, 12.16667)
\psdots[linecolor=darkgreen, linestyle=solid, linewidth=0.3mm, dotstyle=diamond, dotscale=1.3 1.3, linecolor=darkgreen](2.31579, 12.16667)
\rput[l](2.57895, 11.50000){Decode-and-Forward}
\psline[linecolor=red, linestyle=solid, linewidth=0.3mm](2.21053, 11.50000)(2.42105, 11.50000)
\psline[linecolor=red, linestyle=solid, linewidth=0.3mm](2.21053, 11.50000)(2.42105, 11.50000)
\psdots[linecolor=red, linestyle=solid, linewidth=0.3mm, dotstyle=o, dotscale=1.3 1.3, linecolor=red](2.31579, 11.50000)
\rput[l](2.57895, 10.83333){Single-Hop}
\psline[linecolor=blue, linestyle=solid, linewidth=0.3mm](2.21053, 10.83333)(2.42105, 10.83333)
}\end{pspicture}
\endgroup
 
	\caption{Influence of path loss on the achievable rates for $N=1$ and $N=3$ relay nodes which are distributed in equal distances between
	source and destination.  Dashed curves indicate a random transmission schedule and solid lines a fixed schedule.}
	\label{fig:halfduplex:results:path_loss_influence}
      \end{figure}}
      Finally, consider Fig. \ref{fig:halfduplex:results:path_loss_influence} showing the achievable rates depending on the path loss exponent. 
      The performance gain not only increases with the path loss exponent but also the gap between $N=3$ and $N=1$ is increasing in $\PathlossExponent$,
      which underlies that it is highly beneficial to add relay nodes in case of strong shadowing and path loss.

  \section{Conclusions}\label{sec:half.duplex:outro}
    This paper introduced and analyzed different half-duplex protocols using \ac{DF} and \ac{CF} and compared their performance with the
    cut-set bound. In contrast to full-duplex networks, \ac{CF} is able to dominate \ac{DF}. But, the theoretical performance of
    \ac{CF} is not achieved by practical codes yet \cite{Li.Hu.Allerton.2005,Liu.Stankovic.Xiong.2005}. Besides, \ac{DF} can use standard codes such as turbo-codes,
    which are able to closely approach channel capacity. Furthermore, \ac{DF} uses standard encoding and decoding algorithms, which might
    be less complex than the decoding algorithms used for Wyner-Ziv coding. In fading channels \ac{CF} additionally has the problem that the quantization
    levels must be constantly adjusted in order to achieve a reasonable performance, which further limits its applicability due to the high signaling overhead.
    Another important advantage of \ac{DF} is its higher flexibility regarding the number of antennas and deployment, 
    which lets \ac{DF} seem to be favorable over \ac{CF}.  Nonetheless,
    in the case we use \emph{mobile} relay terminals or cooperation on user terminal level, \ac{CF} might be an attractive alternative as the 
    offered performance gains are remarkable.

  \bibliographystyle{IEEEtran}
  \bibliography{IEEEfull,my-references}

  \appendix
  \subsection{Proof of Theorem \ref{theorem:halfduplex:gauss:df}}\label{appendix:proof:theorem:halfduplex:gauss:df}
  In order to prove Theorem \ref{theorem:halfduplex:gauss:df}, we will derive the more general achievable rate region for
  an arbitrary number of relay nodes $N$. The proof is done in two steps: at first we derive the rates for the discrete
  memoryless relay channel and apply then the derived rates to the Gaussian system model introduced in Section \ref{sec:system.model}.
  The achievable rates for the half-duplex discrete memoryless relay channel are an application of \cite[Theorem 3]{Rost.Fettweis.TransIT.2007},
  which derives the \ac{DF} rates for the full-duplex relay channel and are described by the following corollary:
  \begin{corollary}\label{appendix:theorem:half.duplex:1}
    Using the partial decode-and-forward strategy presented in Section \ref{sec:half.duplex:df} we achieve any rate 
    \begin{equation}
      R=\sup\limits_{p\in\Set{P}_\text{DF}}\sum\limits_{m=1}^{N+1}\Message{R}{s}{m},
       \label{appendix:eq:halfduplex:10}
    \end{equation}
    which satisfies
    \begin{align}
      \begin{split}
	\Message{R}{s}{1} & \leq \min\limits_{l\in[1 : N+1]}
	\I\left(\RV{M}_s, \Message{\RV{U}}{s}{1}; \RV{Y}_{l} \big| \left\{\Message{\Vector{\RV{V}}}{[i : N]}{i}\right\}_{i=1}^l, \Vector{\RV{M}}_{[1 : N]}\right) \\
	&\quad{+}\;\sum\limits_{j=1}^{l-1} \I\left(\RV{M}_j,\Message{\RV{V}}{j}{1}; \RV{Y}_{l} \big| \left\{\Message{\Vector{\RV{V}}}{i}{[1;i]}\right\}_{i=j+1}^l, 
	\Message{\Vector{\RV{V}}}{[l : N]}{[1 : l]}, \Vector{\RV{M}}_{[j+1 : N]}\right)
       \end{split}
       \label{appendix:eq:halfduplex:20}\\
      \begin{split}
	\Message{R}{s}{m} & \leq \min\limits_{l\in[m : N+1]}
	\I\left(\Message{\RV{U}}{s}{m}; \RV{Y}_{l} \big| \Message{\Vector{\RV{U}}}{s}{[1 : m-1]}, \left\{\Message{\Vector{\RV{V}}}{[i : N]}{i}\right\}_{i=1}^l, \Vector{\RV{M}}_{[0 : N]}\right) \\
	& \quad {+} \sum\limits_{j=m}^{l-1} \I\left(\Message{\RV{V}}{j}{m}; \RV{Y}_{l} \big| \Message{\Vector{\RV{V}}}{j}{[1 : m-1]}, \left\{\Message{\Vector{\RV{V}}}{i}{[1 : i]}\right\}_{i=j+1}^l, 
	\Message{\Vector{\RV{V}}}{[l : N]}{[1 : l]}, \Vector{\RV{M}}_{[j : N]}\right)
      \end{split}
      \label{appendix:eq:halfduplex:30}
    \end{align}
    for $m\in[2 : N+1]$. The set $\Set{P}_\text{DF}$ is the set of all joint pdf of the form
    \begin{align}
      & p\left(\Vector{\Event{y}}_{[1 : N+1]}, \Message{\Vector{\Event{u}}}{s}{[1 : N+1]}, \Message{\Vector{\Event{v}}}{l\in[1 : N]}{[1 : l]}, \Vector{\Event{m}}_{[0 : N]}\right) =
      p\left(\Vector{\Event{y}}_{[1 : N+1]} \big| \Message{\Vector{\Event{u}}}{s}{[1 : N+1]}, \Message{\Vector{\Event{v}}}{l\in[1 : N]}{[1 : l]}\right)\cdot
      \prod\limits_{l=0}^N p\left(\Event{m}_l \big| \Vector{\Event{m}}_{[l+1 : N]}\right)\nonumber\\
      &\quad{\cdot}\prod\limits_{l=1}^N\prod\limits_{k=1}^l p\left(\Message{\Event{v}}{l}{k} \big| \Message{\Vector{\Event{v}}}{l}{[1 : k-1]}, \Message{\Vector{\Event{v}}}{[l+1 : N]}{k}, \Vector{\Event{m}}_{[l : N]}\right)
      \cdot\prod\limits_{k=1}^{N+1}p\left(\Message{\Event{u}}{s}{k} \big| \Message{\Vector{\Event{u}}}{s}{[1 : k-1]}, \Message{\Vector{\Event{v}}}{l\in[k : N]}{k}, \Vector{\Event{m}}_{\left\{s,[k : N]\right\}}\right).
      \label{appendix:eq:halfduplex:40}
    \end{align}
  \end{corollary}
  \begin{proof}
    Using \cite[Theorem 3]{Rost.Fettweis.TransIT.2007}
    we apply the substitutions $\Message{\RV{U}}{s}{1}\mapsto\left(\Message{\RV{U}}{s}{1}, \RV{M}_s\right)$ and 
    $\Message{\RV{V}}{l}{1}\mapsto\left(\Message{\RV{V}}{l}{1}, \RV{M}_l\right)$ and remove the CF part, yielding the joint pdf in (\ref{appendix:eq:halfduplex:40}). 
    Eq. (\ref{appendix:eq:halfduplex:20}) can be slightly simplified by modifying (\ref{appendix:eq:halfduplex:40}) such that
    the Markov condition $\RV{M}_s\Markov\Message{\RV{U}}{s}{1}\Markov \Message{\Vector{U}}{s}{[2 : N+1]}$ is satisfied (and similarly for all relay messages) which
    yields the results given in \cite{Kramer.2004}.
  \end{proof}

  While the rates for the discrete memoryless channel are easily formulated using a simple modification of the full-duplex channel, the derivation for the
  Gaussian setup is more intricate as the random channel access must be appropriately modeled. Consider the following, more general formulation of
  Theorem \ref{theorem:halfduplex:gauss:df}:
  \begin{theorem}\label{appendix:theorem:half.duplex:20}
    The achievable rate 
    \begin{equation}
      R = \sup\limits_{p\in\Set{P}_\text{DF}}\sum\limits_{k=1}^{N+1} \Message{R}{s}{k}\label{appendix:eq:half.duplex.random:110}
    \end{equation}
    in the Gaussian half-duplex relay network using partial decode-and-forward, a random transmission schedule, and a specific power assignment must satisfy
    \begin{eqnarray}
      \Message{R}{s}{1} & \leq & \min\limits_{l\in[1 : N+1]}
      \MsgPower{Q}{s}{l}{1}\left(\Set{L}_1\right) + \sum\limits_{j=1}^{l-1} \MsgPower{Q}{j}{l}{1}\left(\Set{L}_{j+1}\right) \label{appendix:eq:halfduplex:gauss:120}\\
      \Message{R}{s}{k} & \leq & \min\limits_{l\in[k : N+1]}
      \MsgPower{Q}{s}{l}{k}\left(\Set{L}_0\right) + \sum\limits_{j=k}^{l-1} \MsgPower{Q}{j}{l}{k}\left(\Set{L}_j\right)\label{appendix:eq:halfduplex:gauss:121}
    \end{eqnarray}
    where $\Set{L}_l = [l :  N]$ is the set of nodes for which the state is known. The supremum in (\ref{appendix:eq:half.duplex.random:110}) must be applied
    over all those joint pdf satisfying the individual power constraints and state probabilities as described in Section \ref{sec:half.duplex:df}.
  \end{theorem}
  \begin{proof}
    Let $\RV{Y}\drawniid\CN(0, \sigma_\RV{Y}^2)$ with $\RV{Y} = \RV{A} + j\RV{B}$ such that
    $\RV{A}=\Re(\RV{Y})$ and $\RV{B}=\Im(\RV{Y})$. We can give the pdf of $\RV{Y}$ as follows
    \begin{equation}
      p(\Event{y}) = \frac{1}{\pi\sigma_\RV{Y}^2}\exp\left(-\frac{\Event{a}^2 + \Event{b}^2}{\sigma_\RV{Y}^2}\right).
    \end{equation}
    The entropy of $\RV{Y}$ is given by
    \begin{equation}
      \h\left(\RV{Y}\right) = -\int\limits_{\Set{Y}} p(\Event{y})\log\left(p(\Event{y})\right)\D{\Event{y}}.
    \end{equation}
    Now let $\Event{a} = \Event{r}\cos\varphi$, $\Event{b} = \Event{r}\sin\varphi$, and $r = \sqrt{\Event{a}^2 + \Event{b}^2}$. Using these substitutions we have
    \begin{eqnarray}
      \h\left(\RV{Y}\right) & = & -\int\limits_0^\infty\int\limits_0^{2\pi} p(\Event{y})\log\left(p(\Event{y})\right) \D{\varphi}\D{\Event{r}} \\
      & = & -2\pi\int\limits_0^\infty p(\Event{r})\log\left(p(\Event{r})\right) \D{\Event{r}}
    \end{eqnarray}
    with
    \begin{equation}
      p(\Event{r}) = \frac{1}{\pi\sigma_\RV{Y}^2} \exp\left(-\frac{\Event{r}^2}{\sigma_\RV{Y}^2}\right).
    \end{equation}
    following from the fact the complex Gaussian distribution is circularly symmetric. Now let $\Event{r}' = \Event{r}^2$ we have
    \begin{eqnarray}
      \h\left(\RV{Y}\right) & = & -\pi\int\limits_0^\infty p(\Event{r}')\log\left(p(\Event{r}')\right) \D{\Event{r}'}
    \end{eqnarray}
    with
    \begin{equation}
      p(\Event{r}) = \frac{1}{\pi\sigma_\RV{Y}^2} \exp\left(-\frac{\Event{r}'}{\sigma_\RV{Y}^2}\right).
    \end{equation}
    Now let $\MsgPower{\Vpower}{l}{l'}{k}\left(\Event{\Vector{m}}_{[0;N]}\right)$ denote the overall received power at node $l'$ for message level $k$ sent by node $l$
    as a function of the actual realization of the individual channel states $\Event{\Vector{m}}_{[0;N]}$. 
    Let further $\MsgPower{\sigma}{l}{l'}{k}\left(\Event{\Vector{m}}_{[0;2]}\right)$ denote the variance of
    $\RV{Y}_{l'}$ when all $\Message{\RV{\Vector{V}}}{l}{[1;k]}$ and $\Message{\RV{\Vector{V}}}{j}{[1;l']}$, for $j\in[l+1; N]$, are known and were subtracted from $\RV{Y}_{l'}$:
    \begin{eqnarray}
      \MsgPower{\sigma}{s}{l}{k}\left(\Vector{\Event{m}}_{[0 : N]}\right) & = & \MsgPower{\Vpower}{s}{l}{[k+1 : N+1]}\left(\Vector{\Event{m}}_{[0 : N]}\right) + 
	\MsgPower{\Vpower}{j\in[l+1 : N]}{l'}{[l+1 : j]}\left(\Vector{\Event{m}}_{[0 : N]}\right) + N_l, \label{appendix:eq:halfduplex:160}\\
      \MsgPower{\sigma}{j}{l}{k}\left(\Vector{\Event{m}}_{[0 : N]}\right) & = & \MsgPower{\Vpower}{s}{l}{[1 : N+1]}\left(\Vector{\Event{m}}_{[0 : N]}\right) + 
	\MsgPower{\Vpower}{j'\in[1 : j-1]}{l}{[1 : j']}\left(\Vector{\Event{m}}_{[0 : N]}\right)\nonumber \\ 
	& & {+} \MsgPower{\Vpower}{j}{l}{[k+1 : j]}\left(\Vector{\Event{m}}_{[0 : N]}\right) + \MsgPower{\Vpower}{j'\in[l+1 : N]}{l}{[l+1 : j']}\left(\Vector{\Event{m}}_{[0 : N]}\right) + N_l, \label{appendix:eq:halfduplex:161}\\
      \MsgPower{\Vpower}{j}{l}{k}\left(\Vector{\Event{m}}_{[0 : N]}\right) & = & 
	\left|\sum\limits_{j'\in\left\{s,[k : j]\right\}}
	1\left(\Event{m}_{j'}=T\right)\cdot\left(h_{j',l}\sqrt{\MsgFactor{\Vfactor}{j'}{j}{k} P_{t}}\right)\right|^2.\label{appendix:eq:halfduplex:162}
    \end{eqnarray}
    Then, the differential entropy for the channel output $\RV{Y}_{l'}$
    if the channel states of nodes $\Set{L}$ are known and $\overline{\Set{L}}=[0;N]\setminus\left\{\Set{L}, l\right\}$ are unknown 
    is given by
    \begin{equation}
      \MsgPower{\h}{l}{l'}{k}\left(\Event{\Vector{m}}_\Set{L}\right) = -\pi\int\limits_0^\infty \MsgPower{p}{l}{l'}{k}\left(\Event{y}, \Event{\Vector{m}}_\Set{L}\right)
      \log\left(\MsgPower{p}{l}{l'}{k}\left(\Event{y}, \Event{\Vector{m}}_\Set{L}\right)\right)\D \Event{y}\label{eq:halfduplex:gauss:df:20}
    \end{equation}
    with
    \begin{equation}
      \MsgPower{p}{l}{l'}{k}\left(\Event{y}', \Event{\Vector{m}}_\Set{L}\right) =
	\sum\limits_{\Event{\Vector{m}}_{\overline{\Set{L}}}\in\Set{M}_{\overline{\Set{L}}}}
	p\left(\Event{\Vector{m}}_{\overline{\Set{L}}} | \Event{\Vector{m}}_{\Set{L}}\right)
	\frac{1}{\pi\MsgPower{\sigma}{l}{l'}{k}\left(\Event{\Vector{m}}_{[0;N]}\right)}
	\exp\left(-\frac{\Event{y}'}{\MsgPower{\sigma}{l}{l'}{k}\left(\Event{\Vector{m}}_{[0;N]}\right)}\right).\label{eq:halfduplex:gauss:df:19}
    \end{equation}
    Since $\I\left(\RV{X}; \RV{Y} \big| \RV{Z}\right) = \h\left(\RV{Y} \big| \RV{Z}\right) - \h\left(\RV{Y} \big| \RV{X, Z}\right)$ we can state
    \begin{align}
      \I\left(\RV{U}_s^1; \RV{Y}_l \big| \left\{\RV{V}_{[i : N]}^i\right\}_{i=1}^{l}, \RV{M}_{[1 : N]}\right) & = 
	\mathrm{Q}_{s,l}^1\left(\Set{L}_1\right) \label{eq:half.duplex.random:150} \\
      \I\left(\RV{V}_j^1; \RV{Y}_l \big| \left\{\RV{V}_i^{[1 : i]}\right\}_{i=j+1}^{l}, \RV{\mathbf{V}}_{[l : N]}^{[1 : l]}, \RV{M}_{[j+1 : N]}\right) & = 
	\mathrm{Q}_{j,l}^1\left(\Set{L}_{j+1}\right) \label{eq:half.duplex.random:151} \\
      \I\left(\RV{U}_s^k; \RV{Y}_l \big| \RV{U}_s^{[1 : k-1]}, \left\{\RV{V}_{[i : N]}^i\right\}_{i=1}^{l}, \RV{M}_{\left\{s,[1 : N]\right\}}\right) & =
	\mathrm{Q}_{s,l}^k\left(\Set{L}_0\right) \label{eq:half.duplex.random:152} \\
      \I\left(\RV{V}_j^k; \RV{Y}_l \big| \RV{V}_j^{[1 : k-1]}, \left\{\RV{V}_i^{[1 : i]}\right\}_{i=j+1}^{l}, \RV{\mathbf{V}}_{[l : N]}^{[1;l]}, \RV{M}_{[j : N]}\right) & =
	\mathrm{Q}_{j,l}^k\left(\Set{L}_j\right) \label{eq:half.duplex.random:153},
    \end{align}
    which is sufficient to apply the results of the discrete memoryless half-duplex relay channel in (\ref{appendix:eq:halfduplex:20}) and (\ref{appendix:eq:halfduplex:30}) 
    to the Gaussian half-duplex relay channel. 
  \end{proof}

\subsection{Proof of Theorem \ref{theorem:half.duplex.random:19}}\label{appendix:proof:theorem:halfduplex:gauss:cf}
  In the same way as we proved the previous theorem, we derive again at first the achievable data rates for the discrete memoryless
  channel using the same regular \ac{CF} approach as explained in Section \ref{sec:half.duplex:cf}. Afterwards, we use the derived rates
  and apply them to the Gaussian system model.
  \begin{lemma}\label{appendix:theorem:half.duplex:2}
    The regular encoding \ac{CF} achieves any rate
    \begin{equation}
      R \leq \sup\limits_{p\in\Set{P}_\text{CF}}\I\left(\RV{X}_s; \Vector{\hat{\RV{Y}}}_{[1 : N]}, \RV{Y}_d \big| \Vector{\RV{X}}_{[1 : N]}, \Vector{\RV{M}}_{[0 : N]}\right),\label{appendix:eq:halfduplex:310}
    \end{equation}
    subject to
    \begin{multline}
      \forall l\in[0 :  N-1]: \I\left(\hat{\RV{Y}}_{N-l}; \RV{Y}_{N-l} \big| \Vector{\RV{M}}_{[0 : N]}\right) \leq 
      \I\left(\hat{\RV{Y}}_{N-l}, \RV{X}_{N-l}; \Vector{\hat{\RV{Y}}}_{[N-l+1 :  N]}, \RV{Y}_d \big| \Vector{\RV{X}}_{[N-l+1 :  N]}, \Vector{\RV{M}}_{[0 : N]}\right),
      \label{appendix:eq:halfduplex:311}
    \end{multline}
    and with the supremum over the set $\Set{P}_\text{CF}$ of all joint pdf of the form
    \begin{equation}
      \begin{split}
	& p\left(\Vector{\Event{y}}_{[1 : N+1]}, \Vector{\Event{x}}_{[0 : N]}, \Vector{\hat{\Event{y}}}_{[1 : N]}, \Vector{\Event{m}}_{[0 : N]}\right) =
	p\left(\Vector{\Event{y}}_{[1 : N+1]} \big| \Vector{\Event{x}}_{[0 : N]}, \Vector{\Event{m}}_{[0 : N]}\right)
	{\cdot}\;\prod\limits_{l=1}^N p\left(\hat{\Event{y}}_l \big| \Event{y}_l, \Vector{\Event{m}}_{[0 : N]}\right)\cdot p\left(\Event{x}_l \big| \Vector{\Event{m}}_{[0 : N]}\right).
      \end{split}
      \label{appendix:eq:halfduplex:312}
    \end{equation}
  \end{lemma}
  \begin{proof}
    Let $\Delta_l$ denote the rate of the quantization at node $l$, then we know from rate distortion theory \cite[Ch. 13]{Cover.Thomas.1991} that it is
    lower bounded by
    \begin{equation}
      \Delta_l \geq \I\left(\hat{\RV{Y}}_l; \RV{Y}_l \big| \Vector{\RV{M}}_{[0 : N]}\right).\label{appendix:eq:halfduplex:320}
    \end{equation}
    To decode the quantization index of node $N-l$ corresponding to the destination channel output in block $b-l-1$, the destination
    searches for a quantization that is jointly typical with its channel output, the quantizations of the previous nodes, and the
    broadcast message transmitted by node $N-l$. More formally, it searches for an index
    \begin{align*}
      \exists \RVBlock{\hat q}{N-l}{b-l}: \RVBlock{\hat q}{N-l}{b-l} & =
      \biggl\{\RVBlock{\tilde q}{N-l}{b-l}: 
      \Bigl(\hat{\Event{y}}_{N-l}\left(\RVBlock{\tilde q}{N-l}{b-l}\right),
      \left\{\hat{\Event{y}}_{N-l'}\left(\RVBlock{q}{N-l'}{b-l}\right)\right\}_{l'=0}^{l-1},
      \left\{\Event{x}_{N-l'}\left(\RVBlock{q}{N-l'}{b-l-1}\right)\right\}_{l'=0}^{l},
	\Event{y}_d\left(b-l-1\right)\Bigr)\in\StronglyTypicalSet\biggr\} \\
	& \quad {\cap}\biggl\{\RVBlock{\tilde q}{N-l}{b-l}: 
	\Bigl(\Event{x}_{N-l}\left(\RVBlock{\tilde q}{N-l}{b-l}\right), 
	\left\{\hat{\Event{y}}_{N-l'}\left(\RVBlock{q}{N-l'}{b-l+1}\right)\right\}_{l'=0}^{l-1},
	\left\{\Event{x}_{N-l'}\left(\RVBlock{q}{N-l'}{b-l}\right)\right\}_{l'=0}^{l-1},
	\Event{y}_d\left(b-l\right)\Bigr)\in\StronglyTypicalSet\biggr\},
    \end{align*}
    where $\StronglyTypicalSet$ is the $\epsilon$-strongly typical set. The requirement
    of \emph{strong} typicality arises from the necessity to apply the Markov lemma \cite[Lemma 14.8.1]{Cover.Thomas.1991}
    to prove joint typicality.
    The previous equation can only be fulfilled iff (\ref{appendix:eq:halfduplex:320}) holds and the quantization rate is upper bounded by
    \begin{align*}
      \begin{split}
	\Delta_{N-l} & \leq \I\left(\hat{\RV{Y}}_{N-1}; \Vector{\hat{\RV{Y}}}_{[N-l+1 :  N]}, \RV{Y}_d \big| \Vector{\RV{X}}_{[N-l :  N]}, \Vector{\RV{M}}_{[0 : N]}\right)
	{+}\;\I\left(\RV{X}_{N-l}; \Vector{\hat{\RV{Y}}}_{[N-l+1 :  N]}, \RV{Y}_d \big| \Vector{\RV{X}}_{[N-l+1 :  N]}, \Vector{\RV{M}}_{[0 : N]}\right)
	\end{split}\\
	& \leq \I\left(\hat{\RV{Y}}_{N-l}, \RV{X}_{N-l}; \Vector{\hat{\RV{Y}}}_{[N-l+1 :  N]}, \RV{Y}_d \big| \Vector{\RV{X}}_{[N-l+1 :  N]}, \Vector{\RV{M}}_{[0 : N]}\right).
    \end{align*}
    Similarly the destination decodes in block $b$ the source message transmitted in block
    $b-N$ iff (\ref{appendix:eq:halfduplex:310}) holds. Using standard methods extensively discussed in 
    literature \cite{Cover.Thomas.1991} and in the previous section, (\ref{appendix:eq:halfduplex:311}) and the proof for achievability follow.
  \end{proof}
  Using the previous lemma, we can now derive the achievable rates for the Gaussian system model.
  \begin{theorem}\label{appendix:theorem:half.duplex:21}
    The regular \ac{CF} approach achieves in a Gaussian system model any rate
    \begin{equation}
      R \leq \sup\limits_{p\in\Set{P}_\text{CF}}
      \sum\limits_{\Vector{\Event{m}}_{[0 : N]}\in\Set{M}_{[0 : N]}}p\left(\Vector{\Event{m}}_{[0 : N]}\right)
      \log\left(\frac{\left\|\Matrix{K}_{s, [1 : N]}\left(\Vector{\Event{m}}_{[0 : N]}\right)\right\|}{\left\|\Matrix{K}_{\emptyset, [1 : N]}\left(\Vector{\Event{m}}_{[0 : N]}\right)\right\|}\right),
      \label{eq:half.duplex.random:200}
    \end{equation}
    subject to
    \begin{align}
      & \sum\limits_{\substack{\Vector{\Event{m}}_{[0 : N]}\in\Set{M}_{[0 : N]}:\\\Event{m}_{N-l}=L}}p\left(\Vector{\Event{m}}_{[0 : N]}\right)
      \Biggl[
      \C\left(\frac{\MsgPower{\Vpower}{[N-l : N]}{N-l}{~}}{\MsgPower{\Vpower}{[0 : N-l-1]}{N-l}{~}+ N_{N-l, 1} + N_{N-l}}\right) +\nonumber \\
      & \log\left(\frac{\left\|\Matrix{K}_{[0 :  N-l-1], [N-l : N]}\right\|}{N_{N-l, 1} \left\| \Matrix{K}_{[0 :  N-l-1], [N-l+1 :  N]} \right\|}\right)
      \Biggr]\!\leq\!\sum\limits_{\substack{\Vector{\Event{m}}_{[0 : N]}\in\Set{M}_{[0 : N]}:\\\Event{m}_{N-l}=T}}\hspace{-0.5cm}p\left(\Vector{\Event{m}}_{[0 : N]}\right)
      \log\left(\frac{\left\|\Matrix{K}_{[0 :  N-l], [N-l+1 : N]}\right\|}
      {\left\|\Matrix{K}_{[0 :  N-l-1], [N-l+1 : N]}\right\|}\right)\label{eq:half.duplex.random:210}
    \end{align}
    with the \emph{symmetric} Matrix $\Matrix{K}$ defined as follows:
    \begin{eqnarray}
      \MatrixElement{1}{1}{\Matrix{K}_{\Set{L}, \Set{L}'}\left(\Event{m}_{[0 : N]}\right)} & = & \MsgPower{\Vpower}{\Set{L}}{d}{~} + N_d \label{eq:half.duplex.random:220}\\
      \MatrixElement{j+1}{j+1}{\Matrix{K}_{\Set{L}, \Set{L}'}\left(\Event{m}_{[0 : N]}\right)} & = & \MsgPower{\Vpower}{\Set{L}}{j}{~} + N_{j, 1} + N_j \label{eq:half.duplex.random:221} \\
      \MatrixElement{1}{j+1}{\Matrix{K}_{\Set{L}, \Set{L}'}\left(\Event{m}_{[0 : N]}\right)} & = & \MsgCorrelation{\Vpower}{\Set{L}}{d}{j}{~} \label{eq:half.duplex.random:222} \\
      \MatrixElement{j'+1}{j+1}{\Matrix{K}_{\Set{L}, \Set{L}'}\left(\Event{m}_{[0 : N]}\right)} & = & \MsgCorrelation{\Vpower}{\Set{L}}{j'}{j}{~} \label{eq:half.duplex.random:223}
    \end{eqnarray}
    where $j\in\Set{L}'\setminus\left\{j'\in\Set{L}: \Event{m}_{j'}=T\right\}$.
    The supremum in (\ref{eq:half.duplex.random:200}) is over all $p\in\Set{P}_\text{CF}$ satisfying (\ref{appendix:eq:halfduplex:312}) as well as
    the power constraints given in Section \ref{sec:half.duplex:cf}.
  \end{theorem}
  \begin{proof}
    Eq. (\ref{appendix:eq:halfduplex:310}) can be rewritten as
    \begin{eqnarray}
      R & \leq & \I\left(\RV{X}_s; \Vector{\hat{\RV{Y}}}_{[1 : N]}, \RV{Y}_d \big| \Vector{\RV{X}}_{[1 : N]}, \Vector{\RV{M}}_{[0 : N]}\right) \\
      & = & \Entropy\left(\Vector{\hat{\RV{Y}}}_{[1 : N]}, \RV{Y}_d \big| \Vector{\RV{X}}_{[1 : N]}, \Vector{\RV{M}}_{[0 : N]}\right) - 
      \Entropy\left(\Vector{\hat{\RV{Y}}}_{[1 : N]}, \RV{Y}_d \big| \Vector{\RV{X}}_{[0 : N]}, \Vector{\RV{M}}_{[0 : N]}\right).
    \end{eqnarray}
    The variance of the r.v. in the former term is expressed by $\Matrix{K}_{s, [1 : N]}\left(\Vector{\Event{m}}_{[0 : N]}\right)$ and of the
    second term by $\Matrix{K}_{\emptyset, [1 : N]}\left(\Vector{\Event{m}}_{[0 : N]}\right)$. Hence, if we use both terms in the previous
    equation and sum over all possible joint node states we obtain (\ref{eq:half.duplex.random:200}) defining the maximum achievable rate.
    Consider now the side constraint in (\ref{appendix:eq:halfduplex:311}), which can be reformulated as follows:
    \begin{align}
      \I\left(\hat{\RV{Y}}_{N-l}; \RV{Y}_{N-l} \big| \RV{M}_{[s : N]}\right) & \leq
	\I\left(\hat{\RV{Y}}_{N-l}; \hat{\RV{Y}}_{[N-l+1 :  N]}, \RV{Y}_d \big| \RV{X}_{[N-l :  N]}, \RV{M}_{[s : N]}\right) \nonumber \\
	& \quad{+}\;\I\left(\RV{X}_{N-l}; \hat{\RV{Y}}_{[N-l+1 :  N]}, \RV{Y}_d \big| \RV{X}_{[N-l+1 :  N]}, \RV{M}_{[s : N]}\right) \\
      \I\left(\RV{X}_{N-l}; \hat{\RV{Y}}_{[N-l+1 :  N]}, \RV{Y}_d \big| \RV{X}_{[N-l+1 :  N]}, \RV{M}_{[s : N]}\right) & \geq
      \I\left(\hat{\RV{Y}}_{N-l}; \RV{Y}_{N-l} \big| \RV{M}_{[s : N]}\right) \nonumber \\
      & \quad{-}\;\I\left(\hat{\RV{Y}}_{N-l}; \hat{\RV{Y}}_{[N-l+1 :  N]}, \RV{Y}_d \big| \RV{X}_{[N-l :  N]}, \RV{M}_{[s : N]}\right)
      \label{eq:half.duplex.random:230}
    \end{align}
    Let us pay particular attention to the l.h.s of (\ref{eq:half.duplex.random:230}), which can be reformulated to
    \begin{multline}
      \I\left(\RV{X}_{N-l}; \Vector{\hat{\RV{Y}}}_{[N-l+1 :  N]}, \RV{Y}_d \big| \Vector{\RV{X}}_{[N-l+1 :  N]}, \Vector{\RV{M}}_{[0 : N]}\right) = \\
	\sum\limits_{\substack{\Vector{\Event{m}}_{[0 : N]}\in\Set{M}_{[0 : N]}:\\\Event{m}_{N-l}=T}}p\left(\Vector{\Event{m}}_{[0 : N]}\right)\cdot\biggl(
	\Entropy\left(\Vector{\hat{\RV{Y}}}_{[N-l+1 :  N]}, \RV{Y}_d \big| \Vector{\RV{X}}_{[N-l+1 :  N]}, \Vector{\RV{M}}_{[0 : N]} = \Vector{\Event{m}}_{[0 : N]}\right) - \\
	\Entropy\left(\Vector{\hat{\RV{Y}}}_{[N-l+1 :  N]}, \RV{Y}_d \big| \Vector{\RV{X}}_{[N-l :  N]}, \Vector{\RV{M}}_{[0 : N]} = \Vector{\Event{m}}_{[0 : N]}\right)\biggr).
    \end{multline}
    If we again apply the definitions in (\ref{eq:half.duplex.random:220})-(\ref{eq:half.duplex.random:223}), we obtain the l.h.s of (\ref{eq:half.duplex.random:210}).
    Now, consider the r.h.s. of (\ref{eq:half.duplex.random:230}), which can be reformulated to
    \begin{eqnarray*}
      & & \text{r.h.s of (\ref{eq:half.duplex.random:230})} \\
      & = & \I\left(\hat{\RV{Y}}_{N-l}; \RV{Y}_{N-l} \big| \Vector{\RV{M}}_{[0 : N]}\right) - \I\left(\hat{\RV{Y}}_{N-l}; \Vector{\hat{\RV{Y}}}_{[N-l+1 :  N]}, 
      \RV{Y}_d \big| \Vector{\RV{X}}_{[N-l :  N]}, \Vector{\RV{M}}_{[0 : N]}\right) \\
	& = & \Entropy\left(\hat{\RV{Y}}_{N-l} \big| \Vector{\RV{M}}_{[0 : N]}\right) - \Entropy\left(\hat{\RV{Y}}_{N-l} \big| \RV{Y}_{N-l}, \Vector{\RV{M}}_{[0 : N]}\right)
	- \Entropy\left(\Vector{\hat{\RV{Y}}}_{[N-l+1 :  N]}, \RV{Y}_d \big| \Vector{\RV{X}}_{[N-l :  N]}, \Vector{\RV{M}}_{[0 : N]}\right) \\
	&  & {+}\;\Entropy\left(\Vector{\hat{\RV{Y}}}_{[N-l+1 :  N]}, \RV{Y}_d \big| \Vector{\RV{X}}_{[N-l :  N]}, \hat{\RV{Y}}_{N-l}, \Vector{\RV{M}}_{[0 : N]}\right) \\
	& = & \Entropy\left(\hat{\RV{Y}}_{N-l} \big| \Vector{\RV{M}}_{[0 : N]}\right) - \Entropy\left(\hat{\RV{Y}}_{N-l} \big| \Vector{\RV{X}}_{[N-l :  N]}, \Vector{\RV{M}}_{[0 : N]}\right)
	  - \Entropy\left(\Vector{\hat{\RV{Y}}}_{[N-l+1 :  N]}, \RV{Y}_d \big| \Vector{\RV{X}}_{[N-l :  N]}, \Vector{\RV{M}}_{[0 : N]}\right) \\
	  & & {+}\;\Entropy\left(\Vector{\hat{\RV{Y}}}_{[N-l :  N]}, \RV{Y}_d \big| \Vector{\RV{X}}_{[N-l :  N]}, \Vector{\RV{M}}_{[0 : N]}\right) 
	  - \Entropy\left(\hat{\RV{Y}}_{N-l} \big| \RV{Y}_{N-l}, \Vector{\RV{M}}_{[0 : N]}\right) 
    \end{eqnarray*}
    Again, if we apply the definitions of (\ref{eq:half.duplex.random:220})-(\ref{eq:half.duplex.random:223}), we obtain the r.h.s of (\ref{eq:half.duplex.random:210}).
  \end{proof}

  \clearpage
  \ShowFigureEnd{%
\begin{figure}
  \centering
  \subfigure[Irregular encoding]{\begingroup
\unitlength=1mm
\begin{picture}(87, 40)(0, 0)
  \psset{xunit=1mm, yunit=1mm, linewidth=0.1mm}
  \psset{arrowsize=3pt 4, arrowlength=1.4, arrowinset=.4}\psset{axesstyle=frame}%
  \psset{shadowcolor=gray, shadowsize=1mm}%

  \rput(-3, -10){%
    \rput(5, 30){\rnode{Source}{$\RV{Y}$}}%
    \rput(15, 30){\rnode{Encoder}{\psframebox[framearc=.2, shadow=true]{\rotateleft{Encoder}}}}%
    \ncline{->}{Source}{Encoder}%
    \rput(35, 30){\rnode{Channel}{\psframebox[framearc=.2, shadow=true]{\rotateleft{$p\left(\Event{y}_{1}, \Event{y}_2 | \Event{w}\right)$}}}}%
    \ncline{->}{Encoder}{Channel}\ncput*{$\RV{W}$}%
    \pnode(38.3, 36){Channel1}%
    \pnode(38.3, 24){Channel2}%
%
    \rput(55, 36){\rnode{G11}{\psframebox[framearc=.2, shadow=true]{$g_{2}$}}}%
    \ncline{->}{Channel1}{G11}\ncput*{$\RV{Y}_{2}$}%
    \rput(75, 36){\rnode{G12}{\psframebox[framearc=.2, shadow=true]{$g_{2}'$}}}%
    \ncline{->}{G11}{G12}\ncput*{$\hat{\RV{W}}$}%
    \rput(75, 48){\rnode{Side1}{$\RV{Y}_2$}}%
    \ncline{->}{Side1}{G12}%
    \rput(88, 36){\rnode{Result}{$\hat{\hat{\RV{Y}}}$}}%
    \ncline{->}{G12}{Result}%
    \rput(55, 24){\rnode{G21}{\psframebox[framearc=.2, shadow=true]{$g_{1}$}}}%
    \ncline{->}{Channel2}{G21}\ncput*{$\RV{Y}_{1}$}%
    \rput(75, 24){\rnode{G22}{\psframebox[framearc=.2, shadow=true]{$g_{1}'$}}}%
    \ncline{->}{G21}{G22}\ncput*{$\hat{\RV{W}}$}%
    \rput(75, 12){\rnode{Side2}{$\RV{Y}_1$}}%
    \ncline{->}{Side2}{G22}%
    \rput(88, 24){\rnode{Result}{$\hat{\hat{\RV{Y}}}$}}%
    \ncline{->}{G22}{Result}%
    }
  
\end{picture}
\endgroup}
\subfigure[Regular encoding]{\begingroup
\unitlength=1mm
\begin{picture}(67, 40)(0, 0)
  \psset{xunit=1mm, yunit=1mm, linewidth=0.1mm}
  \psset{arrowsize=3pt 4, arrowlength=1.4, arrowinset=.4}\psset{axesstyle=frame}%
  \psset{shadowcolor=gray, shadowsize=1mm}%

  \rput(-3, -10){%
    \rput(5, 30){\rnode{Source}{$\RV{Y}$}}%
    \rput(15, 30){\rnode{Encoder}{\psframebox[framearc=.2, shadow=true]{\rotateleft{Encoder}}}}%
    \ncline{->}{Source}{Encoder}%
    \rput(35, 30){\rnode{Channel}{\psframebox[framearc=.2, shadow=true]{\rotateleft{$p\left(\Event{y}_{1}, \Event{y}_2 | \Event{w}\right)$}}}}%
    \ncline{->}{Encoder}{Channel}\ncput*{$\RV{W}$}%
    \pnode(38.3, 36){Channel1}%
    \pnode(38.3, 24){Channel2}%
%
    \rput(55, 36){\rnode{G12}{\psframebox[framearc=.2, shadow=true]{$g_{2}$}}}%
    \ncline{->}{Channel1}{G12}\ncput*{$\RV{Y}_{2}$}%
    \rput(55, 48){\rnode{Side1}{$\RV{Y}_2$}}%
    \ncline{->}{Side1}{G12}%
    \rput(68, 36){\rnode{Result}{$\hat{\hat{\RV{Y}}}$}}%
    \ncline{->}{G12}{Result}%
    \rput(55, 24){\rnode{G22}{\psframebox[framearc=.2, shadow=true]{$g_{1}$}}}%
    \ncline{->}{Channel2}{G22}\ncput*{$\RV{Y}_{1}$}%
    \rput(55, 12){\rnode{Side2}{$\RV{Y}_1$}}%
    \ncline{->}{Side2}{G22}%
    \rput(68, 24){\rnode{Result}{$\hat{\hat{\RV{Y}}}$}}%
    \ncline{->}{G22}{Result}%
    }
  
\end{picture}
\endgroup}
  \caption{Two different strategies for \ac{CF} with multiple receivers.}
  \label{fig:halfduplex:regular.CF}
\end{figure}
      \begin{figure}
	\centering
	\begingroup
\unitlength=1mm
\begin{picture}(101, 38)(0, 0)
  \psset{xunit=1mm, yunit=1mm, linewidth=0.1mm}
  \psset{arrowsize=3pt 4, arrowlength=1.4, arrowinset=.4}

  \rput(17, 28)
  {
  \rput[r](15, 0){\textcolor{phase1}{$\begin{array}{c}\text{Phase 1, $[1; n_1]$:}\\\text{($\Event{x}_{s,1}$ with rate $R_{\DF}$)}\end{array}$}}
    \cnodeput(20, 0){S}{$s$}
    \cnodeput(40, 0){R1}{$1$}
    \cnodeput(60, 0){R2}{$2$}
    \cnodeput(80, 0){D}{$d$}
    \ncline[linecolor=phase1]{->}{S}{R1}\nbput{$\Event{x}_{s,1}$}
    \ncline[linestyle=dashed, linecolor=phase2]{<-}{R1}{R2}\nbput{$\Event{x}_2$}
    \ncline[linecolor=phase2]{->}{R2}{D}\nbput{$\Event{x}_2$}
    \ncarc[arcangle=20, linestyle=solid, linecolor=phase1]{->}{S}{D}\naput{$\Event{x}_{s,1}$}
  }

  \rput(17, 8)
  {
  \rput[r](15, 0){\textcolor{phase2}{$\begin{array}{c}\text{Phase 2, $[n_1+1; n]$:}\\\text{($\Event{x}_{s,2}$ with rate $R_{\CF}$)}\end{array}$}}
    \cnodeput(20, 0){S}{$s$}
    \cnodeput(40, 0){R1}{$1$}
    \cnodeput(60, 0){R2}{$2$}
    \cnodeput(80, 0){D}{$d$}
    \ncline[linecolor=phase1]{->}{R1}{R2}\nbput{$\Event{x}_1$}
    \ncarc[arcangle=35, linecolor=phase2, linestyle=solid]{->}{S}{D}\nbput{$\Event{x}_{s,2}$}
    \ncarc[arcangle=20, linecolor=phase2, linestyle=solid]{->}{S}{R2}
    \ncarc[arcangle=-40, linecolor=phase1, linestyle=solid]{->}{R1}{D}
  }
\end{picture}
\endgroup
	\caption[Example for a half-duplex channel with two alternately transmitting relay nodes.]
	{Example for a half-duplex channel with two alternately transmitting relay nodes.
	The solid lines indicate actual information exchange while the dashed line indicates the interfering
	transmission from node $2$ to $1$.}
	\label{figure:halfduplex.random:alternately_transmitting}
      \end{figure}
      \begin{figure}
	\centering
	\begingroup
\unitlength=1mm
\begin{picture}(112, 49)(0, 0)
  \psset{xunit=1mm, yunit=1mm, linewidth=0.1mm}
  \psset{arrowsize=3pt 4, arrowlength=1.4, arrowinset=.4}

  \rput(0, 7){%
    \rput[c](7, 13){Relay $2$:}
    \rput(15, 8)
    {%
      \psframe(0, 0)(40, 10)
      \rput[c](20, 5){\textcolor{phase2}{$\RV{x}_2(\RVBlock{q}{2}{b})$}}
      \psbrace[rot=90, linewidth=0.1mm, ref=t, nodesepB=-2mm](0, 0)(40, 0){\textcolor{phase1}{$\begin{array}{c}n_1 = n\cdot p_1\\\text{(Phase 1)}\end{array}$}}
      \psframe(42, 0)(97, 10)
      \rput[c](69.5, 5){$\RV{y}_2(b)\mapsto\hat{\RV{y}}_2(\RVBlock{q}{2}{b+1})\mapsto\textcolor{phase2}{\RV{x}_2(\RVBlock{q}{2}{b+1})}$}
      \psbrace[rot=90, linewidth=0.1mm, ref=t, nodesepB=-2mm](42, 0)(97, 0){\textcolor{phase2}{$\begin{array}{c}n_2 = n\cdot p_2\\\text{(Phase 2)}\end{array}$}}
    }

    \rput[c](7, 25){Relay $1$:}
    \rput(15, 20)
    {%
      \psframe(0, 0)(40, 10)
      \rput[c](20, 5){$\RV{y}_1(b)\mapsto\textcolor{phase1}{\RV{x}_1(\RVBlock{q}{1}{b})}$}
      \psframe(42, 0)(97, 10)
      \rput[c](69.5, 5){\textcolor{phase1}{$\RV{x}_1(\RVBlock{q}{1}{b})$}}
    }

    \rput[c](7, 37){Source:}
    \rput(15, 32)
    {%
      \psframe(0, 0)(40, 10)
      \rput[c](20, 5){\textcolor{phase1}{$x_{s,1}(\MsgBlock{q}{s}{1}{b})$}}
      \psframe(42, 0)(97, 10)
      \rput[c](69.5, 5){\textcolor{phase2}{$\RV{x}_{s,2}(\MsgBlock{q}{s}{2}{b})$}}
    }
  }

\end{picture}
\endgroup
	\caption[Coding structure for alternatively transmitting relay nodes.]
	{Coding structure for the combined strategy with $N=2$ alternately transmitting relays.}
	\label{figure:halfduplex.random:encoding_mixed_yarp}
      \end{figure}
    \begin{figure}
      \centering
      \begingroup
\unitlength=1mm
\begin{picture}(65, 20)(0, 0)

  \psset{xunit=1mm, yunit=1mm, linewidth=0.2mm}

  \rput(2, 5){\cnodeput(0, 0){Source}{$s$}}
  \rput(60, 5){\cnodeput(0, 0){Destination}{$d$}}
  \rput(20, 5){\cnodeput(0, 0){Relay1}{$1$}}
  \rput(42, 5){\cnodeput(0, 0){Relay2}{$2$}}

  \pnode(2, 18){Source1}
  \pnode(2, 11){Source2}
  \pnode(2, 16){Source3}
  \pnode(60, 18){Destination1}
  \pnode(60, 11){Destination2}
  \pnode(60, 16){Destination3}
  \pnode(20, 13){Relay11}
  \pnode(20, 11){Relay12}
  \pnode(42, 13){Relay21}
  \pnode(42, 11){Relay22}

  \ncline{-}{Source}{Source1}
  \ncline{-}{Destination}{Destination1}
  \ncline{-}{Relay1}{Relay11}
  \ncline{-}{Relay2}{Relay21}

  \ncline{<->}{Source2}{Relay12}\naput{$r$}
  \ncline{<->}{Relay22}{Destination2}\naput{$r$}
  \ncline{<->}{Source3}{Destination3}\naput{$1$}
\end{picture}
\endgroup
      \caption{Setup for our analysis.}
      \label{fig:halfduplex:setup}
    \end{figure}
      \begin{figure}
	\centering
	\subfigure[Coherent transmission]{\begingroup
\unitlength=1mm
\psset{xunit=76.00000mm, yunit=10.66667mm, linewidth=0.1mm}
\psset{arrowsize=2pt 3, arrowlength=1.4, arrowinset=.4}\psset{axesstyle=frame}
\begin{pspicture}(-0.63158, 0.50000)(0.50000, 8.00000)
\rput(-0.02632, -0.75000){%
\psaxes[subticks=0, labels=all, xsubticks=1, ysubticks=1, Ox=-0.5, Oy=2, Dx=0.2, Dy=1]{-}(-0.50000, 2.00000)(-0.50000, 2.00000)(0.50000, 8.00000)%
\multips(-0.30000, 2.00000)(0.20000, 0.0){4}{\psline[linecolor=black, linestyle=dotted, linewidth=0.2mm](0, 0)(0, 6.00000)}
\multips(-0.50000, 3.00000)(0, 1.00000){5}{\psline[linecolor=black, linestyle=dotted, linewidth=0.2mm](0, 0)(1.00000, 0)}
\rput[b](0.00000, 1.25000){distance $r$}
\rput[t]{90}(-0.60526, 5.00000){$R$ in bits per channel use (bpcu)}
\psclip{\psframe(-0.50000, 2.00000)(0.50000, 8.00000)}
\psline[linecolor=blue, plotstyle=curve, linewidth=0.3mm, showpoints=false, linestyle=solid, dotstyle=square, dotscale=1.3 1.3](-0.50000, 3.45943)(-0.40000, 3.45943)(-0.30000, 3.45943)(-0.20000, 3.45943)(-0.10000, 3.45943)(0.00000, 3.45943)(0.10000, 3.45943)(0.20000, 3.45943)(0.23333, 3.45943)(0.26667, 3.45943)(0.30000, 3.45943)(0.33333, 3.45943)(0.36667, 3.45943)(0.40000, 3.45943)(0.43333, 3.45943)(0.46667, 3.45943)(0.50000, 3.45943)
\psline[linecolor=blue, plotstyle=curve, linewidth=0.3mm, showpoints=false, linestyle=solid, dotstyle=square, dotscale=1.3 1.3](-0.50000, 4.95420)(-0.40000, 4.95420)(-0.30000, 4.95420)(-0.20000, 4.95420)(-0.10000, 4.95420)(0.00000, 4.95420)(0.10000, 4.95420)(0.20000, 4.95420)(0.23333, 4.95420)(0.26667, 4.95420)(0.30000, 4.95420)(0.33333, 4.95420)(0.36667, 4.95420)(0.40000, 4.95420)(0.43333, 4.95420)(0.46667, 4.95420)(0.50000, 4.95420)
\psline[linecolor=red, plotstyle=curve, linewidth=0.3mm, showpoints=true, linestyle=dashed, dotstyle=o, dotscale=1.3 1.3](-0.50000, 2.13197)(-0.40000, 2.50654)(-0.30000, 2.94868)(-0.20000, 3.47666)(-0.10000, 4.12514)(0.00000, 5.02260)(0.10000, 5.48108)(0.20000, 5.92645)(0.23333, 6.06518)(0.26667, 6.19272)(0.30000, 6.30683)(0.33333, 6.40046)(0.36667, 6.46772)(0.40000, 6.50174)(0.43333, 6.50446)(0.46667, 6.48355)(0.50000, 6.49679)
\psline[linecolor=red, plotstyle=curve, linewidth=0.3mm, showpoints=true, linestyle=dashed, dotstyle=triangle, dotscale=1.3 1.3](-0.50000, 3.77410)(-0.40000, 3.89566)(-0.30000, 4.04644)(-0.20000, 4.23944)(-0.10000, 4.49818)(0.00000, 5.02217)(0.10000, 5.47937)(0.20000, 5.91468)(0.23333, 6.06443)(0.26667, 6.18219)(0.30000, 6.30141)(0.33333, 6.39214)(0.36667, 6.44929)(0.40000, 6.48725)(0.43333, 6.49267)(0.46667, 6.46667)(0.50000, 6.46472)
\psline[linecolor=red, plotstyle=curve, linewidth=0.3mm, showpoints=true, linestyle=dashed, dotstyle=o, dotscale=1.3 1.3](-0.50000, 1.54770)(-0.40000, 1.81968)(-0.30000, 2.13606)(-0.20000, 2.50209)(-0.10000, 2.92359)(0.00000, 3.40806)(0.10000, 4.30553)(0.20000, 5.19412)(0.23333, 5.44248)(0.26667, 5.64206)(0.30000, 5.77657)(0.33333, 5.83340)(0.36667, 5.81122)(0.40000, 5.72046)(0.43333, 5.57936)(0.46667, 5.40569)(0.50000, 5.21120)
\psline[linecolor=black, plotstyle=curve, linewidth=0.3mm, showpoints=false, linestyle=solid, dotstyle=square, dotscale=1.3 1.3](-0.50000, 4.21501)(-0.40000, 4.39516)(-0.30000, 4.62073)(-0.20000, 4.91064)(-0.10000, 5.29880)(0.00000, 5.91292)(0.10000, 6.20447)(0.20000, 6.65242)(0.23333, 6.85682)(0.26667, 7.10253)(0.30000, 7.37232)(0.33333, 7.61315)(0.36667, 7.77354)(0.40000, 7.83256)(0.43333, 7.79778)(0.46667, 7.68243)(0.50000, 7.49355)
\psline[linecolor=red, plotstyle=curve, linewidth=0.3mm, showpoints=true, linestyle=solid, dotstyle=o, dotscale=1.3 1.3](-0.50000, 2.08571)(-0.40000, 2.44674)(-0.30000, 2.87219)(-0.20000, 3.37997)(-0.10000, 4.00558)(0.00000, 4.90240)(0.10000, 5.24394)(0.20000, 5.56835)(0.23333, 5.66864)(0.26667, 5.76338)(0.30000, 5.85150)(0.33333, 5.93164)(0.36667, 6.00383)(0.40000, 6.06330)(0.43333, 6.11007)(0.46667, 6.15862)(0.50000, 6.26863)
\psline[linecolor=red, plotstyle=curve, linewidth=0.3mm, showpoints=true, linestyle=solid, dotstyle=triangle, dotscale=1.3 1.3](-0.50000, 3.77412)(-0.40000, 3.89573)(-0.30000, 4.04696)(-0.20000, 4.23928)(-0.10000, 4.49863)(0.00000, 4.95324)(0.10000, 5.20055)(0.20000, 5.56722)(0.23333, 5.65899)(0.26667, 5.75155)(0.30000, 5.84895)(0.33333, 5.91151)(0.36667, 5.98412)(0.40000, 6.05857)(0.43333, 6.09815)(0.46667, 6.15085)(0.50000, 6.26529)
\psline[linecolor=red, plotstyle=curve, linewidth=0.3mm, showpoints=true, linestyle=solid, dotstyle=o, dotscale=1.3 1.3](-0.50000, 1.54770)(-0.40000, 1.81968)(-0.30000, 2.13606)(-0.20000, 2.50209)(-0.10000, 2.92359)(0.00000, 3.40733)(0.10000, 4.24970)(0.20000, 4.94804)(0.23333, 5.11232)(0.26667, 5.23370)(0.30000, 5.30899)(0.33333, 5.33668)(0.36667, 5.31727)(0.40000, 5.25337)(0.43333, 5.14931)(0.46667, 5.01156)(0.50000, 4.84903)
\psline[linecolor=darkgreen, plotstyle=curve, linewidth=0.3mm, showpoints=true, linestyle=solid, dotstyle=square, dotscale=1.3 1.3](-0.50000, 4.11060)(-0.40000, 4.27634)(-0.30000, 4.48330)(-0.20000, 4.75148)(-0.10000, 5.12083)(0.00000, 5.76104)(0.10000, 5.87378)(0.20000, 5.98733)(0.23333, 5.98569)(0.26667, 5.92175)(0.30000, 6.07497)(0.33333, 6.29086)(0.36667, 6.49992)(0.40000, 7.23542)(0.43333, 7.49745)(0.46667, 7.43745)(0.50000, 7.31348)
\psline[linecolor=darkgreen, plotstyle=curve, linewidth=0.3mm, showpoints=true, linestyle=solid, dotstyle=diamond, dotscale=1.3 1.3](-0.50000, 3.76980)(-0.40000, 3.88645)(-0.30000, 4.04323)(-0.20000, 4.25923)(-0.10000, 4.56891)(0.00000, 5.07648)(0.10000, 5.42114)(0.20000, 5.88293)(0.23333, 6.07590)(0.26667, 6.27990)(0.30000, 6.45161)(0.33333, 6.49559)(0.36667, 6.30980)(0.40000, 5.91435)(0.43333, 5.47124)(0.46667, 5.33359)(0.50000, 5.21169)
\endpsclip
\psline[linecolor=black, linestyle=solid, linewidth=0.1mm]{->}(0.30000, 2.50000)(0.35000, 3.45943)
\psline[linecolor=black, linestyle=solid, linewidth=0.1mm]{->}(0.30000, 4.00000)(0.35000, 4.95420)
\psline[linecolor=black, linestyle=solid, linewidth=0.1mm]{->}(-0.06000, 2.80000)(-0.30000, 2.87219)
\psline[linecolor=black, linestyle=solid, linewidth=0.1mm]{->}(-0.08000, 2.30000)(-0.20000, 2.50209)
\rput[b](0.30000, 2.50000){\psframebox[linestyle=none, fillcolor=white, fillstyle=solid]{$\SNRSymb_{s,d}=\unit[10]{dB}$}}
\rput[b](0.30000, 4.00000){\psframebox[linestyle=none, fillcolor=white, fillstyle=solid]{$\SNRSymb_{s,d}=\unit[16]{dB}$}}
\rput[l](-0.06000, 2.80000){\psframebox[linestyle=none, fillcolor=white, fillstyle=solid]{Full reuse}}
\rput[l](-0.08000, 2.30000){\psframebox[linestyle=none, fillcolor=white, fillstyle=solid]{No reuse}}
\psframe[linecolor=black, fillstyle=solid, fillcolor=white, shadowcolor=lightgray, shadowsize=1mm, shadow=true](-0.47368, 6.12500)(0.09211, 8.56250)
\rput[l](-0.35526, 8.28125){Upper bound}
\psline[linecolor=black, linestyle=solid, linewidth=0.3mm](-0.44737, 8.28125)(-0.39474, 8.28125)
\rput[l](-0.35526, 7.90625){Combined strategy}
\psline[linecolor=darkgreen, linestyle=solid, linewidth=0.3mm](-0.44737, 7.90625)(-0.39474, 7.90625)
\psline[linecolor=darkgreen, linestyle=solid, linewidth=0.3mm](-0.44737, 7.90625)(-0.39474, 7.90625)
\psdots[linecolor=darkgreen, linestyle=solid, linewidth=0.3mm, dotstyle=square, dotscale=1.3 1.3, linecolor=darkgreen](-0.42105, 7.90625)
\rput[l](-0.35526, 7.53125){Compress-and-Forward}
\psline[linecolor=darkgreen, linestyle=solid, linewidth=0.3mm](-0.44737, 7.53125)(-0.39474, 7.53125)
\psline[linecolor=darkgreen, linestyle=solid, linewidth=0.3mm](-0.44737, 7.53125)(-0.39474, 7.53125)
\psdots[linecolor=darkgreen, linestyle=solid, linewidth=0.3mm, dotstyle=diamond, dotscale=1.3 1.3, linecolor=darkgreen](-0.42105, 7.53125)
\rput[l](-0.35526, 7.15625){Partial DF}
\psline[linecolor=red, linestyle=solid, linewidth=0.3mm](-0.44737, 7.15625)(-0.39474, 7.15625)
\psline[linecolor=red, linestyle=solid, linewidth=0.3mm](-0.44737, 7.15625)(-0.39474, 7.15625)
\psdots[linecolor=red, linestyle=solid, linewidth=0.3mm, dotstyle=triangle, dotscale=1.3 1.3, linecolor=red](-0.42105, 7.15625)
\rput[l](-0.35526, 6.78125){Decode-and-Forward}
\psline[linecolor=red, linestyle=solid, linewidth=0.3mm](-0.44737, 6.78125)(-0.39474, 6.78125)
\psline[linecolor=red, linestyle=solid, linewidth=0.3mm](-0.44737, 6.78125)(-0.39474, 6.78125)
\psdots[linecolor=red, linestyle=solid, linewidth=0.3mm, dotstyle=o, dotscale=1.3 1.3, linecolor=red](-0.42105, 6.78125)
\rput[l](-0.35526, 6.40625){Single Hop}
\psline[linecolor=blue, linestyle=solid, linewidth=0.3mm](-0.44737, 6.40625)(-0.39474, 6.40625)
}\end{pspicture}
\endgroup
 }
	\subfigure[Non-coherent transmission]{\begingroup
\unitlength=1mm
\psset{xunit=76.00000mm, yunit=10.66667mm, linewidth=0.1mm}
\psset{arrowsize=2pt 3, arrowlength=1.4, arrowinset=.4}\psset{axesstyle=frame}
\begin{pspicture}(-0.63158, 0.50000)(0.50000, 8.00000)
\rput(-0.02632, -0.75000){%
\psaxes[subticks=0, labels=all, xsubticks=1, ysubticks=1, Ox=-0.5, Oy=2, Dx=0.2, Dy=1]{-}(-0.50000, 2.00000)(-0.50000, 2.00000)(0.50000, 8.00000)%
\multips(-0.30000, 2.00000)(0.20000, 0.0){4}{\psline[linecolor=black, linestyle=dotted, linewidth=0.2mm](0, 0)(0, 6.00000)}
\multips(-0.50000, 3.00000)(0, 1.00000){5}{\psline[linecolor=black, linestyle=dotted, linewidth=0.2mm](0, 0)(1.00000, 0)}
\rput[b](0.00000, 1.25000){distance $r$}
\rput[t]{90}(-0.60526, 5.00000){$R$ in bits per channel use}
\psclip{\psframe(-0.50000, 2.00000)(0.50000, 8.00000)}
\psline[linecolor=blue, plotstyle=curve, linewidth=0.3mm, showpoints=false, linestyle=solid, dotstyle=square, dotscale=1.3 1.3](-0.50000, 3.45943)(-0.40000, 3.45943)(-0.30000, 3.45943)(-0.20000, 3.45943)(-0.10000, 3.45943)(0.00000, 3.45943)(0.10000, 3.45943)(0.20000, 3.45943)(0.23333, 3.45943)(0.26667, 3.45943)(0.30000, 3.45943)(0.33333, 3.45943)(0.36667, 3.45943)(0.40000, 3.45943)(0.43333, 3.45943)(0.46667, 3.45943)(0.50000, 3.45943)
\psline[linecolor=blue, plotstyle=curve, linewidth=0.3mm, showpoints=false, linestyle=solid, dotstyle=square, dotscale=1.3 1.3](-0.50000, 4.95420)(-0.40000, 4.95420)(-0.30000, 4.95420)(-0.20000, 4.95420)(-0.10000, 4.95420)(0.00000, 4.95420)(0.10000, 4.95420)(0.20000, 4.95420)(0.23333, 4.95420)(0.26667, 4.95420)(0.30000, 4.95420)(0.33333, 4.95420)(0.36667, 4.95420)(0.40000, 4.95420)(0.43333, 4.95420)(0.46667, 4.95420)(0.50000, 4.95420)
\psline[linecolor=red, plotstyle=curve, linewidth=0.3mm, showpoints=true, linestyle=dashed, dotstyle=o, dotscale=1.3 1.3](-0.50000, 1.75614)(-0.40000, 2.07652)(-0.30000, 2.45763)(-0.20000, 2.91489)(-0.10000, 3.47512)(0.00000, 4.21236)(0.10000, 4.87283)(0.20000, 5.56694)(0.23333, 5.79344)(0.26667, 6.00528)(0.30000, 6.19142)(0.33333, 6.34031)(0.36667, 6.44229)(0.40000, 6.49424)(0.43333, 6.50099)(0.46667, 6.48146)(0.50000, 6.49802)
\psline[linecolor=red, plotstyle=curve, linewidth=0.3mm, showpoints=true, linestyle=dashed, dotstyle=triangle, dotscale=1.3 1.3](-0.50000, 3.57681)(-0.40000, 3.63024)(-0.30000, 3.70509)(-0.20000, 3.81198)(-0.10000, 3.96940)(0.00000, 4.23717)(0.10000, 4.87003)(0.20000, 5.56697)(0.23333, 5.79049)(0.26667, 6.00516)(0.30000, 6.19124)(0.33333, 6.33945)(0.36667, 6.44233)(0.40000, 6.48952)(0.43333, 6.49908)(0.46667, 6.47794)(0.50000, 6.49516)
\psline[linecolor=red, plotstyle=curve, linewidth=0.3mm, showpoints=true, linestyle=dashed, dotstyle=o, dotscale=1.3 1.3](-0.50000, 1.54770)(-0.40000, 1.81968)(-0.30000, 2.13606)(-0.20000, 2.50209)(-0.10000, 2.92359)(0.00000, 3.40806)(0.10000, 4.30571)(0.20000, 5.19412)(0.23333, 5.44250)(0.26667, 5.64230)(0.30000, 5.77661)(0.33333, 5.83343)(0.36667, 5.81122)(0.40000, 5.72051)(0.43333, 5.57941)(0.46667, 5.40570)(0.50000, 5.21118)
\psline[linecolor=black, plotstyle=curve, linewidth=0.3mm, showpoints=false, linestyle=solid, dotstyle=square, dotscale=1.3 1.3](-0.50000, 3.93974)(-0.40000, 4.06370)(-0.30000, 4.22906)(-0.20000, 4.45450)(-0.10000, 4.76977)(0.00000, 5.23917)(0.10000, 5.74879)(0.20000, 6.43731)(0.23333, 6.72874)(0.26667, 7.04503)(0.30000, 7.35421)(0.33333, 7.60892)(0.36667, 7.77212)(0.40000, 7.83244)(0.43333, 7.79778)(0.46667, 7.68219)(0.50000, 7.49355)
\psline[linecolor=red, plotstyle=curve, linewidth=0.3mm, showpoints=true, linestyle=solid, dotstyle=o, dotscale=1.3 1.3](-0.50000, 1.75183)(-0.40000, 2.07005)(-0.30000, 2.44784)(-0.20000, 2.89993)(-0.10000, 3.45231)(0.00000, 4.18447)(0.10000, 4.76239)(0.20000, 5.30362)(0.23333, 5.46905)(0.26667, 5.62196)(0.30000, 5.75939)(0.33333, 5.87813)(0.36667, 5.97571)(0.40000, 6.05164)(0.43333, 6.10786)(0.46667, 6.15600)(0.50000, 6.26856)
\psline[linecolor=red, plotstyle=curve, linewidth=0.3mm, showpoints=true, linestyle=solid, dotstyle=triangle, dotscale=1.3 1.3](-0.50000, 3.57671)(-0.40000, 3.63011)(-0.30000, 3.70493)(-0.20000, 3.81178)(-0.10000, 3.96913)(0.00000, 4.23689)(0.10000, 4.76170)(0.20000, 5.30310)(0.23333, 5.46901)(0.26667, 5.62164)(0.30000, 5.75934)(0.33333, 5.87770)(0.36667, 5.97473)(0.40000, 6.05139)(0.43333, 6.10656)(0.46667, 6.15151)(0.50000, 6.26846)
\psline[linecolor=red, plotstyle=curve, linewidth=0.3mm, showpoints=true, linestyle=solid, dotstyle=o, dotscale=1.3 1.3](-0.50000, 1.54770)(-0.40000, 1.81968)(-0.30000, 2.13606)(-0.20000, 2.50209)(-0.10000, 2.92359)(0.00000, 3.40733)(0.10000, 4.24970)(0.20000, 4.94807)(0.23333, 5.11235)(0.26667, 5.23373)(0.30000, 5.30900)(0.33333, 5.33671)(0.36667, 5.31734)(0.40000, 5.25338)(0.43333, 5.14931)(0.46667, 5.01158)(0.50000, 4.84904)
\psline[linecolor=darkgreen, plotstyle=curve, linewidth=0.3mm, showpoints=true, linestyle=solid, dotstyle=square, dotscale=1.3 1.3](-0.50000, 3.75405)(-0.40000, 3.85584)(-0.30000, 3.99663)(-0.20000, 4.19597)(-0.10000, 4.48913)(0.00000, 4.99060)(0.10000, 5.32020)(0.20000, 5.75417)(0.23333, 5.83978)(0.26667, 5.85842)(0.30000, 6.07445)(0.33333, 6.29051)(0.36667, 6.49900)(0.40000, 7.23438)(0.43333, 7.48330)(0.46667, 7.42949)(0.50000, 7.31029)
\psline[linecolor=darkgreen, plotstyle=curve, linewidth=0.3mm, showpoints=true, linestyle=solid, dotstyle=diamond, dotscale=1.3 1.3](-0.50000, 3.76980)(-0.40000, 3.88645)(-0.30000, 4.04323)(-0.20000, 4.25923)(-0.10000, 4.56891)(0.00000, 5.07648)(0.10000, 5.42114)(0.20000, 5.88293)(0.23333, 6.07590)(0.26667, 6.27990)(0.30000, 6.45161)(0.33333, 6.49559)(0.36667, 6.30980)(0.40000, 5.91435)(0.43333, 5.47124)(0.46667, 5.33359)(0.50000, 5.21169)
\endpsclip
\psline[linecolor=black, linestyle=solid, linewidth=0.1mm]{->}(0.30000, 2.50000)(0.35000, 3.45943)
\psline[linecolor=black, linestyle=solid, linewidth=0.1mm]{->}(0.30000, 4.00000)(0.35000, 4.95420)
\psline[linecolor=black, linestyle=solid, linewidth=0.1mm]{->}(-0.06000, 2.80000)(-0.20000, 2.89993)
\psline[linecolor=black, linestyle=solid, linewidth=0.1mm]{->}(-0.08000, 2.30000)(-0.20000, 2.50209)
\rput[b](0.30000, 2.50000){\psframebox[linestyle=none, fillcolor=white, fillstyle=solid]{$\SNRSymb_{s,d}=\unit[10]{dB}$}}
\rput[b](0.30000, 4.00000){\psframebox[linestyle=none, fillcolor=white, fillstyle=solid]{$\SNRSymb_{s,d}=\unit[16]{dB}$}}
\rput[l](-0.06000, 2.80000){\psframebox[linestyle=none, fillcolor=white, fillstyle=solid]{Full reuse}}
\rput[l](-0.08000, 2.30000){\psframebox[linestyle=none, fillcolor=white, fillstyle=solid]{No reuse}}
\psframe[linecolor=black, fillstyle=solid, fillcolor=white, shadowcolor=lightgray, shadowsize=1mm, shadow=true](-0.47368, 6.12500)(0.09211, 8.56250)
\rput[l](-0.35526, 8.28125){Upper bound}
\psline[linecolor=black, linestyle=solid, linewidth=0.3mm](-0.44737, 8.28125)(-0.39474, 8.28125)
\rput[l](-0.35526, 7.90625){Combined strategy}
\psline[linecolor=darkgreen, linestyle=solid, linewidth=0.3mm](-0.44737, 7.90625)(-0.39474, 7.90625)
\psline[linecolor=darkgreen, linestyle=solid, linewidth=0.3mm](-0.44737, 7.90625)(-0.39474, 7.90625)
\psdots[linecolor=darkgreen, linestyle=solid, linewidth=0.3mm, dotstyle=square, dotscale=1.3 1.3, linecolor=darkgreen](-0.42105, 7.90625)
\rput[l](-0.35526, 7.53125){Compress-and-Forward}
\psline[linecolor=darkgreen, linestyle=solid, linewidth=0.3mm](-0.44737, 7.53125)(-0.39474, 7.53125)
\psline[linecolor=darkgreen, linestyle=solid, linewidth=0.3mm](-0.44737, 7.53125)(-0.39474, 7.53125)
\psdots[linecolor=darkgreen, linestyle=solid, linewidth=0.3mm, dotstyle=diamond, dotscale=1.3 1.3, linecolor=darkgreen](-0.42105, 7.53125)
\rput[l](-0.35526, 7.15625){Partial DF}
\psline[linecolor=red, linestyle=solid, linewidth=0.3mm](-0.44737, 7.15625)(-0.39474, 7.15625)
\psline[linecolor=red, linestyle=solid, linewidth=0.3mm](-0.44737, 7.15625)(-0.39474, 7.15625)
\psdots[linecolor=red, linestyle=solid, linewidth=0.3mm, dotstyle=triangle, dotscale=1.3 1.3, linecolor=red](-0.42105, 7.15625)
\rput[l](-0.35526, 6.78125){Decode-and-Forward}
\psline[linecolor=red, linestyle=solid, linewidth=0.3mm](-0.44737, 6.78125)(-0.39474, 6.78125)
\psline[linecolor=red, linestyle=solid, linewidth=0.3mm](-0.44737, 6.78125)(-0.39474, 6.78125)
\psdots[linecolor=red, linestyle=solid, linewidth=0.3mm, dotstyle=o, dotscale=1.3 1.3, linecolor=red](-0.42105, 6.78125)
\rput[l](-0.35526, 6.40625){Single Hop}
\psline[linecolor=blue, linestyle=solid, linewidth=0.3mm](-0.44737, 6.40625)(-0.39474, 6.40625)
}\end{pspicture}
\endgroup
 }
	\caption[Achievable rates for the Gaussian half-duplex two-relay channel.]
	{Achievable rates for the Gaussian half-duplex two-relay channel. Solid curves indicate a fixed transmission strategy
	and dashed lines indicate a random transmission schedule. $\SNRSymb_{s,d}=\unit[16]{dB}$
	again indicates the power-normalized case.}
	\label{fig:halfduplex:results:two_relays}
      \end{figure}
      \begin{figure}[htb]
	\centering
	\subfigure[Half-duplex Network]{\begingroup
\unitlength=1mm
\psset{xunit=38.00000mm, yunit=15.00000mm, linewidth=0.1mm}
\psset{arrowsize=2pt 3, arrowlength=1.4, arrowinset=.4}\psset{axesstyle=frame}
\begin{pspicture}(-1.26316, 1.33333)(1.00000, 6.00000)
\rput(-0.05263, -0.13333){%
\psaxes[subticks=0, labels=all, xsubticks=1, ysubticks=1, Ox=-1, Oy=2, Dx=0.4, Dy=1]{-}(-1.00000, 2.00000)(-1.00000, 2.00000)(1.00000, 6.00000)%
\multips(-0.60000, 2.00000)(0.40000, 0.0){4}{\psline[linecolor=black, linestyle=dotted, linewidth=0.2mm](0, 0)(0, 4.00000)}
\multips(-1.00000, 3.00000)(0, 1.00000){3}{\psline[linecolor=black, linestyle=dotted, linewidth=0.2mm](0, 0)(2.00000, 0)}
\rput[b](0.00000, 1.46667){distance $r$}
\rput[t]{90}(-1.21053, 4.00000){$R$ in bits per channel use (bpcu)}
\psclip{\psframe(-1.00000, 2.00000)(1.00000, 6.00000)}
\psline[linecolor=blue, plotstyle=curve, linewidth=0.3mm, showpoints=false, linestyle=solid, dotstyle=square, dotscale=1.3 1.3](-1.00000, 3.45943)(-0.80000, 3.45943)(-0.60000, 3.45943)(-0.40000, 3.45943)(-0.20000, 3.45943)(0.00000, 3.45943)(0.20000, 3.45943)(0.40000, 3.45943)(0.50000, 3.45943)(0.60000, 3.45943)(0.80000, 3.45943)(1.00000, 3.45943)
\psline[linecolor=blue, plotstyle=curve, linewidth=0.3mm, showpoints=false, linestyle=solid, dotstyle=square, dotscale=1.3 1.3](-1.00000, 4.39232)(-0.80000, 4.39232)(-0.60000, 4.39232)(-0.40000, 4.39232)(-0.20000, 4.39232)(0.00000, 4.39232)(0.20000, 4.39232)(0.40000, 4.39232)(0.50000, 4.39232)(0.60000, 4.39232)(0.80000, 4.39232)(1.00000, 4.39232)
\psline[linecolor=red, plotstyle=curve, linewidth=0.3mm, showpoints=true, linestyle=dashed, dotstyle=o, dotscale=1.3 1.3](-1.00000, 3.46122)(-0.80000, 3.49350)(-0.60000, 3.54870)(-0.40000, 3.65168)(-0.20000, 3.85723)(0.00000, 4.31435)(0.20000, 4.71473)(0.40000, 5.13139)(0.50000, 5.16883)(0.60000, 5.02622)(0.80000, 4.32902)(1.00000, 3.40806)
\psline[linecolor=red, plotstyle=curve, linewidth=0.3mm, showpoints=true, linestyle=dashed, dotstyle=o, dotscale=1.3 1.3](-1.00000, 3.46035)(-0.80000, 3.46035)(-0.60000, 3.46035)(-0.40000, 3.46035)(-0.20000, 3.46035)(0.00000, 3.50717)(0.20000, 4.38783)(0.40000, 5.06273)(0.50000, 5.14888)(0.60000, 5.02228)(0.80000, 4.32899)(1.00000, 3.40806)
\psline[linecolor=black, plotstyle=curve, linewidth=0.3mm, showpoints=false, linestyle=solid, dotstyle=square, dotscale=1.3 1.3](-1.00000, 3.53439)(-0.80000, 3.57378)(-0.60000, 3.64007)(-0.40000, 3.75636)(-0.20000, 3.96967)(0.00000, 4.38653)(0.20000, 4.91220)(0.40000, 5.37590)(0.50000, 5.43785)(0.60000, 5.34604)(0.80000, 4.85526)(1.00000, 4.33525)
\psline[linecolor=red, plotstyle=curve, linewidth=0.3mm, showpoints=true, linestyle=solid, dotstyle=o, dotscale=1.3 1.3](-1.00000, 3.46063)(-0.80000, 3.49162)(-0.60000, 3.54476)(-0.40000, 3.64309)(-0.20000, 3.83838)(0.00000, 4.28260)(0.20000, 4.55484)(0.40000, 4.79827)(0.50000, 4.81400)(0.60000, 4.72462)(0.80000, 4.22576)(1.00000, 3.40733)
\psline[linecolor=red, plotstyle=curve, linewidth=0.3mm, showpoints=true, linestyle=solid, dotstyle=o, dotscale=1.3 1.3](-1.00000, 3.45943)(-0.80000, 3.45943)(-0.60000, 3.45943)(-0.40000, 3.45943)(-0.20000, 3.45943)(0.00000, 3.50606)(0.20000, 4.29242)(0.40000, 4.74569)(0.50000, 4.79675)(0.60000, 4.72041)(0.80000, 4.22571)(1.00000, 3.40733)
\psline[linecolor=darkgreen, plotstyle=curve, linewidth=0.3mm, showpoints=true, linestyle=solid, dotstyle=diamond, dotscale=1.3 1.3](-1.00000, 3.49057)(-0.80000, 3.52730)(-0.60000, 3.59768)(-0.40000, 3.72213)(-0.20000, 3.94425)(0.00000, 4.37730)(0.20000, 4.78289)(0.40000, 5.00492)(0.50000, 4.96886)(0.60000, 4.84611)(0.80000, 4.48474)(1.00000, 4.18854)
\endpsclip
\psline[linecolor=black, linestyle=solid, linewidth=0.1mm]{->}(0.50000, 2.80000)(0.45000, 3.45943)
\psline[linecolor=black, linestyle=solid, linewidth=0.1mm]{->}(0.50000, 3.60000)(0.45000, 4.39232)
\psline[linecolor=black, linestyle=solid, linewidth=0.1mm]{->}(-0.14000, 3.00000)(-0.28000, 3.76027)
\psline[linecolor=black, linestyle=solid, linewidth=0.1mm]{->}(0.18000, 2.70000)(0.04000, 3.66333)
\rput[b](0.50000, 2.80000){\psframebox[linestyle=none, fillcolor=white, fillstyle=solid]{$\SNRSymb_{s,d}=\unit[10]{dB}$}}
\rput[b](0.50000, 3.60000){\psframebox[linestyle=none, fillcolor=white, fillstyle=solid]{$\SNRSymb_{s,d}=\unit[13]{dB}$}}
\rput[c](-0.14000, 3.00000){\psframebox[linestyle=none, fillcolor=white, fillstyle=solid]{Full reuse}}
\rput[c](0.18000, 2.70000){\psframebox[linestyle=none, fillcolor=white, fillstyle=solid]{No reuse}}
\psframe[linecolor=black, fillstyle=solid, fillcolor=white, shadowcolor=lightgray, shadowsize=1mm, shadow=true](-0.94737, 4.93333)(0.18421, 6.13333)
\rput[l](-0.71053, 5.93333){Upper bound}
\psline[linecolor=black, linestyle=solid, linewidth=0.3mm](-0.89474, 5.93333)(-0.78947, 5.93333)
\rput[l](-0.71053, 5.66667){Compress-and-Forward}
\psline[linecolor=darkgreen, linestyle=solid, linewidth=0.3mm](-0.89474, 5.66667)(-0.78947, 5.66667)
\psline[linecolor=darkgreen, linestyle=solid, linewidth=0.3mm](-0.89474, 5.66667)(-0.78947, 5.66667)
\psdots[linecolor=darkgreen, linestyle=solid, linewidth=0.3mm, dotstyle=diamond, dotscale=1.3 1.3, linecolor=darkgreen](-0.84211, 5.66667)
\rput[l](-0.71053, 5.40000){Decode-and-Forward}
\psline[linecolor=red, linestyle=solid, linewidth=0.3mm](-0.89474, 5.40000)(-0.78947, 5.40000)
\psline[linecolor=red, linestyle=solid, linewidth=0.3mm](-0.89474, 5.40000)(-0.78947, 5.40000)
\psdots[linecolor=red, linestyle=solid, linewidth=0.3mm, dotstyle=o, dotscale=1.3 1.3, linecolor=red](-0.84211, 5.40000)
\rput[l](-0.71053, 5.13333){Single Hop}
\psline[linecolor=blue, linestyle=solid, linewidth=0.3mm](-0.89474, 5.13333)(-0.78947, 5.13333)
}\end{pspicture}
\endgroup
 }
	\subfigure[Full-duplex Network]{\begingroup
\unitlength=1mm
\psset{xunit=38.00000mm, yunit=10.00000mm, linewidth=0.1mm}
\psset{arrowsize=2pt 3, arrowlength=1.4, arrowinset=.4}\psset{axesstyle=frame}
\begin{pspicture}(-1.26316, 1.00000)(1.00000, 8.00000)
\rput(-0.05263, -0.20000){%
\psaxes[subticks=0, labels=all, xsubticks=1, ysubticks=1, Ox=-1, Oy=2, Dx=0.4, Dy=1]{-}(-1.00000, 2.00000)(-1.00000, 2.00000)(1.00000, 8.00000)%
\multips(-0.60000, 2.00000)(0.40000, 0.0){4}{\psline[linecolor=black, linestyle=dotted, linewidth=0.2mm](0, 0)(0, 6.00000)}
\multips(-1.00000, 3.00000)(0, 1.00000){5}{\psline[linecolor=black, linestyle=dotted, linewidth=0.2mm](0, 0)(2.00000, 0)}
\rput[b](0.00000, 1.20000){distance $r$}
\rput[t]{90}(-1.21053, 5.00000){$R$ in bits per channel use (bpcu)}
\psclip{\psframe(-1.00000, 2.00000)(1.00000, 8.00000)}
\psline[linecolor=blue, plotstyle=curve, linewidth=0.3mm, showpoints=false, linestyle=solid, dotstyle=square, dotscale=1.3 1.3](-1.49000, 3.45943)(-1.39000, 3.45943)(-1.29000, 3.45943)(-1.19000, 3.45943)(-1.09000, 3.45943)(-0.99000, 3.45943)(-0.89000, 3.45943)(-0.79000, 3.45943)(-0.69000, 3.45943)(-0.59000, 3.45943)(-0.49000, 3.45943)(-0.39000, 3.45943)(-0.29000, 3.45943)(-0.19000, 3.45943)(-0.09000, 3.45943)(0.01000, 3.45943)(0.11000, 3.45943)(0.21000, 3.45943)(0.31000, 3.45943)(0.41000, 3.45943)(0.51000, 3.45943)(0.61000, 3.45943)(0.71000, 3.45943)(0.81000, 3.45943)(0.91000, 3.45943)(1.01000, 3.45943)(1.11000, 3.45943)(1.21000, 3.45943)(1.31000, 3.45943)(1.41000, 3.45943)
\psline[linecolor=blue, plotstyle=curve, linewidth=0.3mm, showpoints=false, linestyle=solid, dotstyle=square, dotscale=1.3 1.3](-1.49000, 4.39232)(-1.39000, 4.39232)(-1.29000, 4.39232)(-1.19000, 4.39232)(-1.09000, 4.39232)(-0.99000, 4.39232)(-0.89000, 4.39232)(-0.79000, 4.39232)(-0.69000, 4.39232)(-0.59000, 4.39232)(-0.49000, 4.39232)(-0.39000, 4.39232)(-0.29000, 4.39232)(-0.19000, 4.39232)(-0.09000, 4.39232)(0.01000, 4.39232)(0.11000, 4.39232)(0.21000, 4.39232)(0.31000, 4.39232)(0.41000, 4.39232)(0.51000, 4.39232)(0.61000, 4.39232)(0.71000, 4.39232)(0.81000, 4.39232)(0.91000, 4.39232)(1.01000, 4.39232)(1.11000, 4.39232)(1.21000, 4.39232)(1.31000, 4.39232)(1.41000, 4.39232)
\psline[linecolor=red, plotstyle=curve, linewidth=0.3mm, showpoints=true, linestyle=solid, dotstyle=o, dotscale=1.3 1.3](-1.49000, 1.59878)(-1.39000, 1.87924)(-1.29000, 2.20512)(-1.19000, 2.58176)(-1.09000, 3.01511)(-0.99000, 3.51225)(-0.89000, 3.55872)(-0.79000, 3.58184)(-0.69000, 3.61187)(-0.59000, 3.65130)(-0.49000, 3.70364)(-0.39000, 3.77387)(-0.29000, 3.86899)(-0.19000, 3.99880)(-0.09000, 4.17666)(0.01000, 4.42023)(0.11000, 4.75158)(0.21000, 5.19668)(0.31000, 5.78442)(0.41000, 6.54730)(0.51000, 7.21738)(0.61000, 6.19424)(0.71000, 5.33457)(0.81000, 4.59876)(0.91000, 3.96186)(1.01000, 3.40733)(1.11000, 2.92359)(1.21000, 2.50209)(1.31000, 2.13606)(1.41000, 1.81968)
\psline[linecolor=darkgreen, plotstyle=curve, linewidth=0.3mm, showpoints=true, linestyle=solid, dotstyle=diamond, dotscale=1.3 1.3](-1.49000, 3.46463)(-1.39000, 3.46711)(-1.29000, 3.47089)(-1.19000, 3.47664)(-1.09000, 3.48537)(-0.99000, 3.49839)(-0.89000, 3.51735)(-0.79000, 3.54402)(-0.69000, 3.58013)(-0.59000, 3.62736)(-0.49000, 3.68786)(-0.39000, 3.76518)(-0.29000, 3.86534)(-0.19000, 3.99786)(-0.09000, 4.17659)(0.01000, 4.42023)(0.11000, 4.75125)(0.21000, 5.18951)(0.31000, 5.72737)(0.41000, 6.25314)(0.51000, 6.46216)(0.61000, 6.16115)(0.71000, 5.61467)(0.81000, 5.09286)(0.91000, 4.67695)(1.01000, 4.36526)(1.11000, 4.13632)(1.21000, 3.96817)(1.31000, 3.84308)(1.41000, 3.74812)
\psline[linecolor=darkgreen, plotstyle=curve, linewidth=0.3mm, showpoints=true, linestyle=solid, dotstyle=square, dotscale=1.3 1.3](-1.49000, 3.46463)(-1.39000, 3.46711)(-1.29000, 3.47089)(-1.19000, 3.47664)(-1.09000, 3.48537)(-0.99000, 3.51225)(-0.89000, 3.55893)(-0.79000, 3.58208)(-0.69000, 3.61212)(-0.59000, 3.65158)(-0.49000, 3.70396)(-0.39000, 3.77421)(-0.29000, 3.86936)(-0.19000, 3.99920)(-0.09000, 4.17708)(0.01000, 4.42065)(0.11000, 4.75197)(0.21000, 5.19708)(0.31000, 5.78477)(0.41000, 6.54758)(0.51000, 6.50377)(0.61000, 6.16293)(0.71000, 5.61467)(0.81000, 5.09286)(0.91000, 4.67695)(1.01000, 4.36526)(1.11000, 4.13632)(1.21000, 3.96817)(1.31000, 3.84308)(1.41000, 3.74812)
\psline[linecolor=black, plotstyle=curve, linewidth=0.3mm, showpoints=false, linestyle=solid, dotstyle=square, dotscale=1.3 1.3](-1.49000, 3.49315)(-1.39000, 3.49908)(-1.29000, 3.50635)(-1.19000, 3.51535)(-1.09000, 3.52658)(-0.99000, 3.54073)(-0.89000, 3.55872)(-0.79000, 3.58184)(-0.69000, 3.61187)(-0.59000, 3.65130)(-0.49000, 3.70364)(-0.39000, 3.77387)(-0.29000, 3.86899)(-0.19000, 3.99880)(-0.09000, 4.17666)(0.01000, 4.42023)(0.11000, 4.75158)(0.21000, 5.19668)(0.31000, 5.78442)(0.41000, 6.54730)(0.51000, 7.31121)(0.61000, 6.37893)(0.71000, 5.65398)(0.81000, 5.09721)(0.91000, 4.67709)(1.01000, 4.36526)(1.11000, 4.13646)(1.21000, 3.96948)(1.31000, 3.84755)(1.41000, 3.75809)
\endpsclip
\psline[linecolor=black, linestyle=solid, linewidth=0.1mm]{->}(0.50000, 2.30000)(0.45000, 3.45943)
\psline[linecolor=black, linestyle=solid, linewidth=0.1mm]{->}(0.50000, 3.50000)(0.45000, 4.39232)
\rput[b](0.50000, 2.30000){\psframebox[linestyle=none, fillcolor=white, fillstyle=solid]{$\SNRSymb_{s,d}=\unit[10]{dB}$}}
\rput[b](0.50000, 3.50000){\psframebox[linestyle=none, fillcolor=white, fillstyle=solid]{$\SNRSymb_{s,d}=\unit[13]{dB}$}}
\psframe[linecolor=black, fillstyle=solid, fillcolor=white, shadowcolor=lightgray, shadowsize=1mm, shadow=true](-0.94737, 6.00000)(0.18421, 7.80000)
\rput[l](-0.71053, 7.50000){Upper bound}
\psline[linecolor=black, linestyle=solid, linewidth=0.3mm](-0.89474, 7.50000)(-0.78947, 7.50000)
\rput[l](-0.71053, 7.10000){Compress-and-Forward}
\psline[linecolor=darkgreen, linestyle=solid, linewidth=0.3mm](-0.89474, 7.10000)(-0.78947, 7.10000)
\psline[linecolor=darkgreen, linestyle=solid, linewidth=0.3mm](-0.89474, 7.10000)(-0.78947, 7.10000)
\psdots[linecolor=darkgreen, linestyle=solid, linewidth=0.3mm, dotstyle=diamond, dotscale=1.3 1.3, linecolor=darkgreen](-0.84211, 7.10000)
\rput[l](-0.71053, 6.70000){Decode-and-Forward}
\psline[linecolor=red, linestyle=solid, linewidth=0.3mm](-0.89474, 6.70000)(-0.78947, 6.70000)
\psline[linecolor=red, linestyle=solid, linewidth=0.3mm](-0.89474, 6.70000)(-0.78947, 6.70000)
\psdots[linecolor=red, linestyle=solid, linewidth=0.3mm, dotstyle=o, dotscale=1.3 1.3, linecolor=red](-0.84211, 6.70000)
\rput[l](-0.71053, 6.30000){Single Hop}
\psline[linecolor=blue, linestyle=solid, linewidth=0.3mm](-0.89474, 6.30000)(-0.78947, 6.30000)
}\end{pspicture}
\endgroup
 }
	\caption[Achievable rates for the Gaussian single-relay channel.]
	{Achievable rates for the Gaussian single-relay channel with non-coherent transmission. Solid curves indicate fixed transmission strategy
	and dashed lines indicate a random transmission schedule.  $\SNRSymb_{s,d}=\unit[13]{dB}$ again indicates the power-normalized case.}
	\label{fig:halfduplex:results:singe_relay}
      \end{figure}
      \begin{figure}
	\centering
	\begingroup
\unitlength=1mm
\psset{xunit=12.33333mm, yunit=5.00000mm, linewidth=0.1mm}
\psset{arrowsize=2pt 3, arrowlength=1.4, arrowinset=.4}\psset{axesstyle=frame}
\begin{pspicture}(-0.97297, 0.00000)(6.00000, 14.00000)
\rput(-0.16216, -0.40000){%
\psaxes[subticks=0, labels=all, xsubticks=1, ysubticks=1, Ox=0, Oy=2, Dx=1, Dy=2]{-}(0.00000, 2.00000)(0.00000, 2.00000)(6.00000, 14.00000)%
\multips(1.00000, 2.00000)(1.00000, 0.0){5}{\psline[linecolor=black, linestyle=dotted, linewidth=0.2mm](0, 0)(0, 12.00000)}
\multips(0.00000, 4.00000)(0, 2.00000){5}{\psline[linecolor=black, linestyle=dotted, linewidth=0.2mm](0, 0)(6.00000, 0)}
\rput[b](3.00000, 0.40000){number of relays $N$}
\rput[t]{90}(-0.81081, 8.00000){$R$ in bits per channel use (bpcu)}
\psclip{\psframe(0.00000, 2.00000)(6.00000, 14.00000)}
\psline[linecolor=blue, plotstyle=curve, linewidth=0.3mm, showpoints=false, linestyle=solid, dotstyle=square, dotscale=1.3 1.3](0.00000, 3.45943)(1.00000, 4.39232)(2.00000, 4.95420)(3.00000, 5.35755)(4.00000, 5.67243)(5.00000, 5.93074)(6.00000, 6.14975)
\psline[linecolor=red, plotstyle=curve, linewidth=0.3mm, showpoints=true, linestyle=dashed, dotstyle=o, dotscale=1.3 1.3](0.00000, 3.45943)(1.00000, 5.17304)(2.00000, 6.36074)(3.00000, 7.22769)(4.00000, 7.95141)(5.00000, 8.52024)(6.00000, 9.07311)
\psline[linecolor=red, plotstyle=curve, linewidth=0.3mm, showpoints=true, linestyle=solid, dotstyle=o, dotscale=1.3 1.3](0.00000, 3.45943)(1.00000, 4.81690)(2.00000, 5.90966)(3.00000, 6.77724)(4.00000, 7.57016)(5.00000, 8.10305)(6.00000, 8.47683)
\psline[linecolor=darkgreen, plotstyle=curve, linewidth=0.3mm, showpoints=true, linestyle=solid, dotstyle=diamond, dotscale=1.3 1.3](0.00000, 3.45943)(1.00000, 4.97685)(2.00000, 6.46622)(3.00000, 7.59841)(4.00000, 8.59006)(5.00000, 9.34856)(6.00000, 10.15919)
\psline[linecolor=black, plotstyle=curve, linewidth=0.3mm, showpoints=false, linestyle=solid, dotstyle=square, dotscale=1.3 1.3](0.00000, 3.45943)(1.00000, 5.43864)(2.00000, 7.66850)(3.00000, 9.37718)(4.00000, 10.84493)(5.00000, 11.96995)(6.00000, 13.02263)
\endpsclip
\psframe[linecolor=black, fillstyle=solid, fillcolor=white, shadowcolor=lightgray, shadowsize=1mm, shadow=true](0.16216, 10.00000)(3.64865, 13.60000)
\rput[l](0.89189, 13.00000){Upper bound}
\psline[linecolor=black, linestyle=solid, linewidth=0.3mm](0.32432, 13.00000)(0.64865, 13.00000)
\rput[l](0.89189, 12.20000){Compress-and-Forward}
\psline[linecolor=darkgreen, linestyle=solid, linewidth=0.3mm](0.32432, 12.20000)(0.64865, 12.20000)
\psline[linecolor=darkgreen, linestyle=solid, linewidth=0.3mm](0.32432, 12.20000)(0.64865, 12.20000)
\psdots[linecolor=darkgreen, linestyle=solid, linewidth=0.3mm, dotstyle=diamond, dotscale=1.3 1.3, linecolor=darkgreen](0.48649, 12.20000)
\rput[l](0.89189, 11.40000){Decode-and-Forward}
\psline[linecolor=red, linestyle=solid, linewidth=0.3mm](0.32432, 11.40000)(0.64865, 11.40000)
\psline[linecolor=red, linestyle=solid, linewidth=0.3mm](0.32432, 11.40000)(0.64865, 11.40000)
\psdots[linecolor=red, linestyle=solid, linewidth=0.3mm, dotstyle=o, dotscale=1.3 1.3, linecolor=red](0.48649, 11.40000)
\rput[l](0.89189, 10.60000){Single-Hop}
\psline[linecolor=blue, linestyle=solid, linewidth=0.3mm](0.32432, 10.60000)(0.64865, 10.60000)
}\end{pspicture}
\endgroup
 
	\caption{Achievable rates depending on the network size.  Solid curves indicate fixed transmission strategy a-priori
	known to all nodes, and dashed lines indicate a random transmission schedule which is chosen randomly at each node. Results for single-hop
	are power-normalized such that the power introduced by additional relay nodes is also used in case of single-hop.}
	\label{fig:halfduplex:results:rates_over_noRelays}
      \end{figure}
      \begin{figure}
	\centering
	\begingroup
\unitlength=1mm
\psset{xunit=19.00000mm, yunit=6.00000mm, linewidth=0.1mm}
\psset{arrowsize=2pt 3, arrowlength=1.4, arrowinset=.4}\psset{axesstyle=frame}
\begin{pspicture}(1.47368, 1.33333)(6.00000, 13.00000)
\rput(-0.10526, -0.33333){%
\psaxes[subticks=0, labels=all, xsubticks=1, ysubticks=1, Ox=2, Oy=3, Dx=1, Dy=2]{-}(2.00000, 3.00000)(2.00000, 3.00000)(6.00000, 13.00000)%
\multips(3.00000, 3.00000)(1.00000, 0.0){3}{\psline[linecolor=black, linestyle=dotted, linewidth=0.2mm](0, 0)(0, 10.00000)}
\multips(2.00000, 5.00000)(0, 2.00000){4}{\psline[linecolor=black, linestyle=dotted, linewidth=0.2mm](0, 0)(4.00000, 0)}
\rput[b](4.00000, 1.66667){path loss exponent $\PathlossExponent$}
\rput[t]{90}(1.57895, 8.00000){$R$ in bits per channel use (bpcu)}
\psclip{\psframe(2.00000, 3.00000)(6.00000, 13.00000)}
\psline[linecolor=blue, plotstyle=curve, linewidth=0.3mm, showpoints=false, linestyle=solid, dotstyle=square, dotscale=1.3 1.3](2.00000, 3.45943)(2.50000, 3.45943)(3.00000, 3.45943)(3.50000, 3.45943)(4.00000, 3.45943)(4.50000, 3.45943)(5.00000, 3.45943)(5.50000, 3.45943)(6.00000, 3.45943)
\psline[linecolor=red, plotstyle=curve, linewidth=0.3mm, showpoints=true, linestyle=solid, dotstyle=o, dotscale=1.3 1.3](2.00000, 4.01428)(2.50000, 4.19805)(3.00000, 4.39471)(3.50000, 4.60171)(4.00000, 4.81690)(4.50000, 5.03851)(5.00000, 5.26512)(5.50000, 5.49563)(6.00000, 5.72921)
\psline[linecolor=red, plotstyle=curve, linewidth=0.3mm, showpoints=true, linestyle=dashed, dotstyle=o, dotscale=1.3 1.3](2.00000, 4.18097)(2.50000, 4.41583)(3.00000, 4.66231)(3.50000, 4.91588)(4.00000, 5.17306)(4.50000, 5.43141)(5.00000, 5.68941)(5.50000, 5.94617)(6.00000, 6.20132)
\psline[linecolor=darkgreen, plotstyle=curve, linewidth=0.3mm, showpoints=true, linestyle=solid, dotstyle=diamond, dotscale=1.3 1.3](2.00000, 4.16950)(2.50000, 4.35000)(3.00000, 4.54681)(3.50000, 4.75671)(4.00000, 4.97685)(4.50000, 5.20478)(5.00000, 5.43855)(5.50000, 5.67666)(6.00000, 5.91794)
\psline[linecolor=black, plotstyle=curve, linewidth=0.3mm, showpoints=false, linestyle=solid, dotstyle=square, dotscale=1.3 1.3](2.00000, 4.56593)(2.50000, 4.76886)(3.00000, 4.98361)(3.50000, 5.20761)(4.00000, 5.43864)(4.50000, 5.67492)(5.00000, 5.91506)(5.50000, 6.15800)(6.00000, 6.40297)
\psline[linecolor=red, plotstyle=curve, linewidth=0.3mm, showpoints=true, linestyle=solid, dotstyle=o, dotscale=1.3 1.3](2.00000, 4.92666)(2.50000, 5.37903)(3.00000, 5.85835)(3.50000, 6.34813)(4.00000, 6.85808)(4.50000, 7.37407)(5.00000, 7.89832)(5.50000, 8.42126)(6.00000, 8.96092)
\psline[linecolor=red, plotstyle=curve, linewidth=0.3mm, showpoints=true, linestyle=dashed, dotstyle=o, dotscale=1.3 1.3](2.00000, 5.17344)(2.50000, 5.68729)(3.00000, 6.21887)(3.50000, 6.76516)(4.00000, 7.31244)(4.50000, 7.85416)(5.00000, 8.40831)(5.50000, 8.95340)(6.00000, 9.50349)
\psline[linecolor=darkgreen, plotstyle=curve, linewidth=0.3mm, showpoints=true, linestyle=solid, dotstyle=diamond, dotscale=1.3 1.3](2.00000, 5.25542)(2.50000, 5.78830)(3.00000, 6.35745)(3.50000, 6.96787)(4.00000, 7.60939)(4.50000, 8.26889)(5.00000, 8.94044)(5.50000, 9.61900)(6.00000, 10.30290)
\psline[linecolor=black, plotstyle=curve, linewidth=0.3mm, showpoints=false, linestyle=solid, dotstyle=square, dotscale=1.3 1.3](2.00000, 6.65233)(2.50000, 7.30983)(3.00000, 7.97993)(3.50000, 8.67260)(4.00000, 9.38101)(4.50000, 10.07435)(5.00000, 10.75966)(5.50000, 11.42964)(6.00000, 12.12546)
\endpsclip
\rput(5, 9.2){\psellipse(0, 0)(0.2, 2)}
\rput(4.8, 11.4){$N=3$}
\rput(5, 5.5){\psellipse(0, 0)(0.1, 0.9)}
\rput(5.2, 4.3){$N=1$}
\psframe[linecolor=black, fillstyle=solid, fillcolor=white, shadowcolor=lightgray, shadowsize=1mm, shadow=true](2.10526, 10.33333)(4.36842, 13.33333)
\rput[l](2.57895, 12.83333){Upper bound}
\psline[linecolor=black, linestyle=solid, linewidth=0.3mm](2.21053, 12.83333)(2.42105, 12.83333)
\rput[l](2.57895, 12.16667){Compress-and-Forward}
\psline[linecolor=darkgreen, linestyle=solid, linewidth=0.3mm](2.21053, 12.16667)(2.42105, 12.16667)
\psline[linecolor=darkgreen, linestyle=solid, linewidth=0.3mm](2.21053, 12.16667)(2.42105, 12.16667)
\psdots[linecolor=darkgreen, linestyle=solid, linewidth=0.3mm, dotstyle=diamond, dotscale=1.3 1.3, linecolor=darkgreen](2.31579, 12.16667)
\rput[l](2.57895, 11.50000){Decode-and-Forward}
\psline[linecolor=red, linestyle=solid, linewidth=0.3mm](2.21053, 11.50000)(2.42105, 11.50000)
\psline[linecolor=red, linestyle=solid, linewidth=0.3mm](2.21053, 11.50000)(2.42105, 11.50000)
\psdots[linecolor=red, linestyle=solid, linewidth=0.3mm, dotstyle=o, dotscale=1.3 1.3, linecolor=red](2.31579, 11.50000)
\rput[l](2.57895, 10.83333){Single-Hop}
\psline[linecolor=blue, linestyle=solid, linewidth=0.3mm](2.21053, 10.83333)(2.42105, 10.83333)
}\end{pspicture}
\endgroup
 
	\caption{Influence of path loss on the achievable rates for $N=1$ and $N=3$ relay nodes which are distributed in equal distances between
	source and destination.  Dashed curves indicate a random transmission schedule and solid lines a fixed schedule.}
	\label{fig:halfduplex:results:path_loss_influence}
      \end{figure}
}

\end{document}